\documentclass[11pt,preprint]{aastex}

\usepackage{color}
\usepackage{natbib}
\usepackage{amsmath}
\tightenlines
\slugcomment{}

\newcommand \abs        {{\rm abs}}

\newcommand \beq        {\begin{equation}}
\newcommand \beqa	{\begin{eqnarray}}

\newcommand \cm         {\,{\rm cm}}

\newcommand \dust       {{\rm dust}}
\newcommand \eeq	{\end{equation}}
\newcommand \eeqa	{\end{eqnarray}}

\newcommand \erg	{\,{\rm erg}}

\newcommand \eV 	{\,{\rm eV}}
\newcommand \gm         {\,{\rm g}}

\newcommand \gtsim	{\gtrsim}		 
\newcommand \Ha 	{{\rm H}}

\newcommand \HH	        {{\rm H}_2}

\newcommand \K  	{\,{\rm K}}
\newcommand \kms	{\,{\rm km~s}^{-1}}
\newcommand \kpc	{\,{\rm kpc}}

\newcommand \Lsol	{L_{\odot}}
\newcommand \ltsim	{\lesssim}		 

\newcommand \Md         {M_{\rm d}}
\newcommand \MH         {M_{\rm H}}

\newcommand \MJy        {\,{\rm MJy}}

\newcommand \Msol	{M_{\odot}}

\newcommand \pc  	{\,{\rm pc}}

\newcommand \qpah       {q_{\rm PAH}}
\newcommand \s	        {\,{\rm s}}
\newcommand \sr  	{\,{\rm sr}}

\newcommand \Td         {T_{\rm d}}

\newcommand \Umin       {U_{\rm min}}

\newcommand \mm         {\,{\rm mm}}

\newcommand \Tdchar     {T_{\rm d,char}}

\newcommand \SigLd      {\Sigma_{L{\rm d}}}
\newcommand \SigLdmin   {\Sigma_{L{\rm d,min}}}
\newcommand \SigMd      {\Sigma_{M{\rm d}}}

\newcommand{\oldtext}[1]{}
\newcommand{\newtext}[1]{{#1}}
\newcommand{\newnewtext}[1]{{#1}}

\newcommand{\figwidth}{8.0cm}

\countdef\decade=200
\advance\decade by \year
\countdef\hours=201
\advance\hours by \time
\divide\hours by 60
\countdef\mins=202
\advance\mins by \hours
\multiply\mins by 60
\multiply\hours by 100
\countdef\miltime=203
\advance\miltime by \hours
\advance\miltime by \time
\advance\miltime by -\mins

\begin{document}

\title{%
        \vspace*{-3.0em}
        {\normalsize\rm {\it The Astrophysical Journal}, {\bf 780}, 172 (2014 Jan.\ 10)}\\ 
        \vspace*{1.0em}
        Andromeda's Dust
	}

\author{B.\ T.\ Draine\altaffilmark{1,2},
G.\ Aniano\altaffilmark{1,3}, 
Oliver Krause\altaffilmark{4},
Brent Groves\altaffilmark{4},
Karin Sandstrom\altaffilmark{4},
Robert Braun\altaffilmark{5},
Adam Leroy\altaffilmark{6},
Ulrich Klaas\altaffilmark{4},
Hendrik Linz\altaffilmark{4},
Hans-Walter Rix\altaffilmark{4},
Eva Schinnerer\altaffilmark{4},
Anika Schmiedeke\altaffilmark{4}, and
Fabian Walter\altaffilmark{4}}
\altaffiltext{1}{%
Princeton University Observatory, Peyton Hall, Princeton, NJ 08544-1001, USA;
draine@astro.princeton.edu}
\altaffiltext{2}{%
Osservatorio Astrofisico Arcetri, Largo E. Fermi 5, I-50125 Firenze, Italy}
\altaffiltext{3}{%
Institut d'Astrophysique Spatiale, F-91405 Orsay, France;
ganiano@ias.u-psud.fr}
\altaffiltext{4}{%
Max-Planck-Institut fur Astronomie, Konigstuhl 17, D-69117 Heidelberg, Germany}
\altaffiltext{5}{%
CSIRO -- Astronomy and Space Science, PO Box 76, Epping, NWS 1710, Australia} 
\altaffiltext{6}{%
National Radio Astronomy Observatory, 
520 Edgemont Road, Charlottesville, VA 22903, USA}

\begin{abstract}
{\it Spitzer Space Telescope} and {\it Herschel
Space Observatory} imaging of M31 is used, 
with a physical dust model, to construct maps of
dust surface density, dust-to-gas ratio, starlight heating intensity,
and polycyclic aromatic hydrocarbon (PAH) abundance,
out to $R\approx 25\kpc$.
The global dust mass is $\Md=5.4\times10^7\Msol$, 
the global dust/H mass ratio is $\Md/\MH=0.0081$,
and the global PAH abundance is 
$\langle\qpah\rangle=0.039$.
The dust surface density has
an inner ring at $R=5.6\kpc$, 
a maximum at $R=11.2\kpc$, and an outer ring
at $R\approx 15.1\kpc$.
The dust/gas ratio varies
from $\Md/\MH\approx 0.026$ at the
center to $\sim$$0.0027$ at $R\approx 25\kpc$.
From the dust/gas ratio, we estimate the interstellar mediu (ISM) metallicity 
to vary by a factor $\sim$10, from
$Z/Z_\odot\approx3$ at $R=0$ to
$\sim0.3$ at $R=25\kpc$.
The dust heating rate parameter $\langle U\rangle$ peaks at the center,
with $\langle U\rangle\approx 35$,
declining to $\langle U\rangle\approx 0.25$ at $R=20\kpc$.
Within the central kiloparsec, the starlight heating intensity inferred from
the dust modeling is close to
what is estimated from the stars in the bulge.
The PAH abundance reaches a peak $\qpah\approx 0.045$ at
$R\approx 11.2\kpc$.
When allowance is made for the different
spectrum of the bulge stars, $\qpah$ for the dust in the central kiloparsec
is similar to the overall value of $\qpah$ in the disk.
%
The silicate--graphite--PAH dust model used here
is generally able to reproduce the
observed dust spectral energy distribution across M31, but
overpredicts $500\micron$ emission at $R\approx2$--$6\kpc$, 
suggesting that at $R=2$--$6\kpc$, the dust opacity
varies more steeply with frequency (with $\beta\approx 2.3$
between 200 and 600$\micron$)
than in the model.
\end{abstract}


\section{Introduction
         \label{sec:intro}}

The Andromeda galaxy, M31, is the nearest large spiral galaxy.
At a distance $D=744\kpc$\footnote{%
     All radii, luminosities and masses in this paper have been corrected to
     $D=744\kpc$.}
\citep{Vilardell+Ribas+Jerdi+etal_2010}, 
M31 provides an opportunity
to study the dust and gas in an external star-forming 
galaxy with spatial resolution
that is surpassed only for the Magellanic Clouds.
The structure of the stellar spheroid, disk, and halo of M31 is the subject
of ongoing investigations, now being carried out using photometry of
large numbers of individual stars 
\citep[e.g.,][]{Dalcanton+Williams+Lang+etal_2012}.

The isophotal major radius is $R_{25}=95\arcmin = 20.6\kpc$\,@\,744\,kpc
\citep{deVaucoleurs+deVaucoleurs+Corwin+etal_1991}.
M31 is classified as a SA(s)b spiral
\citep{deVaucoleurs+deVaucoleurs+Corwin+etal_1991},
but the gas and dust do not conform to a
regular spiral pattern.
Images of both \ion{H}{1} 
\citep{Braun+Thilker+Walterbos+Corbelli_2009,
       Chemin+Carignan+Foster_2009}
and infrared emission
\citep{Haas+Lemke+Stickel+etal_1998,
       Barmby+Ashby+Bianchi+etal_2006,
       Gordon+Bailin+Engelbracht+etal_2006,
       Fritz+Gentile+Smith+etal_2012,
       Smith+Eales+Gomez+etal_2012}
show structure that appears as much ring-like as spiral in character.
The centers of the rings are often
offset significantly from the dynamical center of M31.\footnote{%
    For example, in model P1 of
    \citet{Corbelli+Lorenzoni+Walterbos+etal_2010}
    the center of the ring at 10--12$\kpc$ is offset from the dynamical
    center by $0.6\kpc$.}
The off-center ring-like structure has been attributed to
a nearly head-on collision with M32
\citep{Block+Bournaud+Combes+etal_2006}.


Previous studies of far-infrared (FIR) emission from the dust in M31 include
maps made with
{\it IRAS} \citep{Habing+Miley+Young+etal_1984,Devereux+Price+Wells+Duric_1994},
{\it Infrared Space Observatory} ({\it ISO})
\citep{Haas+Lemke+Stickel+etal_1998},
and
{\it Spitzer} \citep{Gordon+Bailin+Engelbracht+etal_2006}.
The total infrared luminosity was well-measured, but the
limited angular resolution of 
{\it IRAS} 60 and 100$\micron$ ($105\arcsec$ FWHM), 
{\it ISO} 175$\micron$ ($110\arcsec$ FWHM), and
{\it Spitzer} $160\micron$ ($39\arcsec$ FWHM)
allowed only a relatively coarse image of the dust distribution.

The present study takes advantage of the high sensitivity and angular
resolution of {\it Herschel Space Observatory}
\citep{Pilbratt+Riedinger+Passvogel+etal_2010} to characterize the
dust in M31 on angular scales as small as 
$\sim$$25\arcsec$ ($=90\pc\,@\,744\kpc$),
and at wavelengths as long as $500\micron$, thereby capturing the emission
from whatever cold dust may be present.
The {\it Herschel} Exploitation of Local Galaxy Andromeda 
\citep[HELGA;][]{Fritz+Gentile+Smith+etal_2012,Smith+Eales+Gomez+etal_2012} 
recently employed {\it Herschel}
imaging to study the dust distribution in M31, using single-temperature
modified blackbody fits to the $70-500\micron$ emission from each pixel.
The present study differs from HELGA in two ways.
First, we use an independent set of {\it Herschel} 
observations
\citep[][O.\ Krause et al.\ 2013, in preparation]
{Groves+Krause+Sandstrom+etal_2012},
somewhat deeper than those obtained by HELGA.
Secondly, we use a physical dust model \citep{Draine+Li_2007} to
model the spectral energy distribution (SED) 
from $6-500\micron$, and to estimate the dust mass
surface density, intensities of starlight heating the dust,
and the polycyclic aromatic hydrocarbon (PAH) abundance, 
using methods recently developed by 
\citet{Aniano+Draine+Calzetti+etal_2012} for studying the galaxies
in the KINGFISH sample \citep{Kennicutt+Calzetti+Aniano+etal_2011}.

The organization of the paper is as follows:
the observational data are described in
Section \ref{sec:data}, and
the dust-model fitting methods are outlined in Section \ref{sec:modeling}.
Results are presented in Sections \ref{sec:Ldust}--\ref{sec:pah},
where we estimate the total dust luminosity and mass,
the dust/gas ratio, the metallicity as a function of radius,
the characteristics of the starlight heating the dust, and the PAH
abundance.
Evidence for variation of the dust properties is discussed in
Section \ref{sec:dust_properties}.
The results are 
summarized in Section \ref{sec:summary}.

Appendix \ref{app:PACS_vs_MIPS} examines the inconsistency between
PACS and MIPS photometry of M31.
The starlight contribution from the
M31 bulge stars is calculated in Appendix \ref{app:bulge}.

\section{Observational Data
         \label{sec:data}}

M31 has been mapped 
using the IRAC 
\citep{Fazio+Hora+Allen+etal_2004} 
and MIPS 
\citep{Rieke+Young+Engelbracht+etal_2004}
cameras on {\it Spitzer Space Telescope}
\citep{Werner+Roellig+Low+etal_2004}.
More recently, maps have been obtained
by the PACS
\citep{Poglitsch+Waelkens+Geis+etal_2010}
and SPIRE
\citep{Griffin+Abergel+Abreu+etal_2010}
cameras on {\it Herschel}
\citep{Pilbratt+Riedinger+Passvogel+etal_2010}.

The present analysis uses IRAC data from 
\citet{Barmby+Ashby+Bianchi+etal_2006}\footnote{%
   IRAC images in bands 1--4 were multiplied by extended source
   calibration factors 0.91, 0.94, 0.66, 0.74
   \citep{Reach+Megeath+Cohen+etal_2005}.
   }
and MIPS data from
\citet{Gordon+Bailin+Engelbracht+etal_2006}.\footnote{%
   MIPS images were generated by the Mips\_enhancer v3.10 pipeline
   on 2007 Jul 3.
   }
The PACS and SPIRE imaging was done in parallel mode at medium
scan speed
(20$\arcsec \s^{-1}$) for a total time
of $\sim$24\,hr
(O.\ Krause et al.\ 2013, in preparation),
and are the same data as used by \citet{Groves+Krause+Sandstrom+etal_2012},
except for small changes in calibration.
We use the most recent calibrations of PACS and SPIRE.\footnote{
   The PACS and SPIRE images were processed by HIPE v9,
   and the Level 1 HIPE images were then processed by
   Scanamorphos v18.0 \citep{Roussel_2013}.
   We used the calibration files in HIPE v9 (version 42 for PACS,
   and version 10.0 for SPIRE).
   Intensities in the SPIRE bands were obtained by
   dividing the HIPE v9 flux density per beam by
   effective beam solid angles $\Omega=(1.103,1.944,4.183)\times10^{-8}\sr$
   for SPIRE250, 350, and 500, as recommended by
   \citet{Griffin+North+Schulz+etal_2013}.
   }
We refer to the different images by the camera name and nominal
wavelength in microns: IRAC3.6, IRAC4.5, IRAC5.8, IRAC8.0, MIPS24, MIPS70,
MIPS160, 
PACS70, PACS100, PACS160, SPIRE250, SPIRE350, and SPIRE500.

The center of M31 is located at $\alpha_{{\rm J}2000}=10.685^\circ$,
$\delta_{{\rm J}2000}=41.269^\circ$
\citep{Crane+Dickel+Cowan_1992}, or
Galactic coordinates
$\ell=121^\circ,b=-22^\circ$.
Because M31 is only 22$^\circ$ from the Galactic plane,
removal of the Galactic foreground ``cirrus'' emission 
is challenging, particularly in view of the large angular extent
of M31.
We are helped by
the high inclination $i\approx78^\circ$ of M31, which 
raises the surface brightness, and increases the contrast with
foreground and background emission.

Subtraction of foreground and background emission 
has been carried out following
methods described in \citet{Aniano+Draine+Calzetti+etal_2012}, with 
automatic identification of background pixels and fitting of a
``tilted plane'' background model (with three parameters -- zero point,
tilt, and tilt orientation) for all bands except IRAC5.8 and IRAC8.0.
For IRAC5.8 and IRAC8.0, it was found that the simple ``tilted plane''
background model left obvious large-scale residuals, presumably
a consequence of the large angular extent of M31.
For IRAC5.8 and IRAC8.0, it was found necessary to use more
complex curved surfaces,
rather than simple tilted planes, to model the background.
Even so, the background estimation for IRAC5.8 and IRAC8.0 seems
to be less successful than for other bands, for reasons that are 
further discussed in Section \ref{sec:pah}.

The foreground cirrus has significant structure on $\sim$1$^\circ$
scales
(see, e.g., the {\it IRAS} $100\micron$ 
or LAB 21$\cm$ map of the area around M31),
and the simple tilted plane model (or low-order curved surface in the case
of IRAC5.8 and IRAC8.0) will not remove all of the foreground cirrus.
This will be problematic in the low surface brightness outer regions
of M31.
However, after the foreground- and 
background-subtracted FIR images have been fitted by
dust models, comparison of the resulting dust map with maps of
the \ion{H}{1} at the radial velocity of M31 will allow us to assess
the effects of imperfect foreground subtraction.
Except for IRAC5.8 and IRAC8.0, the present tilted plane 
foreground model appears adequate for
$R\ltsim25\kpc$ (see Section \ref{sec:Mdust}).

\begin{figure}[t]
\begin{center}
\includegraphics[width=\figwidth,angle=0,
                 clip=true,trim=0.6cm 0.6cm 1.6cm 8.1cm]
                {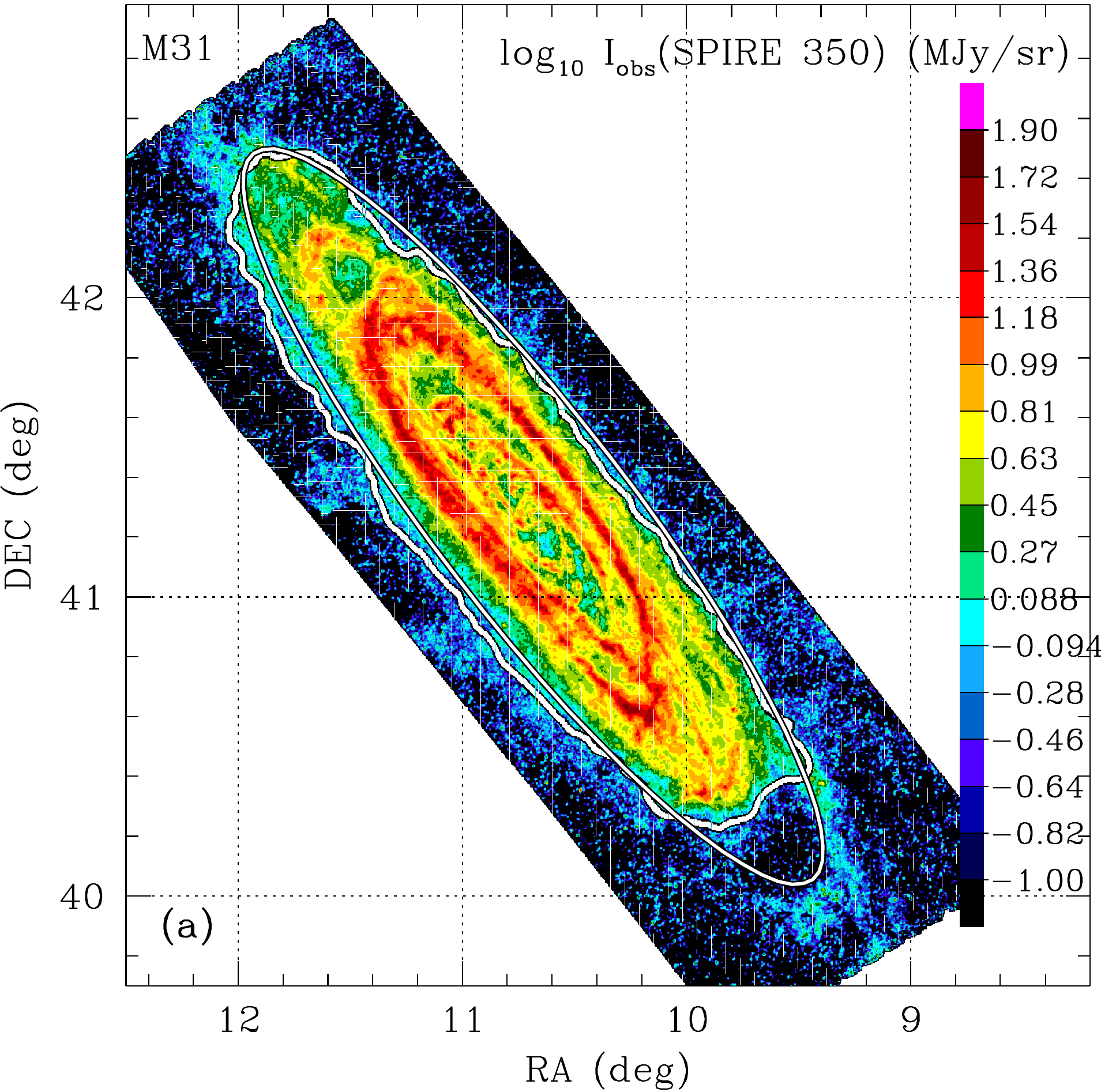}
\includegraphics[width=\figwidth,angle=0,
                 clip=true,trim=0.6cm 0.6cm 1.6cm 8.1cm]
                 {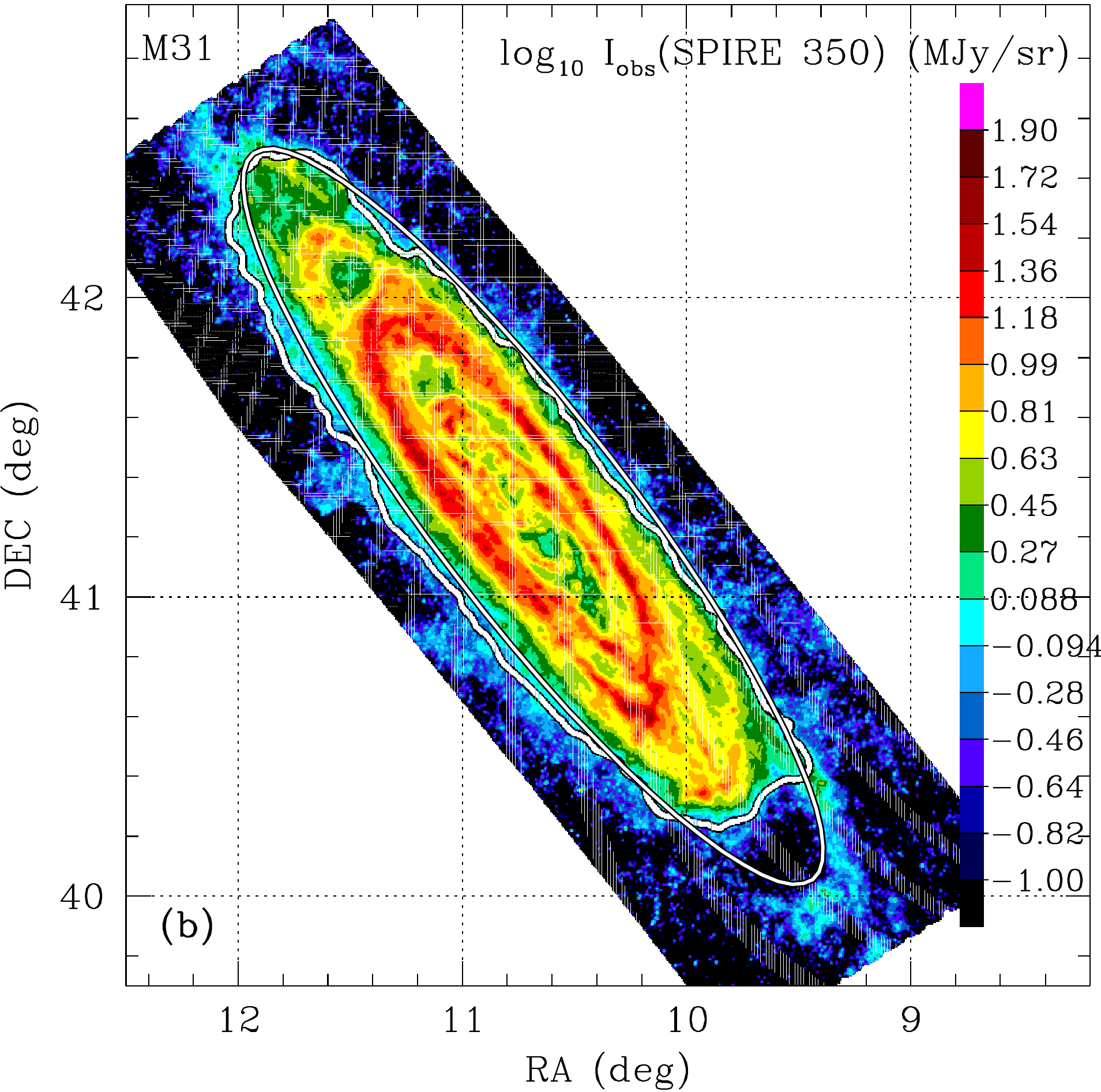}
\caption{\label{fig:S350}
         \footnotesize
         Background-subtracted $350\micron$ images of M31.
         Left: native resolution of SPIRE350.
         Right: SPIRE350 image convolved to the MIPS160 PSF.
         It is seen that most of the structure visible at S350
         resolution remains at M160 resolution, with improved signal/noise
         in low surface brightness regions.
         The contour delineates the ``galaxy mask'': the
         region within which the signal-to-noise ratio (S/N) 
         is high enough that dust modeling
         can be done pixel-by-pixel (see text).
         The ellipse shown is a tilted circle of radius 
         $R=5500\arcsec = 19.8\kpc$\,@\,$744\kpc$
         centered on the
         dynamical center, with inclination $i=77.7^\circ$ and 
         ${\rm P.A.}=37.7^\circ$ \citep{Corbelli+Lorenzoni+Walterbos+etal_2010}.
        }
\end{center}
\end{figure}

For modeling the dust in resolved systems, it is essential that
multiband imaging be convolved to a common point spread function (PSF).
The present study is carried out at two
angular resolutions: that of the SPIRE350 camera
(FWHM=$24.9\arcsec=90\pc\,@\,744\kpc$, henceforth referred to as 
{\it S350 resolution}) and
that of the MIPS160 camera (FWHM=$39\arcsec=141\pc\,@\,744\kpc$, 
henceforth referred to
as {\it M160 resolution}).
Image convolutions were carried out using kernels obtained as
described by \citet{Aniano+Draine+Gordon+Sandstrom_2011}.
For studies at S350 resolution, we are unable to use
SPIRE500 or MIPS160 imaging, because the PSF of those cameras is
too broad.
The convolved images at S350 resolution were sampled with 
$10\arcsec\times10\arcsec$ pixels.
Images convolved to M160 resolution were sampled with
$18\arcsec\!\times\!18\arcsec$ pixels.

Figure \ref{fig:S350} shows SPIRE350 images of M31; the left image
is at S350 resolution; the image on the right
is convolved to M160 resolution.  The lower resolution of M160
is evident when the images are compared, but most details of the image that
are seen at S350 resolution are also identifiable at M160 resolution.
Analysis of M31 at M160 resolution -- benefiting from the improved
signal/noise of the larger M160 pixels as well as being able to use
both SPIRE500 and MIPS160 data to constrain the models -- will, therefore,
not lose important structural features.

M31 is highly inclined.
Using DRAO \ion{H}{1} 21cm data,
\citet{Chemin+Carignan+Foster_2009} estimated an average inclination
$i=74.3^{\circ}$ and ${\rm P.A.}=37.7^\circ$ for the disk at $R=6-27\kpc$. 
From WSRT \ion{H}{1} 21cm observations,
\citet{Corbelli+Lorenzoni+Walterbos+etal_2010} obtain an inclination
$i=77.7^\circ$ and
position angle ${\rm P.A.}=37.7^\circ$ for $R=10-13\kpc$.
\citet{Corbelli+Lorenzoni+Walterbos+etal_2010} showed
that the \ion{H}{1} kinematics can be described using
circular rings with centers offset
from the dynamical center of M31,
and with inclinations between 75$^\circ$ and 79$^\circ$
for $10<R<25\kpc$).

We will assume the disk to have a single inclination $i=77.7^\circ$ and
position angle ${\rm P.A.}=37.7^\circ$.
Figure \ref{fig:S350} shows ellipses with major radius 5500$^{\prime\prime}$
corresponding to $R=19.8\kpc$ at $D=744\kpc$, with
$i=77.7^\circ$ and ${\rm P.A.}=37.7^\circ$.
Radial trends will be studied by averaging in annuli centered on
the location of the presumed supermassive black hole,
$\alpha_{{\rm J}2000}=10.6847^\circ$, $\delta_{{\rm J}2000}=41.2690^\circ$
\citep{Crane+Dickel+Cowan_1992}.

The S350 PSF provides excellent spatial resolution
(FWHM = 90\,pc along the major axis).  We will compare results obtained
at S350 resolution with results obtained using the M160 PSF.
While the dust models at M160 resolution smooth out dust structures
smaller than $\sim$$140\pc$ along the major axis (or $\sim$$700\pc$ along
the minor axis), working at M160 resolution permits (1) extending the
wavelength coverage to $500\micron$ (by including SPIRE500 photometry),
and (2) using MIPS160 photometry, which is somewhat deeper than the PACS160
imaging.
In addition,
the larger pixels
provide improved signal/noise in low surface brightness regions.
At S350 resolution, the models are constrained by photometry in
11 bands (4 IRAC, 2 MIPS, 3 PACS, 2 SPIRE).  At M160 resolution,
MIPS160 and SPIRE500 are added, giving a total of 13 photometric
constraints for each pixel.
Comparison of model results at S350 and M160 resolution
will provide insight into the reliability of the method.

\section{Modeling the Dust and Starlight Heating
         \label{sec:modeling}}

\subsection{Dust Model}

We model the dust in M31 using the dust models of
\citet[][hereafter DL07]{Draine+Li_2007}.
The dust is assumed to be a mixture of carbonaceous
grains and amorphous silicate grains with a size distribution that is
consistent with studies of the wavelength-dependent extinction and infrared
emission produced by dust in the diffuse interstellar medium (ISM) 
in the solar neighborhood
\citep[][hereafter WD01]{Weingartner+Draine_2001a}.
We treat the PAH abundance as variable, proportional to the
parameter $\qpah$, 
\newtext{where $\qpah$ is}
defined to be the fraction of the total dust mass
that is contributed by PAHs containing fewer than $10^3$ C atoms.

The optical properties of the
dust mixture
are based on an adopted grain size distribution, dielectric functions
for the graphite and silicate materials, and adopted absorption cross sections
per C atom for the PAHs.
Heat capacities are also required to calculate the temperature distribution
function for the smaller carbonaceous and silicate grains
\newtext{\citep{Draine+Li_2001}}.

By fitting the dust model to the extinction curve per H
in the solar neighborhood, the grain
volume per H is constrained.
We assume the size distribution for the local Milky Way dust
with $R_V=3.1$ obtained by
WD01, but with all grain abundances relative
to H lowered by
the factor 0.93 recommended by
\citet{Draine_2003a}.

\begin{table}[h]
\begin{center}
\caption{\label{tab:dustvol}
         Grain Mass per H in the Solar Neighborhood}
\begin{tabular}{l c c c}
\hline\\
Material & $V (\cm^3/\Ha)$ & $\Md/\MH$ & $\Md/\MH$        \\
         & DL07            & DL07      & Renormalized DL07\\
\hline
Amorphous silicate   & $3.68\times10^{-27}$ & 0.00836$^a$ & 0.00682$^b$ \\
Carbonaceous          & $2.11\times10^{-27}$ & 0.00277$^c$ & 0.00226$^d$ \\
Total                & $5.79\times10^{-27}$ & 0.0111\newtext{$^{a,c}$} 
                                                           & ~0.00908$^{b,d}$ \\
Observed toward $\zeta$Oph &                 &             & ~0.0091$^e$~~~~\\
\hline
\multicolumn{4}{l}{$^a$ For $\rho_{\rm sil}=3.8\gm\cm^{-3}$.}\\
\multicolumn{4}{l}{$^b$ For $\rho_{\rm sil}=3.1\gm\cm^{-3}$.}\\ 
\multicolumn{4}{l}{$^c$ For $\rho_{\rm carb.}=2.2\gm\cm^{-3}$.}\\
\multicolumn{4}{l}{$^d$ For $\rho_{\rm carb.}=1.8\gm\cm^{-3}$.}\\
\multicolumn{4}{l}{$^e$ From Table 23.1 of \citet{Draine_2011a}.}\\
\hline
\end{tabular}
\end{center}
\end{table}
\begin{table}[h]
\begin{center}
\caption{\label{tab:renormalized_DL07}
         Some Properties of Renormalized DL07 Dust Model}
\begin{tabular}{l c c c}
\hline\\
Property & Silicates & Carbonaceous & {\bf Total} \\
\hline
$A_V/\SigMd$ (mag/($\Msol\pc^{-2}$)) & ~5.403~ & 13.39~~ & ~7.394~ \\
$\kappa(60\micron)$  ($\cm^2\gm^{-1}$) & 102.5~~~ & 106.3~~~ & 103.4~~~ \\
$\kappa(70\micron)$  ($\cm^2\gm^{-1}$) & ~73.62~~ & ~72.24~~ & ~73.28~~ \\
$\kappa(100\micron)$ ($\cm^2\gm^{-1}$) & ~34.59~~ & ~29.02~~ & ~33.20~~ \\
$\kappa(160\micron)$ ($\cm^2\gm^{-1}$) & ~13.11~~ & ~10.69~~ & ~12.51~~ \\
$\kappa(250\micron)$ ($\cm^2\gm^{-1}$) & ~~5.263~ & ~~3.981~ & ~~4.941~ \\
$\kappa(350\micron)$ ($\cm^2\gm^{-1}$) & ~~2.448~ & ~~2.060~ & ~~2.351~ \\
$\kappa(500\micron)$ ($\cm^2\gm^{-1}$) & ~~1.198~ & ~~1.062~ & ~~1.164~ \\
$\kappa(850\micron)$ ($\cm^2\gm^{-1}$) & ~~0.4865 & ~~0.4202 & ~~0.4699 \\
$\kappa(1.28\mm)$    ($\cm^2\gm^{-1}$) & ~~0.2224 & ~~0.1866 & ~~0.2135 \\
$\kappa(2.10\mm)$    ($\cm^2\gm^{-1}$) & ~~0.1130 & ~~0.0914 & ~~0.1076 \\
\hline
\multicolumn{4}{l}{$\kappa=$ absorption cross section per unit dust mass}
\end{tabular}
\end{center}
\end{table}

\subsection{\label{sec:renormalize_Md}
            Renormalization of the DL07 Dust Mass}

The dust volume/H in the DL07 dust model 
is determined by fitting the observed extinction in the
solar neighborhood; 
the dust mass/H can then be calculated if solid densities are assumed
for the dust materials.
For the silicates, DL07 assumed a density $\rho=3.8\gm\cm^{-3}$,
and for the carbonaceous grains the density contributed by
carbon alone was taken to be
$\rho_{\rm C}=2.2\gm\cm^{-3}$.
With these adopted densities, the DL07 dust model that reproduces the
extinction per H in the solar neighborhood
has $\Md/\MH=0.0112$ (see Table \ref{tab:dustvol}).
The dust mass estimates in \citet{Aniano+Draine+Calzetti+etal_2012}
are based on the original DL07 value
$\Md/\MH=0.0112$ for dust in the solar neighborhood.
 
However, as noted by \citet{Draine+Dale+Bendo+etal_2007},
this dust mass/H is in mild 
conflict with elemental abundances in the 
\newtext{solar-neighborhood}
ISM: the total mass of
elements ``depleted'' from the gas in the diffuse
\oldtext{interstellar medium}
\newtext{ISM}
is estimated to be less
than what would be calculated for the
WD01 dust model with the above solid densities.
For the well-studied sightline toward $\zeta$ Oph,
the observed depletions imply 
$\Md/\MH=0.0091\pm0.0006$
\citep[][Table 23.1]{Draine_2011a}.\footnote{%
   This value involves
   an assumption about the oscillator strength for \ion{C}{2}]2325\AA\ and
   a second assumption about the depletion of oxygen -- 
   see \citet{Draine_2011a}.}

In the present paper we will assume solar neighborhood dust to have
$\Md/\MH=0.0091$.
This can be reconciled with the volume of grain material required
by the WD01 extinction model if the grain solid densities are reduced by
a factor $\sim$0.81.
Here we define the ``renormalized DL07'' model to be the same
grain size distributions (i.e., same volume of grain material) 
and same dielectric functions as used by DL07, but with the
silicate density $\rho_{\rm sil}$ reduced
from $3.8\gm\cm^{-3}$ to $3.1\gm\cm^{-3}$, and the density of 
carbonaceous grain material reduced from $2.1\gm\cm^{-3}$ to $1.8\gm\cm^{-3}$.
Table \ref{tab:renormalized_DL07} gives $A_V/\SigMd$ for the
renormalized DL07 model, as well as opacities at selected FIR
wavelengths.

With this renormalization,
all dust masses and dust/H ratios in 
\citet{Aniano+Draine+Calzetti+etal_2012}
should be reduced by 
a factor $f_M=0.00908/0.01113=0.816$.
In this renormalized model for dust in the diffuse ISM, 
75.1\% of the dust mass is provided by
the amorphous silicates, and 24.9\% by the carbonaceous component (PAHs
included).


\subsection{Dust Masses and Starlight Intensities}

\newtext{
The present study seeks to estimate dust masses in M31 by modeling
the observed FIR and submillimeter emission.  The dust is assumed
to be heated by starlight, and the modeling has the
freedom to adjust the starlight intensities such that the grains
are heated to temperatures such that their emission spectrum
is consistent with the observed shape of the SED.
Then, with the dust temperatures set by the starlight heating rates, 
the dust mass is proportional to the observed emission.
Note that there is no single temperature characterizing the dust -- 
even in a single radiation field,
grains of different size and composition have different temperatures,
and the very small grains undergo temperature fluctuations due to single-photon
heating.
Additionally, a single ``pixel'' -- which may be several hundred parsecs in
transverse dimension -- may include subregions with different starlight
intensities.
}

\newnewtext{For a given starlight intensity $U$, grain size $a$ and
composition, we solve for the temperature probability distribution
$(dP/dT)_{U,a,{\rm comp}}$.  For large grains $dP/dT$ can be approximated by
a $\delta$-function, but for small grains $dP/dT$ can be very broad, and
must be solved for as described by \citet{Li+Draine_2001b}.
The time-averaged emission spectrum for a grain is then
\beq
(p_\nu)_{U,a,{\rm comp}} = 
4\pi \int dT \, (dP/dT)_{U,a,{\rm comp}} \, 
(C_{\abs,\nu})_{a,{\rm comp}}
B_\nu(T)
~~~,
\eeq
where $(C_{\abs,\nu})_{a,{\rm comp}}$ is the absorption cross section at
frequency $\nu$.
The emission per unit mass for a dust mixture exposed to starlight $U$ is
\beq
\left(\frac{dL_\nu}{dM_d}\right)_U
=
\frac{\sum_{\rm comp} \int da 
(dn/da)_{\rm comp} (p_\nu)_{U,a,{\rm comp}}}
{\sum_{\rm comp}\int da (dn/da)_{\rm comp} 
(4\pi/3)a^3 \rho_{\rm comp}}
~~~,
\eeq
where $\rho_{\rm comp}$ is the solid density, and
$(dn/da)_{\rm comp}$ is the size distribution.  The luminosity of
a region $j$ is obtained by summing over the starlight distribution:
\beq
L_{\nu,j} = \int dU \frac{dM_{{\rm d},j}}{dU}
\left(\frac{dL_\nu}{dM_d}\right)_U
~~~.
\eeq%
}%
For a region $j$ of solid angle $\Omega_j$,
\newtext{with dust mass $M_{{\rm d},j}=\Sigma_{M{\rm d},j}\Omega_j D^2$},   
the dust mass $dM_{{\rm d},j}$ exposed to starlight intensities
in $[U,U+dU]$ is assumed to be given by the
simple parameterization proposed by DL07:
\beq \label{eq:dM/dU}
\frac{dM_{{\rm d},j}}{dU} = \Sigma_{M{\rm d},j}\Omega_j D^2
\left[ (1-\gamma_j)\delta(U-U_{{\rm min},j}) + 
\gamma_j \frac{(\alpha_j-1)U^{-\alpha_j}}{U_{{\rm min},j}^{1-\alpha_j}
-U_{\rm max}^{1-\alpha_j}}
\right] ~~~,
\eeq
\oldtext{for $\alpha_j>1$,} 
where $\Sigma_{M{\rm d},j}$ 
is the total dust mass surface density
in region $j$, \oldtext{and} $\delta$ is the Dirac $\delta$-function,
\newtext{and $\alpha_j>1$ is a power-law index characterizing the
distribution of starlight intensities}.
A fraction $(1-\gamma_j)$ of the dust mass is heated by starlight intensity
$U=U_{{\rm min},j}$, with the remaining fraction $\gamma_j$ exposed to
starlight with intensities $U_{{\rm min},j}<U\leq U_{\rm max}$,
with a power-law distribution $dM_{\rm d}/dU\propto U^{-\alpha_j}$.
For NGC~628 and NGC~6946, 
\citet{Aniano+Draine+Calzetti+etal_2012} found that the parameter
$U_{\rm max}$ could be fixed at $U_{\rm max}=10^7$ without significantly
degrading the quality of the fits, hence we also fix $U_{\rm max}=10^7$.
Thus, for each region $j$,
we have five adjustable parameters characterizing the dust:
$\{\Sigma_{M{\rm d},j},q_{{\rm PAH},j},U_{{\rm min},j},\alpha_j,\gamma_j\}$.\newtext{\footnote{\newtext{Maps of the best-fit values of
$\Sigma_{M{\rm d},j}$, $q_{{\rm PAH},j}$, $U_{{\rm min},j}$, $\alpha_j$, and
$\gamma_j$ at M160 and S350 resolution are available from
http://www.astro.princeton.edu/$\sim$draine/m31dust}
}}
We require $\Sigma_{M{\rm d},j}\geq0$, $0\leq\gamma_j\leq1$, and
$1<\alpha_j<3$.
We use one additional parameter to characterize the contribution of
direct starlight (taken to have the spectrum of a 5000K blackbody) to the
photometry for pixel $j$.  
Thus each region has 6 adjustable parameters, and either 11 or
13 data, depending on whether we use S350 or M160 resolution.
Because MIPS70 and PACS70 cover essentially the same wavelengths,
and likewise for MIPS160 and PACS160, the effective number of
constraints is 10 (at S350 resolution) or 11 (at M160 resolution).

The fitting procedure assumes the DL07 model dust to be heated by
starlight with the solar-neighborhood spectrum found by
\citet{Mathis+Mezger+Panagia_1983}.
In Sections \ref{sec:starlight} and \ref{sec:pah} we examine
the effects of changes in the starlight
spectrum when the bulge stars make a significant contribution to the
dust heating.

\subsection{Systematic Uncertainties in Dust Mass Estimation}

\oldtext{
The present study seeks to estimate dust masses in M31 by modeling
the observed far-infrared and submm emission.  The modeling has the
freedom to adjust the starlight intensities such that the grains
are heated to temperatures consistent with the shape of the SED.
Then, with the dust temperature set, the dust mass is proportional
to the observed emission.%
}
\newtext{The model SED depends on the modeled distribution of
dust temperatures {\it and} on the wavelength-dependence of the
dust opacity $\kappa_\nu$.}
The wavelength-dependence of \oldtext{the model dust opacity}
\newtext{$\kappa_\nu$}
is tested by whether the model can reproduce the observed
shape of the SED, but if \oldtext{the model dust opacity}
\newtext{$\kappa_\nu$} is in error by some
constant factor $A$, the derived dust masses will be off by
a factor $1/A$.

The DL07 dust model uses adopted dielectric functions for
the amorphous silicate and carbonaceous grains which, with the assumption of
spherical shape, allows absorption cross sections to be calculated.
For solar-neighborhood abundances, 
the absorption cross section per H is given in Table \ref{tab:DL07_vs_Planck}
for $\lambda=100\micron$, $250\micron$, and $500\micron$.

\citet{Planck_dust_2011} examined the FIR
emission in the diffuse ISM, finding that
the SED for low-velocity \ion{H}{1} was consistent with
$\kappa\propto\nu^\beta$, with $\beta\approx 1.8$,
and $\tau(250\micron)/N_\Ha=(1.0\pm0.3)\times10^{-25}\cm^2/\Ha$.
As seen in Table \ref{tab:DL07_vs_Planck}, the {\it Planck}
estimate for $\tau/N_\Ha$ agrees well with the DL07 value at $100\micron$,
but at $250$ and $500\micron$ the DL07 opacities are smaller, by
a factor 1.3 at $250\micron$, and a factor 1.5 at $\lambda=500\micron$.
However, we will see below (Section \ref{sec:dust_properties})
that the shape of the M31 SED seems more consistent with the
$\beta\approx 2.1$ of the DL07 model than the $\beta\approx1.8$
favored by \citep{Planck_dust_2011}.

If the {\it Planck} opacities are correct, then modeling the FIR emission
using the DL07 model will tend to overestimate the dust mass.
The present study does not use data longward of $500\micron$, so
we would not expect this error to be larger than a factor $\sim$1.3 (the ratio
of the opacities at 250$\micron$), but the possibility of a systematic
error in the dust mass estimate should be kept in mind.
Future studies employing the full {\it Planck} intensity data 
can be expected to
shed further light on dust opacities 
\citep{Planck_dust_gal_plane_2013,
       Planck_overview_2013,
       Planck_component_sep_2013}.

\begin{table}[h]
\begin{center}
\caption{\label{tab:DL07_vs_Planck}
         Far-infrared Absorption/H}
\begin{tabular}{l c c}
\hline
                   & \multicolumn{2}{c}{$\tau(\lambda)/N_\Ha  ~~(10^{-25}\cm^2/\Ha)$}\\
$\lambda(\micron)$ & DL07 Model & Planck Observations$^a$ \\
\hline
$100$       & 5.0    & $5.2\pm1.6\,^b$ \\
$250$       & 0.75   & $1.0\pm0.3$ \\
$500$       & 0.18   & $0.28\pm0.08\,^b$ \\
\hline
\multicolumn{3}{l}{$^a$~~\citet{Planck_dust_2011}.}\\
\multicolumn{3}{l}{$^b$~~Obtained from $\tau(250\micron)/N_\Ha$
assuming $\beta=1.8$}\\
\hline
\end{tabular}
\end{center}
\end{table}

\subsection{Single Pixels versus Annuli}

Determination of the dust {\it mass}
requires that the shape of the dust SED be well-determined, so that
the dust temperatures can be constrained.
With the signal-to-noise properties of the present observations, 
we do not attempt to 
estimate the dust mass in a single $10\arcsec\!\times\!10\arcsec$
pixel unless
the dust luminosity per unit area (on the sky plane)
$\SigLd>\SigLdmin=1.5\!\times\!10^6\Lsol\kpc^{-2}$
(IR intensity $I > 4.8\times10^{-5}\erg\s^{-1}\cm^{-2}\sr^{-1}$).
The irregular contour in Figure \ref{fig:S350} and the other images
bounds the contiguous region with $\SigLd>\SigLdmin$
(there are additional pixels with $\SigLd>\SigLdmin$ 
outside this contour).\footnote{%
   The boundary contour is established by modeling at M160 resolution
   (18\arcsec pixels),
   using all 13 bands to estimate $\SigLd$.  When modeling at higher
   resolution (e.g., S350, with 10\arcsec pixels) we continue to
   use the contours established at M160 resolution to define the
   ``galaxy mask''.}
For $\SigLd\ltsim\SigLdmin$ the single-pixel dust mass
estimates are quite uncertain, especially at S350 resolution where
the pixels are smaller, and SPIRE500 and MIPS160 cannot be used.
At M160 resolution
the larger ($18\arcsec\!\times\!18\arcsec$) pixels and inclusion of
SPIRE500 and MIPS160 help, but even at M160 resolution the single-pixel 
dust mass estimates near $\SigLdmin=1.5\!\times\!10^6\Lsol\kpc^{-2}$ 
have substantial uncertainties.

To study radial dependences, we will sometimes take the
dust properties obtained by single-pixel modeling and average them
over annuli with widths $10\arcsec$ along the minor axis,
and $\Delta R=46.9\arcsec$ ($169\pc$) along the major axis.
While these annuli are not resolved along the minor axis (the MIPS160
FWHM=39\arcsec), they are resolved along the major axis.\footnote{%
   If the center of a pixel falls within an annulus, we include the
   entire pixel in the sums for that annulus -- we do not divide
   pixels that overlap annular boundaries.}

In order to study the dust at $R>17\kpc$, where a large fraction of
the pixels have $\SigLd<\SigLdmin$ (see Figure \ref{fig:Ldust}(c)),
we use the
background-subtracted images to extract the flux in each band from
larger concentric annuli with widths $40\arcsec$ along the minor axis, and
$\Delta R=188\arcsec$ ($677\pc$) along the major axis.
These $\Delta R=677\pc$ annuli are fully resolved, even by MIPS160.
The dust mass and starlight intensity distribution in each
annulus are then estimated by fitting the SED of the annulus.
By integrating over the many pixels in each annulus, the 
random noise is substantially suppressed.
We will see that 
reliable estimates of the mean dust mass surface density
are possible out to $R\approx25\kpc$.

\section{Dust Luminosity
         \label{sec:Ldust}}

\begin{figure}[t]
\begin{center}
\includegraphics[width=\figwidth,angle=0,
                 clip=true,trim=0.6cm 0.6cm 1.6cm 8.1cm]
                {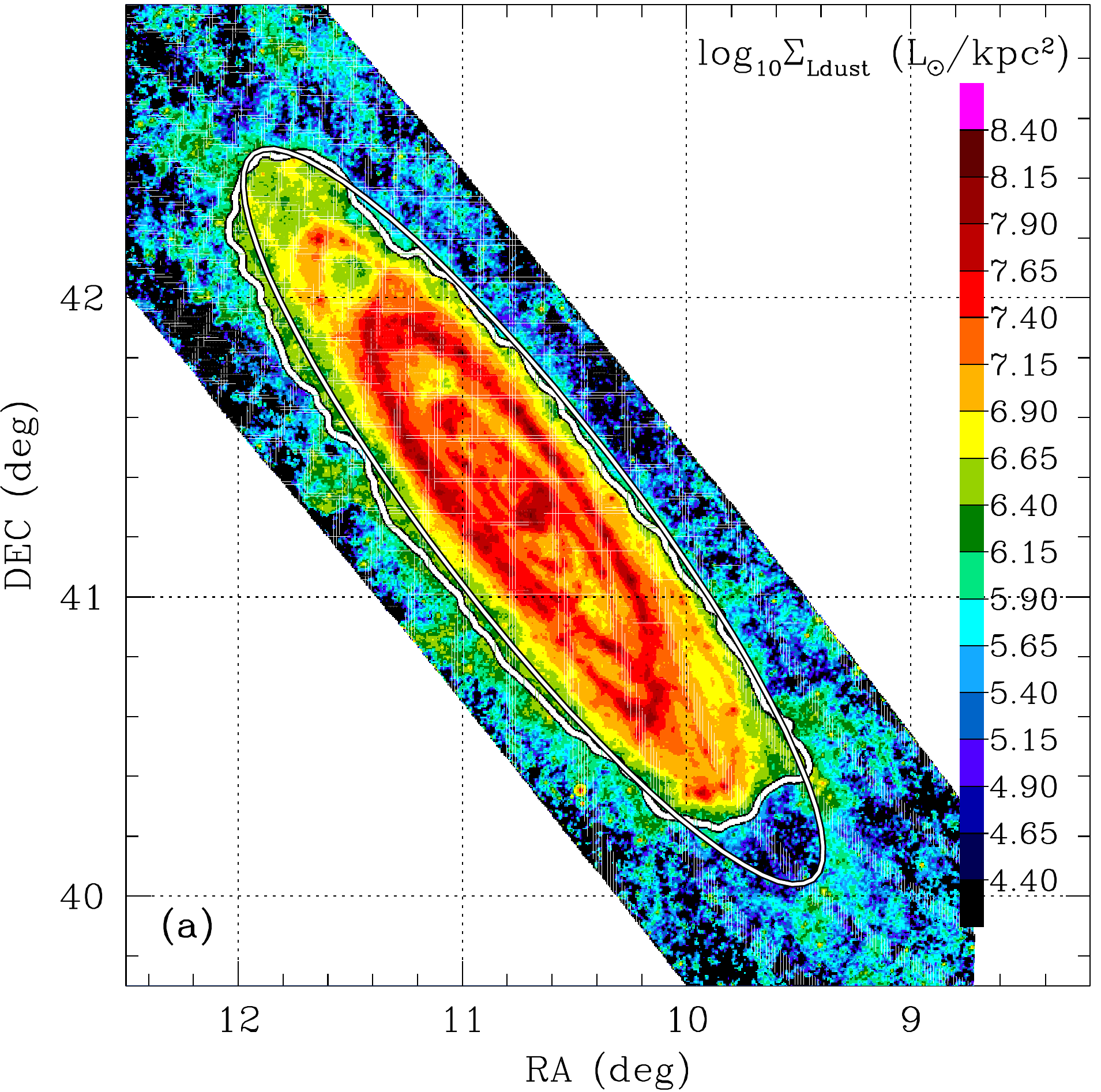}
\includegraphics[width=\figwidth,angle=0,
                 clip=true,trim=0.5cm 4.5cm 0.5cm 2.8cm]
                {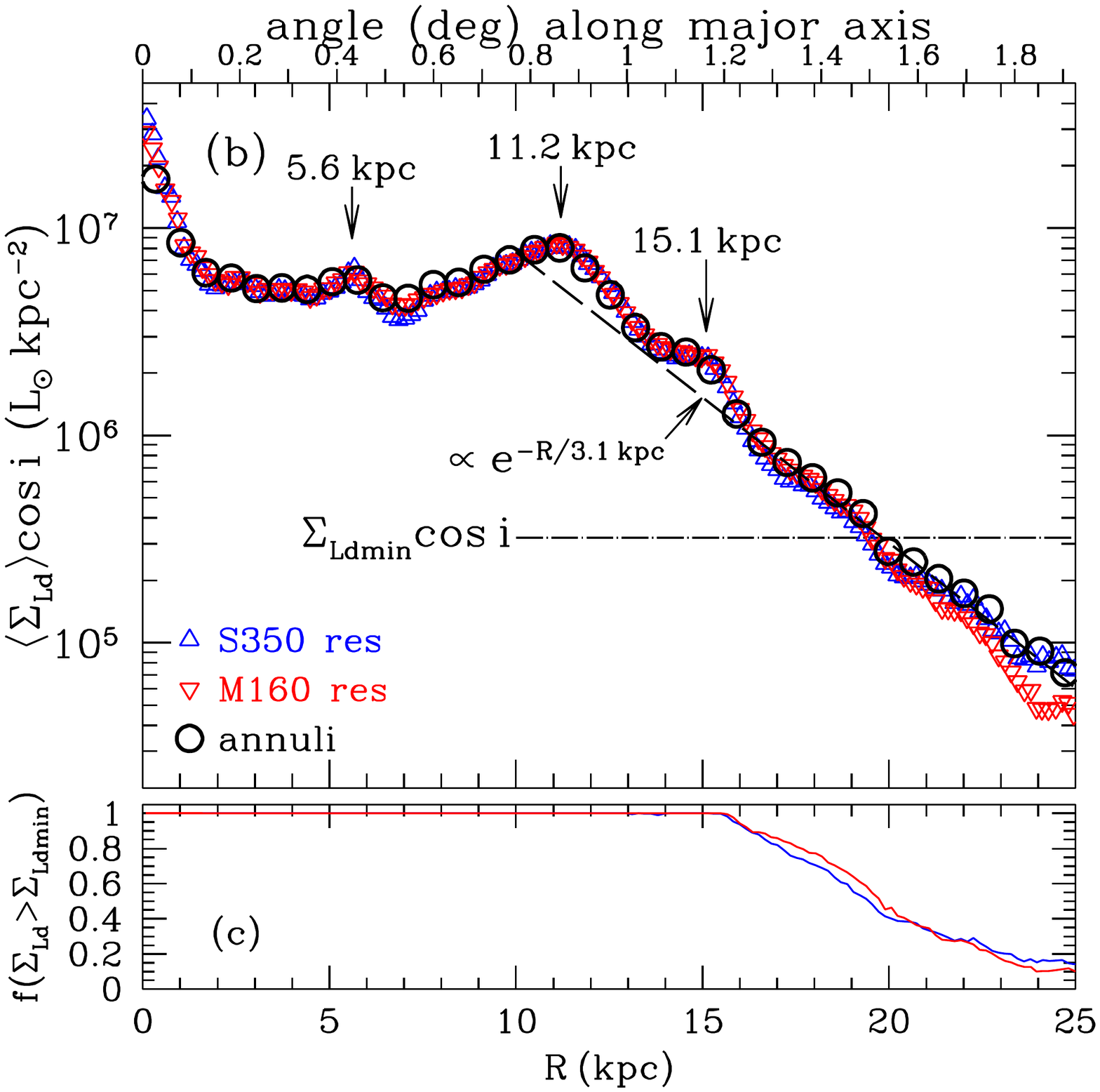}

\caption{\label{fig:Ldust}
         \footnotesize
         (a) Dust luminosity per area on the sky plane $\SigLd$ at M160
         resolution.
         (b) 
         $\SigLd$ as a function of radius $R$.
         Triangles: \oldtext{and squares:} 
         annular average of single-pixel modeling 
         at S350 \oldtext{resolution} and M160 resolution. 
         \oldtext{, respectively.}
         Circles: $\SigLd$ based on modeling of SED of $\Delta R=677\pc$
         annuli, using all photometric bands.
         Horizontal dot-dash line: surface brightness below which
         we do not attempt to model individual pixels.
         Dashed line: exponential fit (\ref{eq:SigLdfit})
         for $R\gtsim 10\kpc$, with $3.1\kpc$ scale length.
         (c) Fraction of pixels at each $R$ that have
         $\SigLd > \SigLdmin$.  The drop below 100\% for
         $R > 15.6\kpc$ is due to the low $\SigLd$ area in the SW.
        }
\end{center}
\end{figure}

As described in \citet{Aniano+Draine+Calzetti+etal_2012}, the
background-subtracted multiwavelength images are first 
used to estimate $\SigLd$, 
the total dust luminosity per projected
area on the plane of
the sky.
The luminosity is obtained by fitting a DL07 dust model to the
observed fluxes from the pixel, and then integrating over the model SED.
While the model dust mass in individual pixels
is unreliable for $\SigLd\ltsim\SigLdmin$, the
pixel luminosity estimate is reliable down to considerably
lower surface brightness.
Figure \ref{fig:Ldust}(a) shows the dust luminosity surface
density $\SigLd$ at M160 resolution.
From the image it can be seen by eye that the dust luminosity is
detectable \oldtext{at least} down to $\SigLd\approx 10^{5.9}\Lsol\kpc^{-2}$.

The mean luminosity per area as a function of radius is found
by averaging the single-pixel luminosities over the
$\Delta R=177\pc$ annuli.
Figure \ref{fig:Ldust}(b) shows $\SigLd\cos i$,
the dust luminosity surface density on the disk plane,
averaged over the $\Delta R=177\pc$ annuli,
and plotted against the annular radius $R$.
Triangles show \newtext{annular averages of single-pixel} 
results obtained at M160 resolution (using all cameras)
and at S350 resolution (not using SPIRE500 or MIPS160).
\newtext{Circles show results of modeling the SED of $\Delta R=677\pc$
annuli.}
The smooth decline in $\SigLd$ with increasing $R$
out to $R\approx23\kpc$
is an indication that background subtraction has been reasonably successful
at least down to surface brightnesses 
$\SigLd\approx4\times10^5\Lsol\kpc^{-2}$
($\SigLd\cos i\approx 8\times10^4\Lsol\kpc^{-2}$).

Figure \ref{fig:Ldust}(c) shows the 
fraction $f$ of the pixels
at each $R$ that have $\SigLd>\SigLdmin$.
The fraction $f=1$ for $R<15\kpc$, but at $R\approx15.6\kpc$ the fraction
begins to fall.  At $R\approx24\kpc$ the fraction has fallen to
$f\approx \oldtext{0.08}\newtext{0.1}$; 
some fraction of these pixels may be raised
above $\SigLdmin$ by unresolved background galaxies.

The dust luminosity surface density $\SigLd$ 
derived from the SEDs of $\Delta R=677\pc$
annuli, shown in Figure \ref{fig:Ldust}(b), is in good agreement with 
$\SigLd$ obtained from pixel-based modeling.
From the annular SEDs, the dust luminosity surface density 
for $17\kpc<R<25\kpc$ is
\beqa \label{eq:SigLdfit}
\SigLd\cos i &\approx& 8.0\times10^5\exp[-(R-17\kpc)/3.1\kpc]\Lsol\kpc^{-2}
~~~.
\eeqa
For purposes of estimating the integrated dust properties from
the pixel-based modeling, we sum
over the M160 pixel-based modeling for $R < 17\kpc$,
and use the annular SED modeling
for $R>17\kpc$ (see Table \ref{tab:results}).
We estimate the $R<25\kpc$ dust luminosity to be
$L_d=4.26\times10^9\Lsol$. 

\begin{table}[t]
\begin{center}
\caption{\label{tab:results}
        Global Quantities for M31$^a$}
\begin{tabular}{l c c c c}
\hline
Quantity               & $R<17\kpc$       & $17-20\kpc$      & $20-25\kpc$ 
                       & Total \\
\hline
$L_{\rm d}$ ($\Lsol$)  & $3.98\times10^9$ & $1.76\times10^8$ & $1.04\times10^8$
                       & $4.26\times10^9$
\\ 
$\Md$       ($\Msol$)  & $4.50\times10^7$ & $5.25\times10^6$ & $3.83\times10^6$
                       & $5.41\times10^7$
\\
$\MH$ ($\Msol$)        & $4.87\times10^9$ & $8.61\times10^8$ & $9.83\times10^8$
                       & $6.71\times10^9$
\\
$\langle \qpah\rangle_{\rm MMP83}$ & 0.0364  &          &    \\
$\langle \qpah\rangle_{\rm corr}$  & 0.0391  &          &   \\
$\Md/\MH$              & 0.00924          &  0.00610    &  0.00390 
                       & 0.00806
\\
\hline
\multicolumn{3}{l}{$^a$ $D=744\kpc$, inclination $i=77.7^o$}
\end{tabular}
\end{center}
\end{table}
\begin{figure}[t]
\begin{center}
\includegraphics[width=\figwidth,angle=0,
                 clip=true,trim=0.6cm 0.6cm 1.6cm 8.1cm]
                {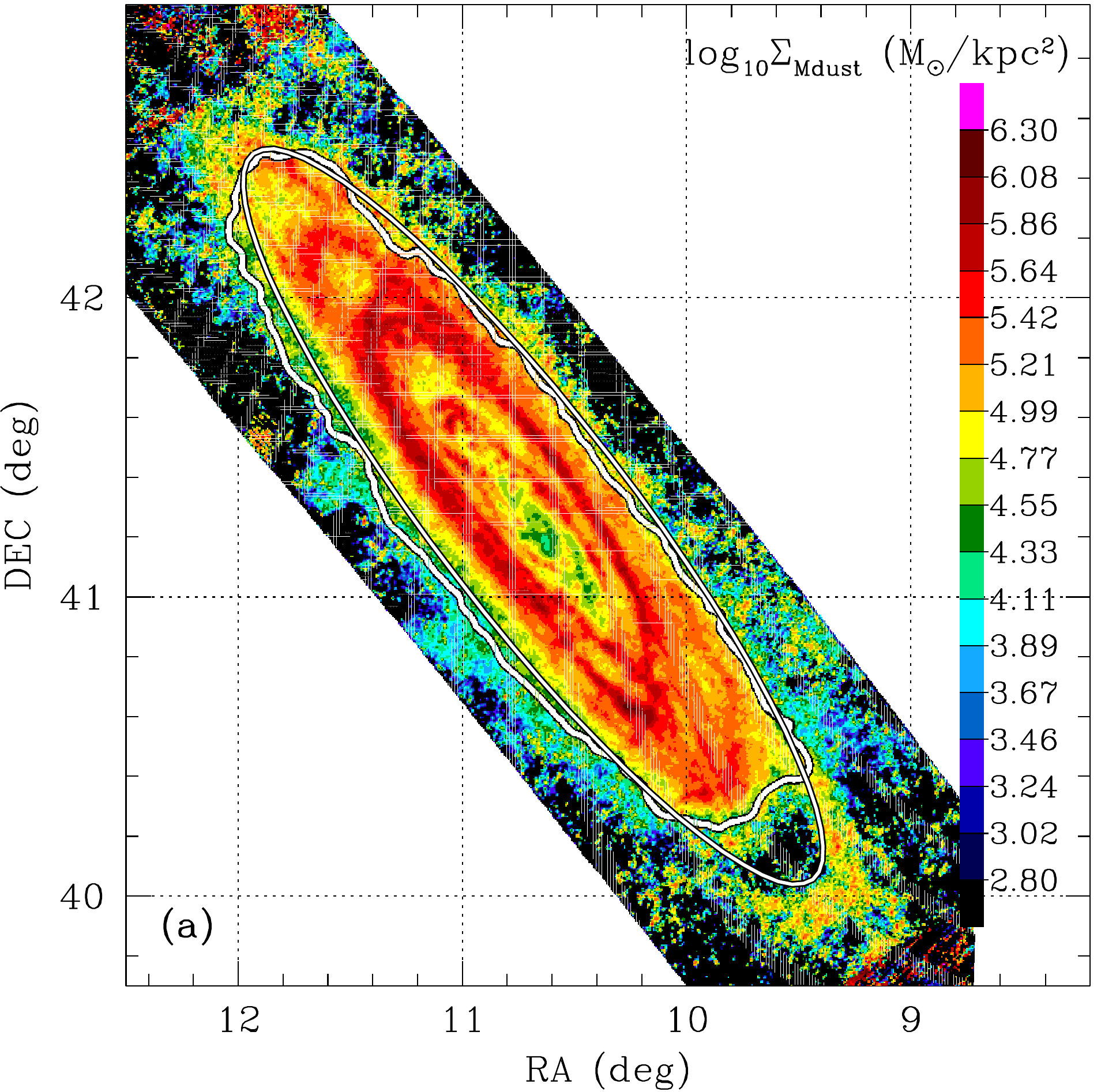}
\includegraphics[width=\figwidth,angle=0,
                 clip=true,trim=0.5cm 4.5cm 0.5cm 2.8cm]
                {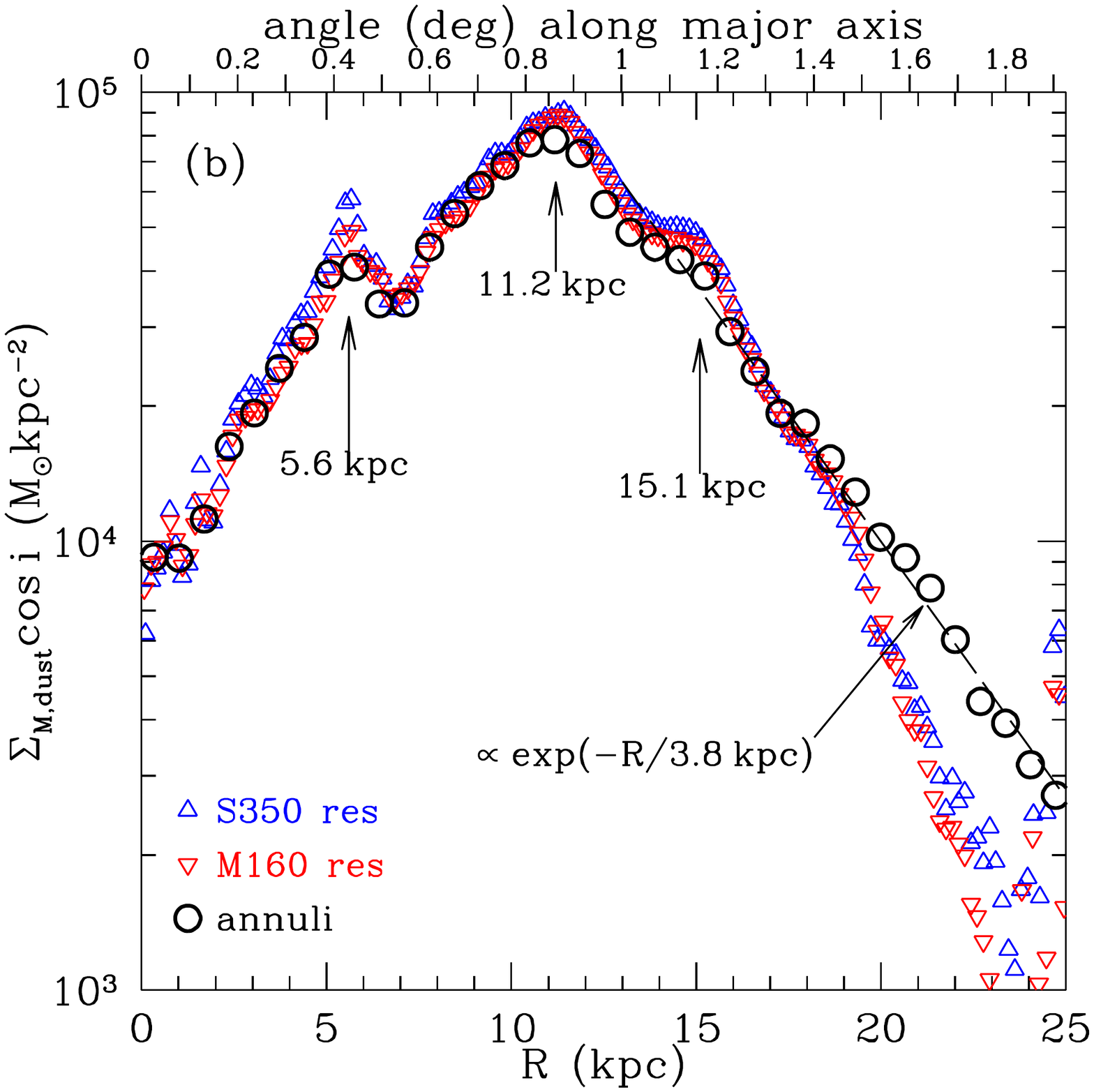}
\caption{\label{fig:sigma_dust}
         \footnotesize
         (a) Map of dust mass surface density $\SigMd$ at M160
         resolution.  The irregular contour delineates the contiguous
         region with $\SigLd>\SigLdmin$; outside of this contour, the
         dust map is unreliable.
         (b) Radial profile of $\SigMd$ projected onto
         the M31 disk plane, for $\cos i=0.213$.
         \newtext{Triangles: annular average of $\SigMd$ from single-pixel
         modeling at S350 and M160 resolution.}  
         For each $R$, $\SigMd$ is the average for
         pixels with $\SigLd>\SigLdmin$.
         \newtext{Circles: $\SigMd$ obtained
         by modeling SED of $\Delta R=677\pc$ annuli, using all bands.}
        }
\end{center}
\end{figure}
\begin{figure}[t]
\begin{center}
\includegraphics[width=\figwidth,angle=0,
                 clip=true,trim=0.5cm 4.5cm 0.5cm 2.8cm]
                {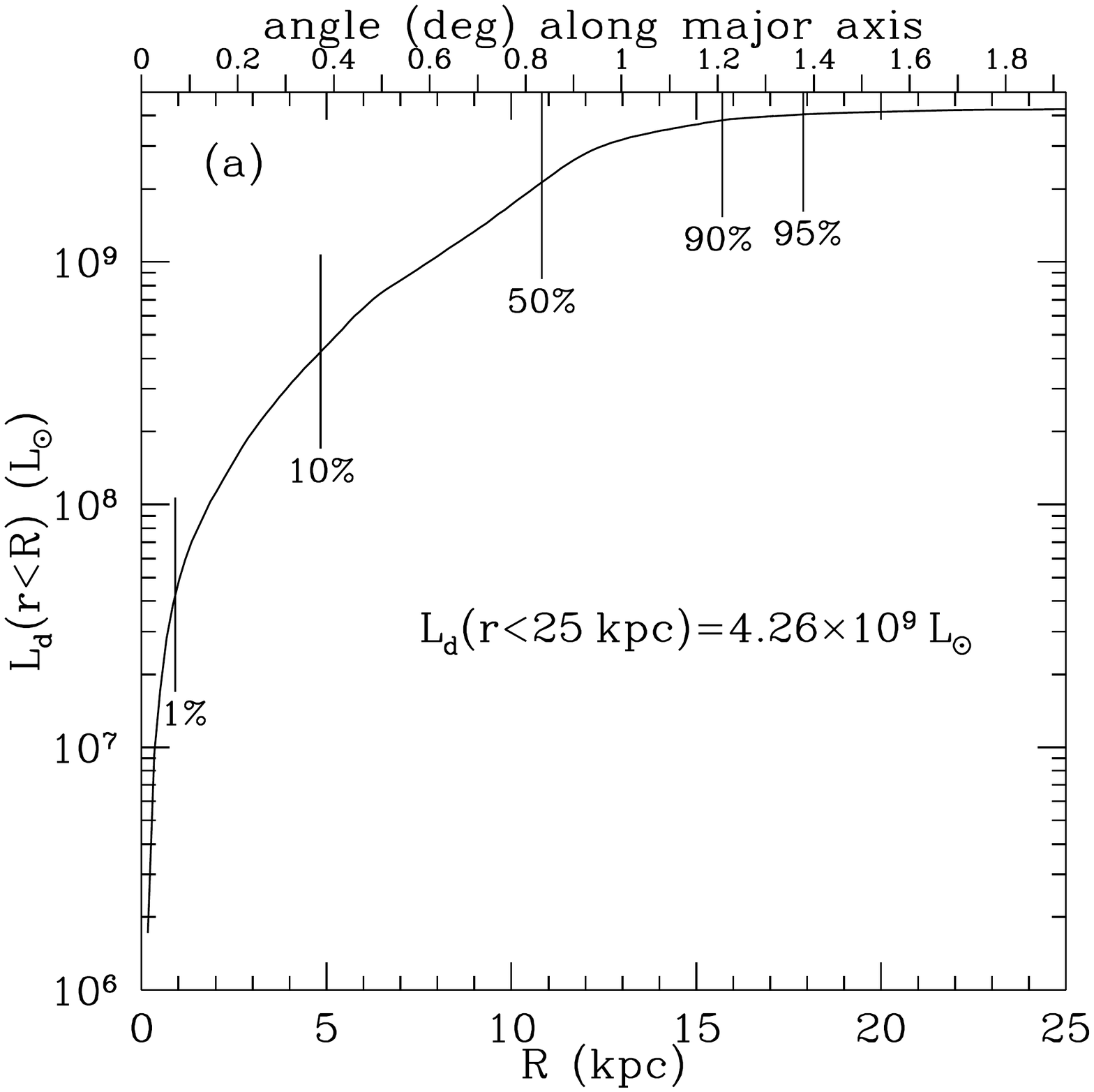}
\includegraphics[width=\figwidth,angle=0,
                 clip=true,trim=0.5cm 4.5cm 0.5cm 2.8cm]
                {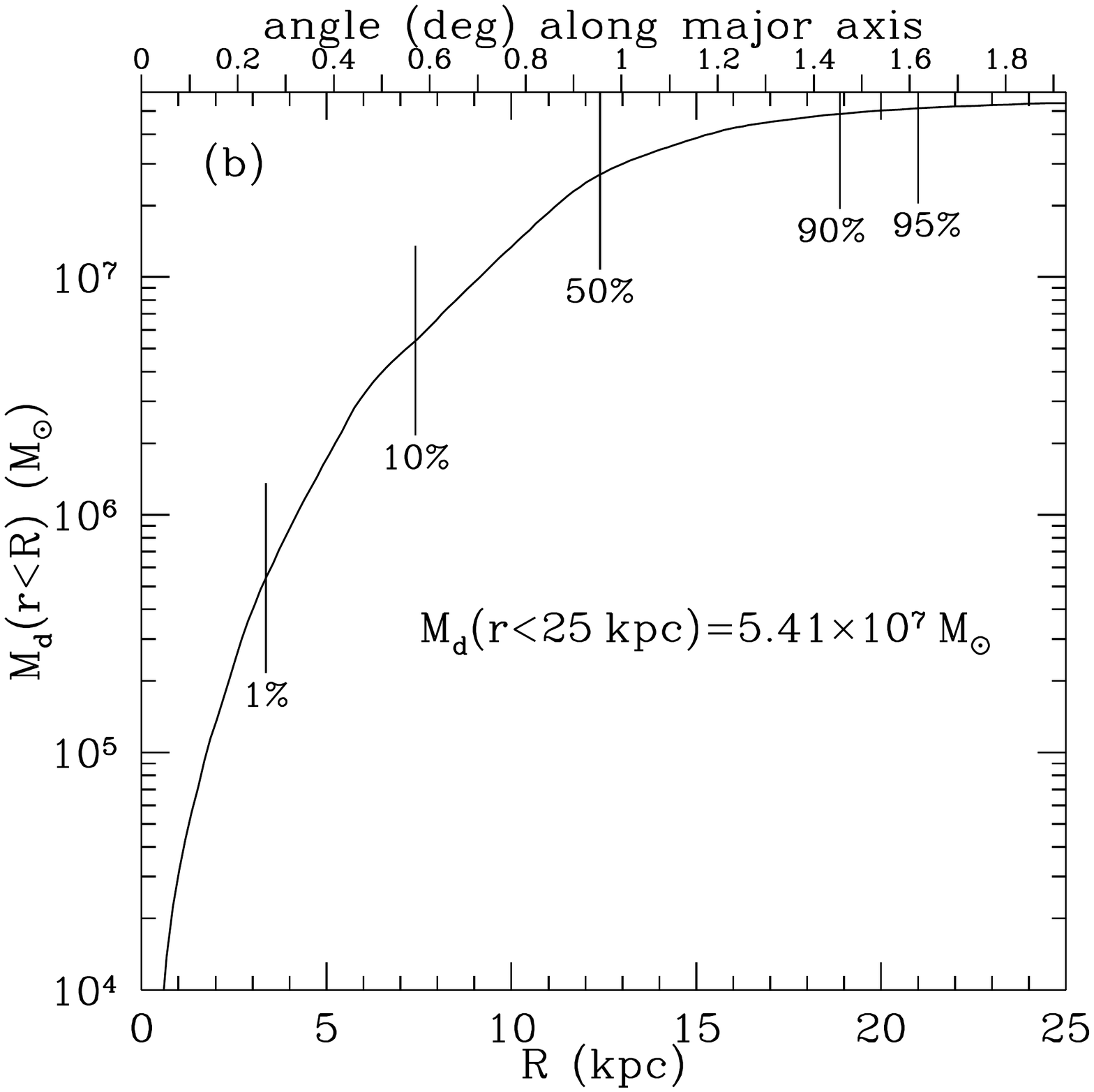}
\caption{\label{fig:sum_Ld_and_Md}
         \footnotesize
         (a) Dust luminosity interior to radius $R$, showing
         radii containing 1\%, 10\%, 50\%, 90\%, and 95\% of the total dust
         luminosity within $25\kpc$.
         (b) Dust mass interior to radius $R$, showing
         radii containing 1\%, 10\%, 50\%, 90\%, and 95\% of the total dust
         mass interior to $25\kpc$.
         For $R<17\kpc$ $L_d(r<R)$ and $M_d(r<R)$ are based on
         single-pixel modeling at M160 resolution; for $17\kpc<R<25\kpc$
         the results are based on modeling the SEDs of $\Delta R=677\pc$
         annuli.
        }
\end{center}
\end{figure}

Because $\nu L_\nu$ peaks near $100\micron$, the 
global dust luminosity of M31 was already reasonably
well-determined by the low-resolution
observations by {\it IRAS} (60, 100$\micron$) and 
{\it COBE} (140, 240$\micron$).
Adding $175\micron$ imaging by {\it ISO},
\citet{Haas+Lemke+Stickel+etal_1998} obtained
$L_d\approx 3.8\times10^9\Lsol$.\footnote{%
   All luminosities have been corrected to $D=744\kpc$}
Using MIPS photometry,
\citet{Gordon+Bailin+Engelbracht+etal_2006} found
$L_d\approx 4.0\times10^9\Lsol$, consistent, within the uncertainties, 
with the present value of
$L_d=4.26\times10^9\Lsol$.

There is considerable structure in the dust luminosity image in Figure
\ref{fig:Ldust}(a), with close similarities to the SPIRE $350\micron$ image
in Figure \ref{fig:S350}.
While the structure is not purely ``circular'' (on the plane of the galaxy),
the dust luminosity density extracted in (inclined) 
circular annuli centered on
the dynamical center, shown in
Figure \ref{fig:Ldust}(b), exhibits
a strong central peak (due to small amounts of relatively warm dust),
two clear rings (at $R=5.6\kpc$ and $R=11.2\kpc$) 
and indications of a third ring
(at $R\approx15.1\kpc$).
The $11.2\kpc$ ring was evident in the original {\it IRAS} imaging
\citep{Habing+Miley+Young+etal_1984},
and all three rings were noted
in the {\it ISO} $175\micron$ imaging
by \citet{Haas+Lemke+Stickel+etal_1998}.

\section{Dust Mass, Dust-to-Gas Ratio, and Metallicity
         \label{sec:Mdust}}

\subsection{Dust Mass}

Before the launch of {\it Herschel Space Observatory}, 
the mass of dust in M31 had been estimated by
\citet{Xu+Helou_1996}, 
\citet{Haas+Lemke+Stickel+etal_1998},
and
\citet{Gordon+Bailin+Engelbracht+etal_2006}
(see Table \ref{tab:history}).
With {\it Herschel} imaging out to 500$\micron$, we are now able
to obtain improved estimates for the total dust mass, and to
map the dust out to $\sim$$20\kpc$.
Figure \ref{fig:sigma_dust}(a)
shows the dust mass surface density $\SigMd$
on the plane of the sky, obtained by modeling at M160 resolution.

In addition to constructing maps showing the dust properties in every
pixel, we also
average the dust properties over rings that are circular on the
disk plane in order to study radial trends.
For each annulus,
we calculate
the mean dust mass surface density obtained by summing the dust masses 
in pixels with $\SigLd>\SigLdmin$, dividing by the total area of
the annulus.  For $R<15.6\kpc$ all pixels have $\SigLd>\SigLdmin$,
but this procedure will underestimate
the actual mean surface density for $R>16\kpc$.
Figure \ref{fig:sigma_dust}(b)
shows the dust mass surface density projected onto the M31 disk plane,
as a function of
galactocentric radius $R$.
\newtext{Triangles show annular averages of single-pixel modeling}
\oldtext{for models} at S350 resolution (MIPS160 and SPIRE500
not used) and \oldtext{also for models} at M160 resolution (using all the data).
Dust mass surface densities obtained at S350 resolution (not using 
MIPS160 and SPIRE500) agree to within $\sim$10\% with dust mass estimates
obtained at M160 resolution, using all cameras.

Dust mass 
surface densities $\SigMd$ estimated by fitting SEDs of $\Delta R=677\pc$
annuli are \oldtext{also} shown \newtext{(open circles)}
in Figure \ref{fig:sigma_dust}(b),
and are seen to agree to within $\sim$10\% with the results of the
pixel-based modeling for $R<15\kpc$ where the signal-to-noise is high.
For $R>16\kpc$, where a growing fraction of pixels has
$\SigLd<\SigLdmin$, the annular SED is the best way to estimate
the dust mass.
The dust mass surface density at $R>17\kpc$ is seen to be approximated by
the broken line in Figure \ref{fig:sigma_dust}b:
\beqa
\SigMd\cos i &\approx& 2.2\times10^4\exp[-(R-17\kpc)/3.8\kpc]\Msol\kpc^{-2}
~~~~~{\rm for}~R>17\kpc.
\eeqa

The dust mass surface density in Figure \ref{fig:sigma_dust}(b)
shows two distinct peaks, corresponding to
rings at $R=5.6\kpc$ and $R=11.2\kpc$, with a third ring
at $R=15.1\kpc$.
The $11.2\kpc$ ring coincides with a peak in the \ion{H}{1}
21cm emission, a ring of star formation
\citep[seen in H$\alpha$;][]{Devereux+Price+Wells+Duric_1994},
and a $\sim$40\% overdensity of stars with ages $>1\,{\rm Gyr}$
\citep{Dalcanton+Williams+Lang+etal_2012}.
The image of the dust mass surface density in Figure \ref{fig:sigma_dust}(a)
shows a conspicuous deficiency of dust between $\sim$16--20$\kpc$ on
the SW side of the disk.  As will be discussed further below, this
is not an artifact -- the \ion{H}{1} gas shows a similar deficiency
in the same region.

The DL07 dust model has\footnote{ 
  The coefficient 0.74 differs from the value 0.67 in 
  \citet{Aniano+Draine+Calzetti+etal_2012} because of the mass 
  renormalization discussed in Section \ref{sec:renormalize_Md}.}
\beq
A_V=0.74 \left(\frac{\SigMd}{10^5\Msol\kpc^{-2}}\right)~{\rm mag}
~~~.
\eeq
At
$R\approx20\kpc$, the dust mass surface density 
(projected onto the disk of M31)
$\SigMd\cos i\approx 9\times10^3\Msol\kpc^{-2}$,
corresponds to a visual extinction $A_V\approx 0.07\,{\rm mag}$ normal
to the disk of M31, or $A_V\approx 0.3\,{\rm mag}$ along
our line-of-sight.
The peak dust mass surface density occurs at $R=11.2\kpc$ with a dust
surface density, projected onto the M31 disk, 
$\SigMd\cos i\approx 7\times10^4\Msol\kpc^{-2}$, corresponding to
$A_V\approx 0.5\,{\rm mag}$
normal to the disk ($A_V\approx 2.4\,{\rm mag}$ along our line-of-sight).
Many pixels in Figure \ref{fig:sigma_dust} have higher dust surface
densities, reaching $\SigMd\approx10^{6.0}\Msol\kpc^{-2}$,
corresponding to $A_V\approx 7\,{\rm mag}$.  In these pixels the dust
is presumably distributed inhomogeneously.
{\it Hubble Space Telescope} observations of M31
\citep{Dalcanton+Williams+Lang+etal_2012} may allow measurement
of reddening toward many individual stars in M31; it will be of
great interest to compare the stellar reddening values with the
present maps of dust surface density.

We consider our best estimate for the dust mass to be the dust mass
obtained using all cameras
(including MIPS160). 
For $R<17\kpc$ we use the M160 resolution single-pixel estimates for
$\SigMd$, while for $R>17\kpc$ we use the annular SEDs to estimate
the dust mass in each annulus.
We find a total dust mass
$\Md=5.4\times10^7\Msol$ within $R=25\kpc$ (see Table \ref{tab:results}).



It is difficult to estimate objectively the uncertainty in the
estimate for $\Md$.  Calibration uncertainties in the photometry itself
are at least 10\% for each of the MIPS, PACS, and SPIRE bands.
The inconsistencies between MIPS and
PACS at $70\micron$ and $160\micron$ 
are larger than expected (see Appendix \ref{app:PACS_vs_MIPS}), 
and, in addition, there are
difficult-to-assess errors arising from assumptions made in the
modeling about the physical properties of the dust, and simplified
treatments of the starlight intensity distribution.
Overall, we adopt a tentative uncertainty estimate of 20\% for the global
dust mass given in Tables \ref{tab:results} and \ref{tab:history}.

Figure \ref{fig:sum_Ld_and_Md} shows the cumulative dust luminosity
and dust mass as a function of $R$.
We find that 
half of the dust mass in M31 lies at $R>12.3\kpc$, and $>10\%$ lies beyond
$19\kpc$.

\begin{table}[h]
\begin{center}
\caption{\label{tab:history}
         Estimates for the Dust Mass in M31}
\begin{tabular}{c l l}
\hline
$\Md (10^7\Msol)^a$ & Data used & Reference \\
\hline

$2.2\pm0.7$             & {\it IRAS}, extinction, H\,I 
                        & \citet{Xu+Helou_1996}
\\
$3.5\pm1.0$ & {\it IRAS}100, COBE-DIRBE 140,240, ISO 175$\micron$ 
                        & \citet{Haas+Lemke+Stickel+etal_1998}
\\
$4$           & {\it IRAS}, COBE-DIRBE, {\it ISO}, MIPS 
                        & \citet{Gordon+Bailin+Engelbracht+etal_2006}
\\
$5.05\pm0.45$ & PACS 100,160; SPIRE 250,350,500; annuli
                        & HELGA\,I \citet{Fritz+Gentile+Smith+etal_2012}
\\
$2.6$         & PACS 100,160; SPIRE 250,350,500; S500 pixels
                        & HELGA\,II \citet{Smith+Eales+Gomez+etal_2012}
\\
${\bf 5.4\pm 1.1}$      & {\bf MIPS, PACS, SPIRE}; M160 pixels and annuli
                        & {\bf This work}
\\
\hline
\multicolumn{3}{l}{$^a$ For $D=744\kpc$.}\\
\end{tabular}
\end{center}
\end{table}

The HELGA survey
\citep{Fritz+Gentile+Smith+etal_2012,Smith+Eales+Gomez+etal_2012} 
obtained PACS and SPIRE imaging of M31 at a scan speed of
$60\arcsec s^{-1}$ (three times faster than the scan speed for
the PACS and SPIRE imaging used in the present study).
HELGA\,I \citep{Fritz+Gentile+Smith+etal_2012} measured the fluxes 
in a central circular aperture and five concentric annuli extending out
to $20\kpc$.
Except for the central circle, the annular boundaries were ellipses.
The $\lambda\geq 100\micron$ SED of each annulus 
was fit with a modified blackbody $F_\nu \propto \nu^\beta
B_\nu(T)$, with fixed $\beta=1.8$.
The total dust mass $\Md=(5.05\pm0.45)\times10^7\Msol$
found within 20 kpc by HELGA\,I 
is close to the present estimate 
$\Md(R<20\kpc)=(5.0\pm1.0)\times10^7\Msol$.
 
HELGA\,II \citep{Smith+Eales+Gomez+etal_2012} 
restricted their study to pixels
satisfying their criterion of $5\sigma$ detection in each of six bands
(MIPS70, PACS100, PACS160, and the SPIRE bands).
The MIPS70 photometry
was used only as an upper limit: a modified blackbody $A\nu^\beta B_\nu(T)$
was used, with $A$, $\beta$, and $T$ adjusted to fit the 100--500$\micron$
photometry for each pixel; models for the cool dust
were acceptable provided they
did not exceed the MIPS70 photometry.
In the regions included in their study, HELGA\,II found a total dust
mass $\Md=2.6\times10^7\Msol$, but
\citet{Smith+Eales+Gomez+etal_2012} note that the pixels satisfying
their $5\sigma$ criterion accounted for only $\sim$50\% of the global
$500\micron$ emission from M31.

\begin{figure}[t]
\begin{center}
\includegraphics[width=\figwidth,angle=0,
                 clip=true,trim=0.6cm 0.6cm 1.6cm 8.1cm]
                {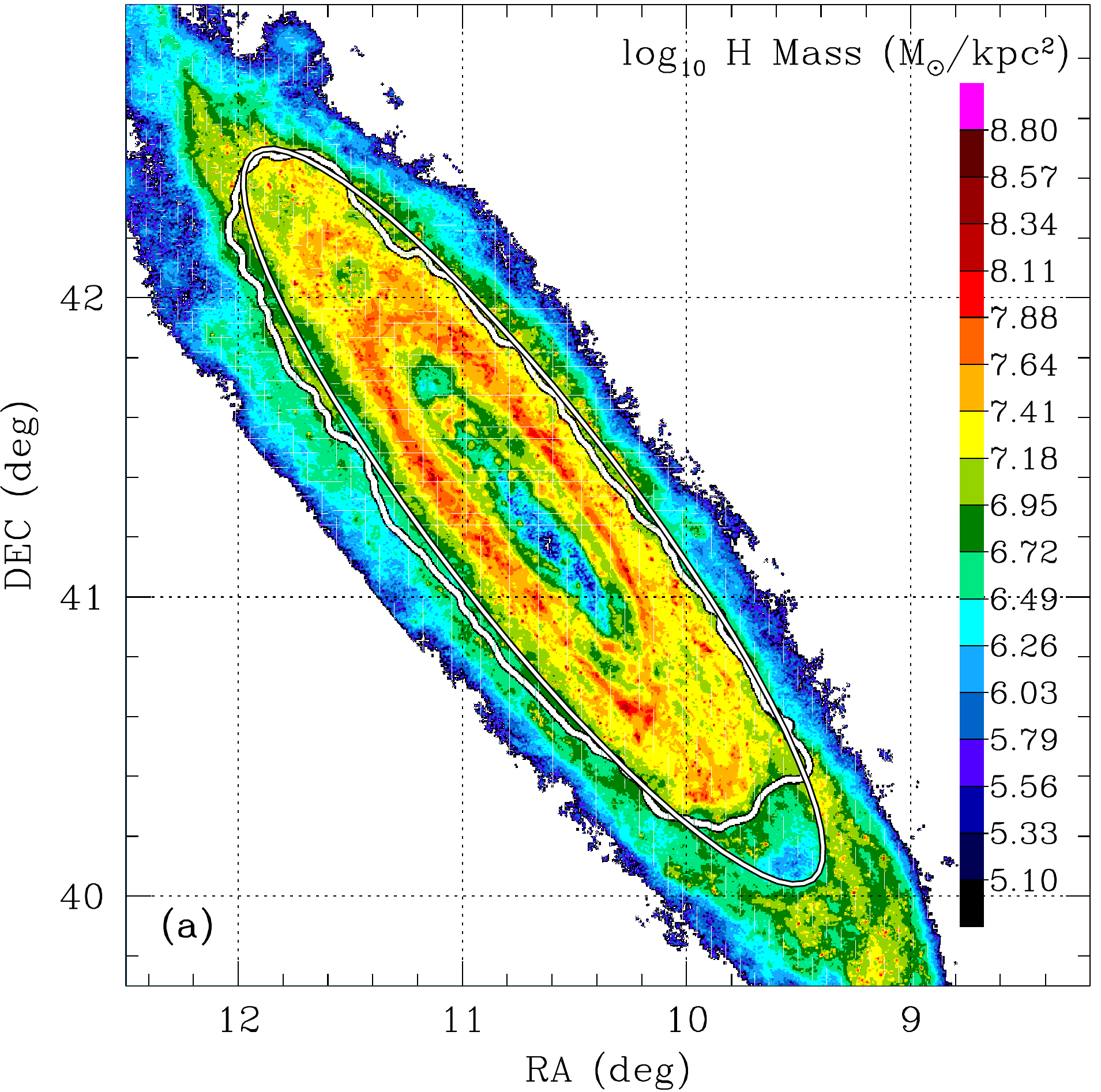}
\includegraphics[width=\figwidth,angle=0,
                 clip=true,trim=0.6cm 0.6cm 1.6cm 8.1cm]
                {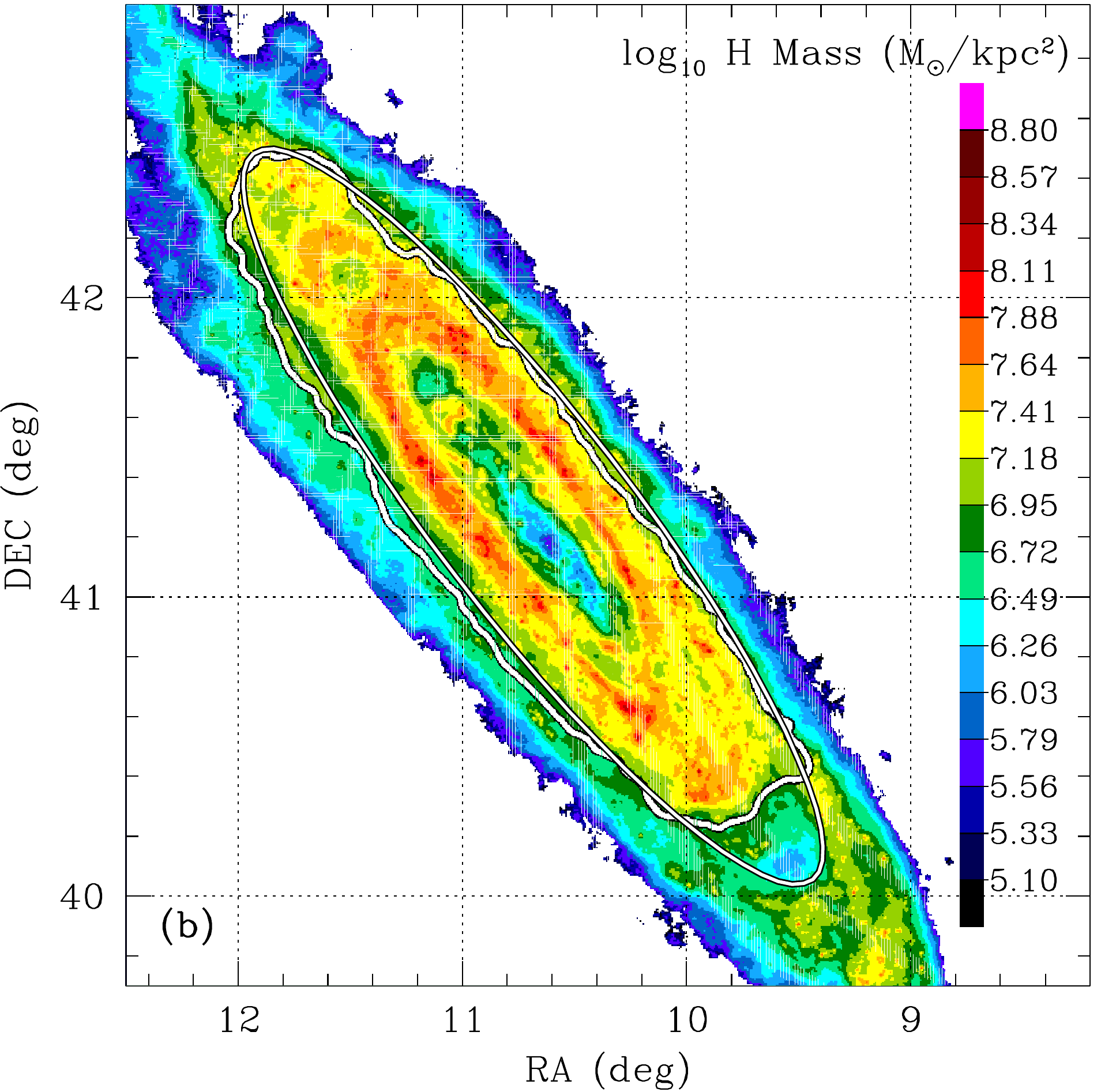}
\caption{\label{fig:gas}
         \footnotesize
         Surface density $\Sigma_{\rm H}=\Sigma({\rm H\,I})+\Sigma({\rm H}_2)$
         from H\,I\,21cm \citep{Braun+Thilker+Walterbos+Corbelli_2009}
         and CO\,1-0 
         \citep[][using $X_{\rm CO}=2\times10^{20}\cm^{-2}(\K\kms)^{-1}$]
                 {Nieten+Neininger+Guelin+etal_2006} at
         S350 resolution (left) and M160 resolution (right).
         Most of the structure seen at S350 resolution survives at
         M160 resolution.
         Note the deficiency of gas in the SW at radii $15\kpc<R<20\kpc$.
         }
\end{center}
\end{figure}
\begin{figure}[t]
\begin{center}
\includegraphics[width=\figwidth,angle=0,
                 clip=true,trim=0.6cm 0.6cm 1.6cm 8.1cm]
                {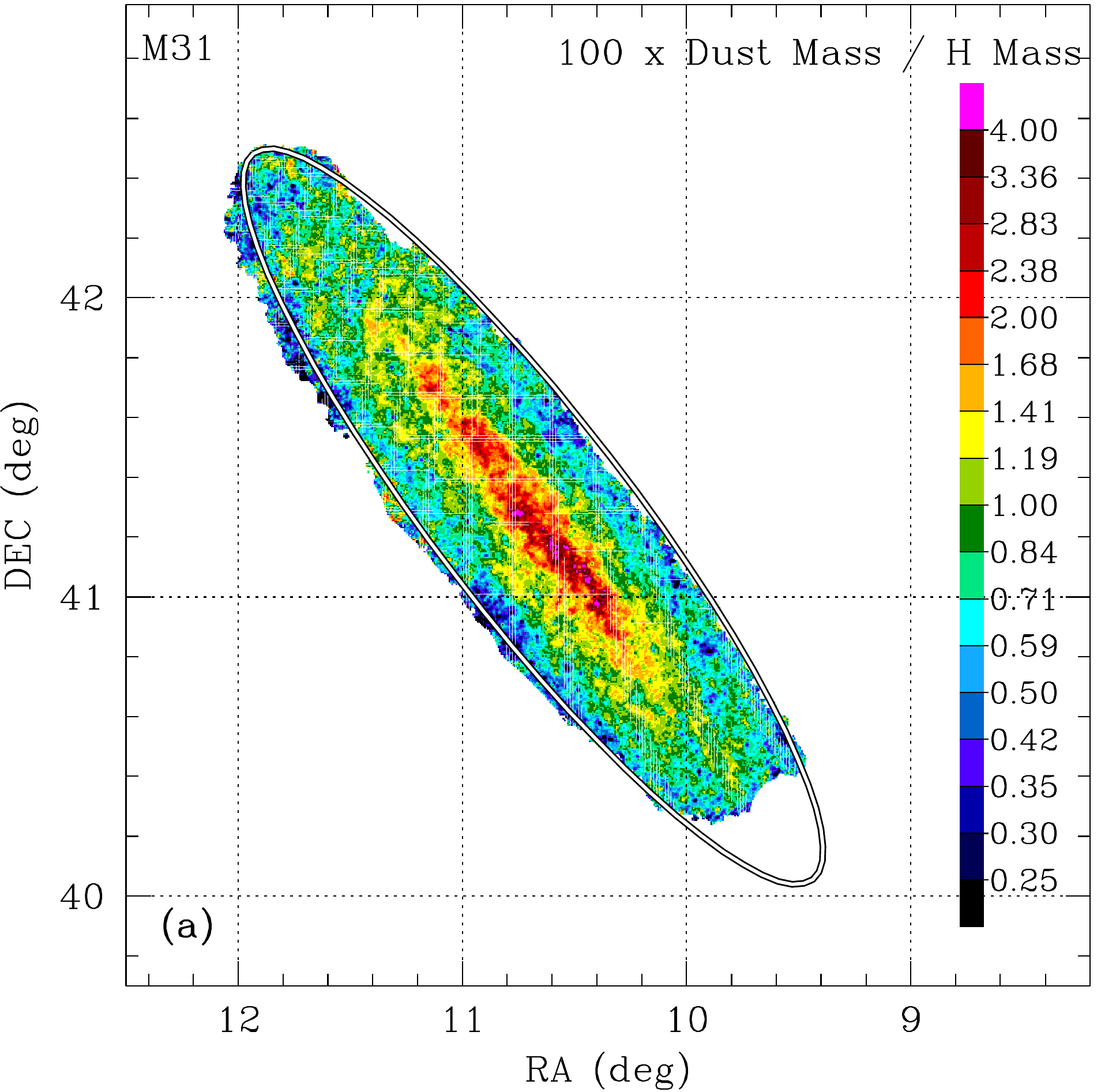}
\includegraphics[width=\figwidth,angle=0,
                 clip=true,trim=0.6cm 1.2cm 3.5cm 9.4cm]
                {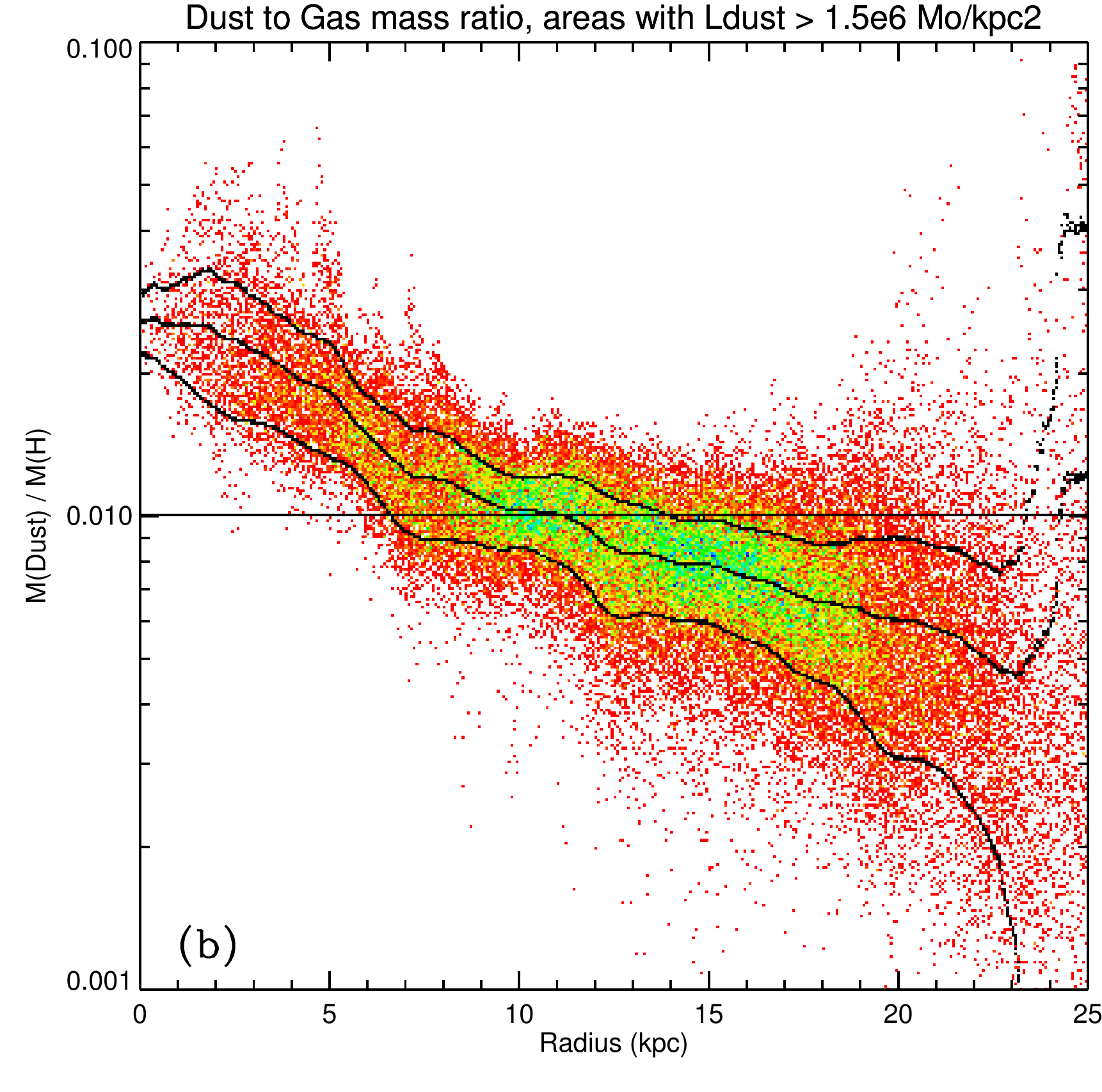}

\caption{\label{fig:dust_to_gas_pixels}
         \footnotesize
         (a) Dust-to-H mass ratio at M160 resolution, for pixels
         within the $\SigLd=\SigLdmin$ contour. 
         (b) Dust-to-H mass ratio for each M160 resolution pixel
         with $\SigLd>\SigLdmin$.
         The central curve is the mean in each radial bin; the
         other curves are the mean $\pm\sigma$, 
         where $\sigma^2$ is the estimated variance of the distribution.
         }
\end{center}
\end{figure}

\subsection{Dust-to-gas Ratio and Metallicity}

\citet{Braun+Thilker+Walterbos+Corbelli_2009} mapped the \ion{H}{1}
21\,cm emission out to $\sim$25$\kpc$ from the center of M31.
After correcting for self-absorption, they obtained
a map of the \ion{H}{1} surface density, which we use here.
Integrating this map out to
$R=25\kpc$ yields an 
\ion{H}{1} mass $M({\rm H\,I})=6.38\times10^9\Msol$
(88\% of the total \ion{H}{1} in their map). 
CO (1--0) 
has been mapped out to $\sim$12$\kpc$, 
with the associated $\HH$ mass estimated\footnote{
Using a conversion factor 
$X_{\rm CO}(J=1\rightarrow0)=2\times10^{20}{\rm H}_2\cm^{-2}(\K\kms)^{-1}$
\citep{Bolatto+Wolfire+Leroy_2013}.}
to be $M({\rm H}_2)\approx 3.4\times10^8\Msol$
\citep{Nieten+Neininger+Guelin+etal_2006}.
We take the total H surface density to be
$\Sigma_{\rm H}\approx\Sigma({\rm H\,I})+\Sigma({\rm H}_2)$,
where $\Sigma({\rm H}_2)$ is obtained from the observed CO 1-0 emission
assuming a constant $X_{\rm CO}=2\times10^{20}\HH\cm^{-2}(\K\kms)^{-1}$.

We may be underestimating $\Sigma_{\rm H}$.
$\HH$ that is ``CO dark'' (i.e., not associated with CO 1-0 emission)
is assumed to be a small fraction of the total $\HH$ mass.
More importantly, we do not include \ion{H}{2} gas in our estimate for
$\Sigma_{\rm H}$.  The ionized gas mass in bright \ion{H}{2} regions is
small, but the mass in low-density diffuse \ion{H}{2} may not
be negligible.  For our Galaxy, it is estimated that 
diffuse \ion{H}{2} accounts for
$\sim$23\% of the total ISM mass at $R<20\kpc$ \citep{Draine_2011a}.
\newnewtext{The center of M31 has
$\Sigma_{\rm H} \approx 1.8\times10^6\Msol\kpc^{-2}$
averaged over $R<1\kpc$.  
The extinction-corrected surface brightness of 
H$\alpha$ $\sim$$2\times10^{-5}\erg\cm^{-2}\s^{-1}\sr^{-1}$
\citep{Devereux+Price+Wells+Duric_1994, Tabatabaei+Berkhuijsen_2010}
corresponds to an \ion{H}{2} surface density
$6\times10^6 (\cm^{-3}/\langle n_e\rangle) T_4^{0.94} \Msol\kpc^{-2}$,
where $\langle n_e\rangle$ is the electron density within the emitting
regions.
The density-sensitive [\ion{S}{2}]6716/[\ion{S}{2}]6731 line ratio 
indicates $\langle n_e\rangle > 1.2\times10^2\cm^{-3}$
for $R\ltsim 0.2\kpc$
\citep{Ciardullo+Rubin+Jacoby+etal_1988}, hence the mass
in $\sim$$10^4\K$ plasma appears to be small compared to that in neutral gas.
X-ray observations indicate that 
$R\ltsim1\kpc$ is filled with hot gas
with $T\approx 4\times10^6\K$ and a mean surface density
$\sim$$2\times10^5\Msol\kpc^{-2}$
\citep{Bogdan+Gilfanov_2008}, only $\sim$10\% of the $R<1\kpc$ \ion{H}{1} 
mass.}

It is also possible that we may have overestimated $\Sigma_{\rm H}$ in
the center --
\citet{Leroy+Bolatto+Gordon+etal_2011} find that $X_{\rm CO}$ in the center
\newnewtext{($R<2\kpc$)}
of M31 is lower by about a factor two.
\oldtext{, in which case the
dust/gas ratio in the center may be even higher than shown in Figure
\ref{fig:dust_to_gas_pixels}.}
\newnewtext{However, there is little CO emission at $R\ltsim2\kpc$
\citep{Nieten+Neininger+Guelin+etal_2006}, hence the total gas mass there 
is insensitive to uncertainties in $X_{\rm CO}$.
}

Figure \ref{fig:gas} shows $\Sigma_{\rm H}$ at S350 and M160 resolution.
Note the deficiency of gas in the SW for $R>15.6\kpc$; the same
region is also seen (see Figure \ref{fig:Ldust}) to be deficient
in emission from dust.
The reason for the deficiency in dust and gas in the SW corner of M31
(between $R=16$ and $R\approx20\kpc$)
is not known, but this part of the M31 disk has the appearance 
of having been affected by some recent event.
\citet{Block+Bournaud+Combes+etal_2006} argued that a nearly
head-on collision with M32 $\sim$210 Myr ago can account for the observed offset
of the center of the 11 kpc ring, but such an encounter would not seem
likely to produce the observed deficiency of \ion{H}{1} and dust in the
SW at $R\approx16-20\kpc$.
It may be the result of an encounter with another member of the Local Group
within the past Gyr.
\citet{Lewis+Braun+McConnachie+etal_2013} discuss the Giant Stellar Stream, 
extending from the SW side of the disk toward M33.
\ion{H}{1}\,21\,cm observations also show a diffuse gaseous filament connecting
M31 and M33, including a feature extending from the SW side of the M31
disk toward M33.
The origin of the Giant Stellar Stream and the \ion{H}{1} filamentary
structure is uncertain, but strongly suggestive of a close passage
of M33, possibly accounting for the deficiency of interstellar matter
on the SW side of the disk, $\sim$18$\kpc$ from the center of M31.

Figure \ref{fig:dust_to_gas_pixels}(a) 
is a map of the dust-to-H mass ratio at M160 resolution.
Some azimuthal structure is evident, but the main feature
is a conspicuous radial trend, with the dust-to-H ratio peaking at the
center and declining with $R$.
Figure \ref{fig:dust_to_gas_pixels}(b) shows results for every pixel
(at M160 resolution)
satisfying the condition $\SigLd > \SigLdmin$.
Because of the low signal/noise ratio (S/N) in individual pixels, the scatter in
the derived $\Md/\MH$ is quite pronounced for $R\gtsim17\kpc$.
However, the mean dust/H ratio in each annulus
exhibits a clear trend, decreasing with increasing $R$.
The central dust-to-H ratio is
$\Md/\MH\approx 0.028$, 
declining to $\sim$0.005 at $R\approx 20\kpc$: a factor of $\sim$5
change in the dust-to-H ratio moving from the center to $R\approx 20\kpc$.

\begin{figure}[t]
\begin{center}
\includegraphics[width=\figwidth,angle=0,
                 clip=true,trim=0.5cm 4.5cm 0.5cm 2.8cm]
                {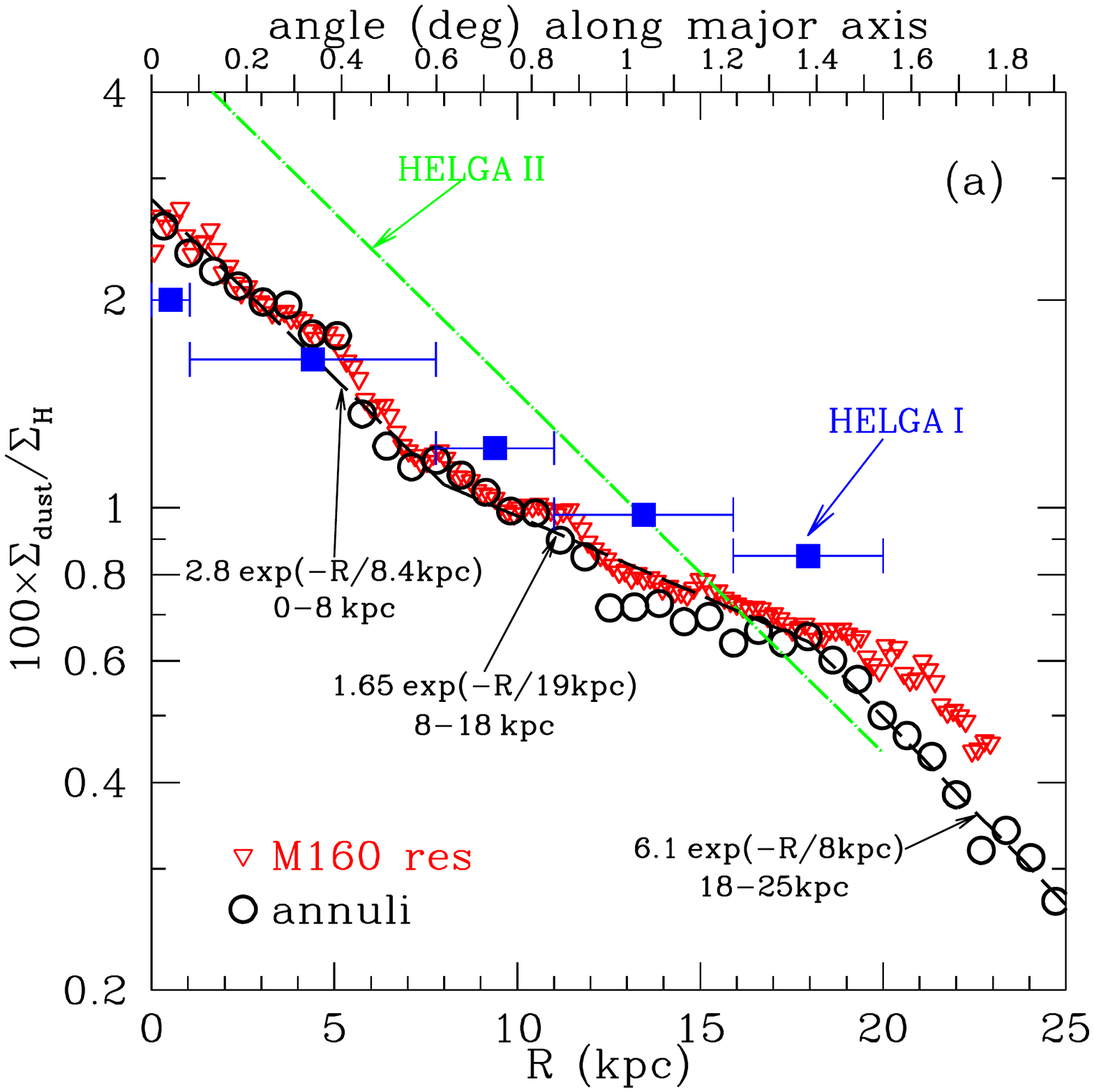}
\includegraphics[width=\figwidth,angle=0,
                 clip=true,trim=0.5cm 4.5cm 0.5cm 2.8cm]
                {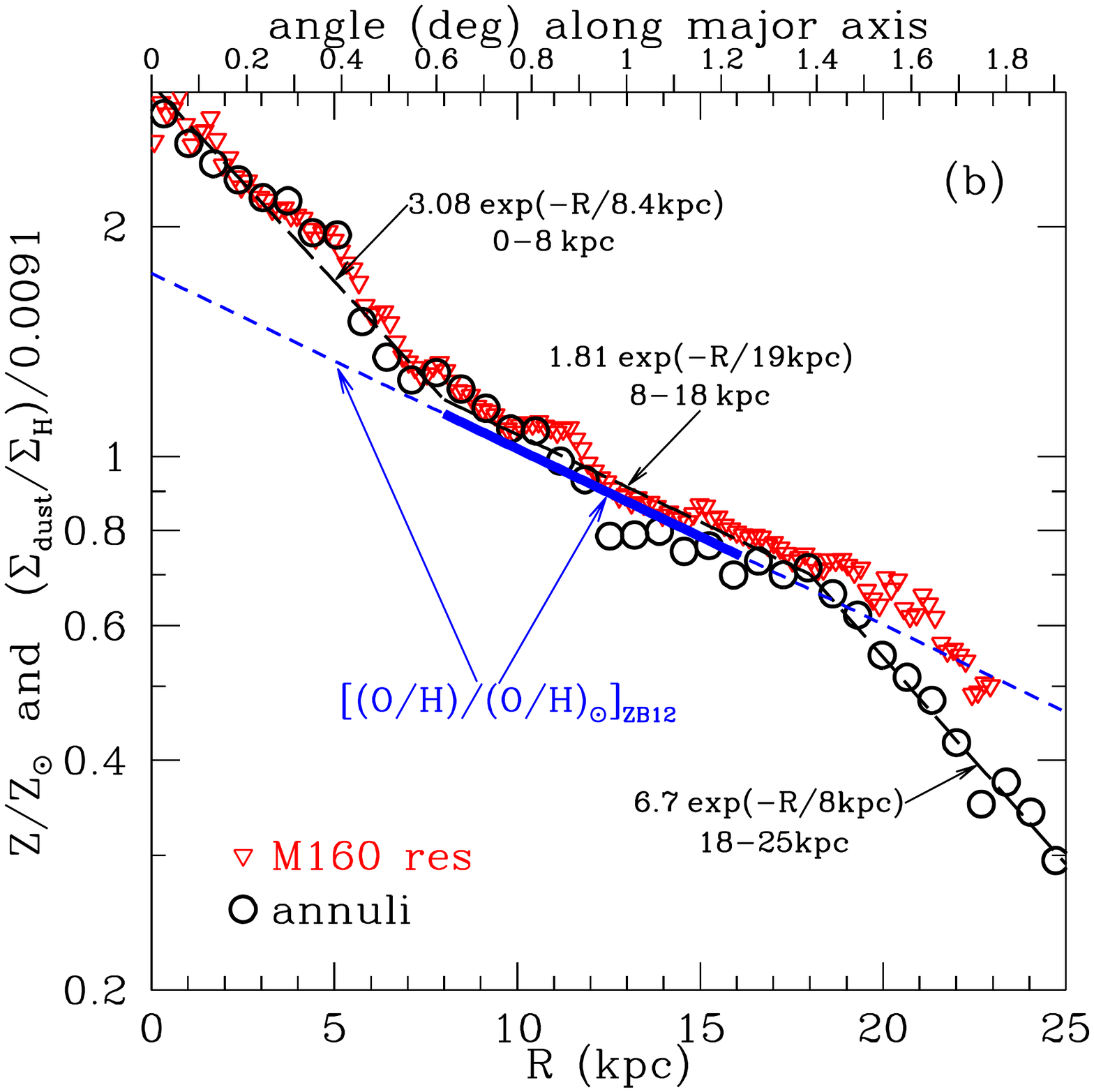}
\caption{\label{fig:dust_to_gas_vs_R}
\footnotesize
(a) 100$\times$dust/H mass ratio in M31 as a function of
galactocentric radius $R$.
Triangles: present work, modeling at M160 resolution, using only
pixels with $\SigLd>\SigLdmin$.
Circles: present work, modeling the SED for $\Delta R=677\pc$ annuli.
Squares: HELGA\,I \citep{Fritz+Gentile+Smith+etal_2012}.
Dot-dash line: HELGA\,II \citep{Smith+Eales+Gomez+etal_2012}.
(b) Metallicity relative to solar, based on our dust
modeling (triangles \newtext{and circles, as in (a)}), 
and H\,II region oxygen abundances \citep{Zurita+Bresolin_2012};
the line is shown as solid in the
8--16~kpc range where the oxygen abundances are from the
``direct'' method.
         }
\end{center}
\end{figure}

Figure \ref{fig:dust_to_gas_vs_R} (triangles) 
shows the radial profile of the dust-to-H ratio
estimated at M160 resolution.
Each triangle in Figure \ref{fig:dust_to_gas_vs_R} is obtained by
summing the M160 resolution dust and gas within rings for $R<23\kpc$, 
including only pixels
satisfying the criterion $\SigLd > \SigLdmin$,
and calculating the ratio of
(total dust)/(total gas) for that ring.
The dust/gas ratio therefore applies only to the pixels satisfying
the $\SigLd > \SigLdmin$ cut.
From Figure \ref{fig:Ldust}(b) we see that 
essentially 100\% of the pixels at $R<16\kpc$ satisfy the
surface brightness cut.
However, at $R=20\kpc$, $\sim$50\% of the pixels
in the annulus do not satisfy the cut.  This incompleteness is due in part
to the
general radial decline in surface brightness, but in part it reflects
the conspicuous deficit of both gas and dust in the SW corner of M31.
Also shown in Figure \ref{fig:dust_to_gas_vs_R} (circles) is the dust-to-H mass
ratio estimated for $\Delta R=677\pc$ annuli, where the dust mass is
obtained by modeling the SED of each annulus, and the gas mass is the total
mass of gas within the annulus.
The dust-to-gas ratio is well-behaved out to $25\kpc$.
For $R<18\kpc$ the 
dust-to-H mass ratio estimates for the $\Delta R=677\pc$
annuli at $R=16-23\kpc$ are in good agreement with the dust-to-H mass ratios
determined only for the pixels with $\SigLd>\SigLdmin$.
For $R>18\kpc$ the dust/gas ratios for the pixels with
$\SigLd>\SigLdmin$ is somewhat higher than the result from the annular
photometry.  This bias is attributable to the fact that for $\SigLd$
just below $\SigLdmin$, noise can raise the
pixel above the threshold, leading to overestimation of the dust mass.
In this regime, dust mass estimates from the 
annular photometry and modeling should be more reliable.

We find that the dust/H ratio for 0--25$\kpc$
declines monotonically with increasing $R$,
with
\beq \label{eq:dust2gas_vs_R}
\frac{\Md}{\MH} \approx
\left\{ \begin{array}{l l} 
0.0280 \exp(-R/8.4\kpc)  & R< 8\kpc\\
0.0165 \exp(-R/19\kpc)   & 8\kpc < R < 18\kpc\\
0.0605 \exp(-R/8\kpc)    & 18\kpc < R \ltsim 25\kpc ~~~,\\
\end{array}
\right.
\eeq
shown as a dashed line in Figure \ref{fig:dust_to_gas_vs_R}.
Also shown in Figure \ref{fig:dust_to_gas_vs_R}(a) is the dust/gas
ratio estimated by 
HELGA\,I \citep{Fritz+Gentile+Smith+etal_2012} for each of 5 radial
zones.
The HELGA\,I results are in fair agreement with our findings 
interior to $\sim$8$\kpc$, but at larger radii tend to exceed our
dust mass estimates.

HELGA\,II \citep{Smith+Eales+Gomez+etal_2012} conclude that the dust/H ratio 
follows an exponential profile in M31, with
${M_\dust}/{M_\Ha} \approx 0.049 \exp(-R/8.7\kpc)$;
this is plotted in Figure \ref{fig:dust_to_gas_vs_R}(a).
Our central dust/H ratio $\sim$0.028 is only 60\% of the value found
by HELGA\,II.

\section{\label{sec:metallicity}
         Metallicity of the ISM in M31}

If depletions are similar to the local Milky Way, we expect
\beq
\frac{\Md}{\MH} \approx 0.0091 \frac{Z}{Z_\odot}
~~~.
\eeq
where $Z$ is the mass fraction of elements other than H and He.
This allows us to estimate the metallicity from our measured
dust/H mass ratio (\ref{eq:dust2gas_vs_R}):
\beq \label{eq:Z_vs_R}
\frac{Z}{Z_\odot} \approx
\left\{ \begin{array}{l l} 
3.08 \exp(-R/8.4\kpc)  & R< 8\kpc\\
1.81 \exp(-R/19\kpc)   & 8\kpc < R < 18\kpc\\
6.65 \exp(-R/8\kpc)    & 18\kpc < R \ltsim 25\kpc~~~.
\end{array}
\right.
\eeq
This relation is plotted in Figure \ref{fig:dust_to_gas_vs_R}(b).

\citet[][hereafter ZB12]{Zurita+Bresolin_2012} 
measured elemental abundances in M31 
\ion{H}{2} regions using ``direct'' methods
between 8 and 16 kpc.
When they allow for depletion of oxygen into grains and a bias against
\ion{H}{2} regions with high oxygen abundances, they
estimate that
\beqa
{\rm (O/H)} &\approx& 1.8 {\rm (O/H)}_\odot \exp(-R/19\kpc)
~~~.
\eeqa
However, it is important to note that the 
\ion{H}{2} region abundance determinations
have appreciable uncertainties,
and do not agree well with metallicity determinations in the
atmospheres of B supergiants.
Our result (Equation (\ref{eq:Z_vs_R})) for the metallicity is in excellent
agreement with the ZB12 metallicity
over the $8\kpc<R<16\kpc$ range where the ZB12 \ion{H}{2} region
metallicities were based on direct determinations of the gas temperature,
and are therefore most reliable.

According to Equation (\ref{eq:Z_vs_R}),
the ISM in M31 has supersolar abundances for $R \ltsim 11\kpc$.
This is consistent with the high WC/WN stellar ratio observed in M31
\citep{Neugent+Massey+Georgy_2012}.

\section{Dust Temperature and Starlight Properties
         \label{sec:starlight}}

\begin{figure}[t]
\begin{center}
\includegraphics[width=\figwidth,angle=0,
                 clip=true,trim=0.6cm 0.6cm 1.6cm 8.1cm]
                {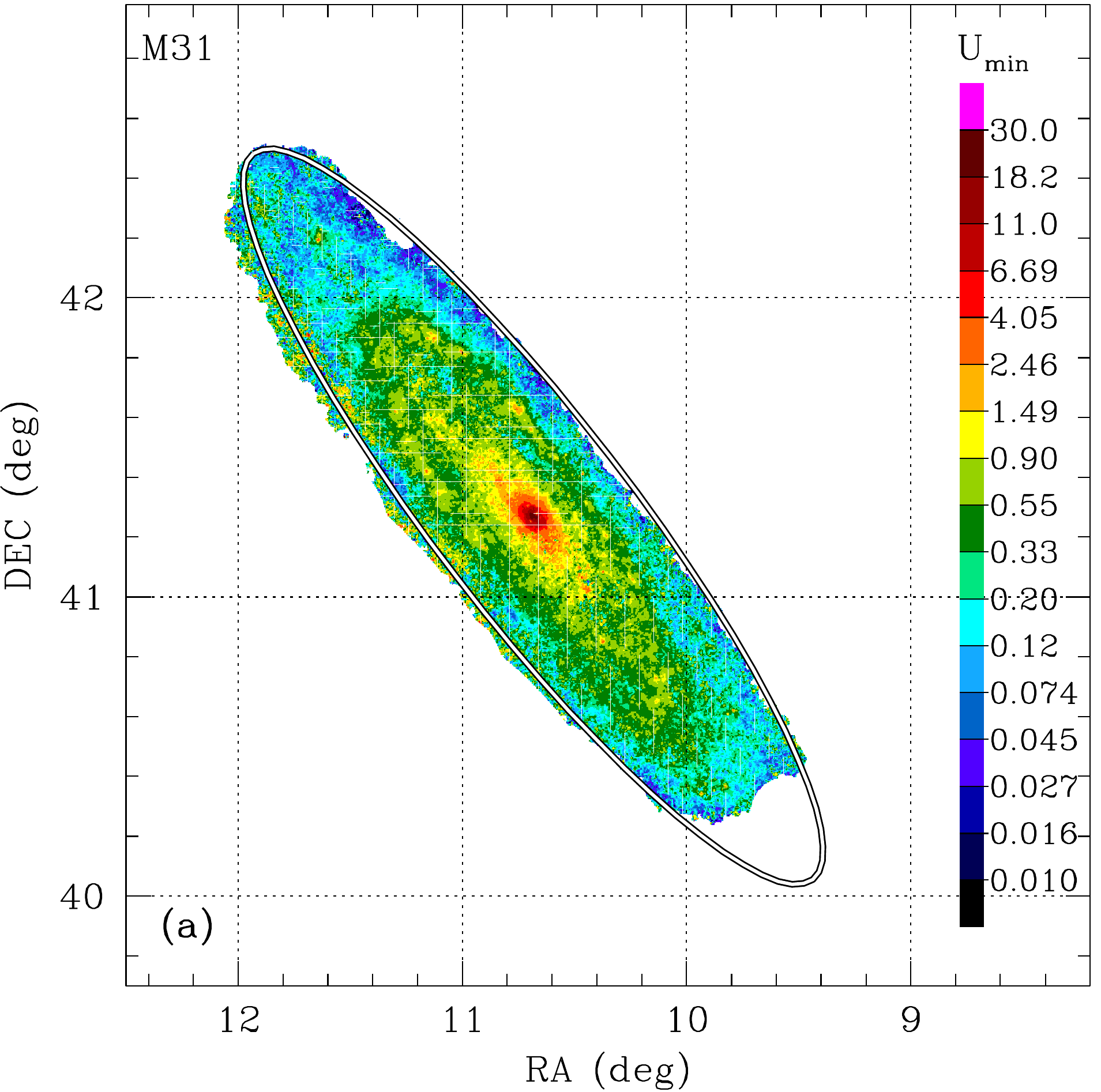}
\includegraphics[width=\figwidth,angle=0,
                 clip=true,trim=0.5cm 4.5cm 0.5cm 2.8cm]
                {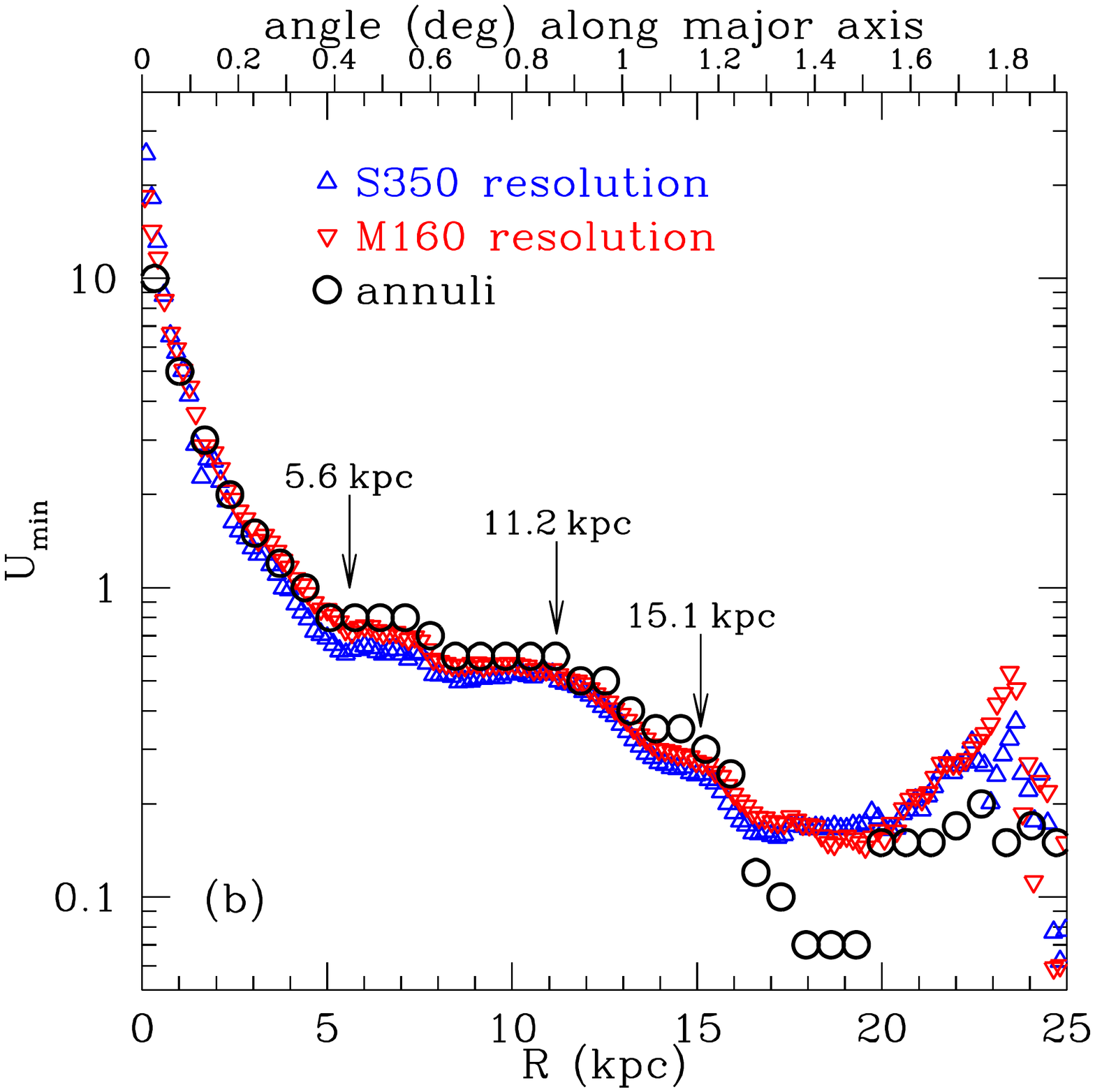}
\caption{\label{fig:Umin}
         \footnotesize
         (a) Starlight heating rate parameter $\Umin$ 
         in M31 at S350 resolution.
         (b) Radial profile of $\Umin$, estimated at S350 resolution
         and M160 resolution (triangles) and using annular
         photometry (circles).
         Locations of dust mass surface density maxima are indicated.
        }
\end{center}
\end{figure}

If the distribution of both stars and dust were known, the
intensity and spectrum of the starlight heating the dust could
be obtained from the equations of radiative transfer
\citep[see, e.g.,][]{Popescu+Tuffs_2013}.
In dusty star-forming galaxies this is a formidable problem,
because of the complex and correlated spatial distributions of
both stars and dust.

To model the infrared emission from dust, we take a much
simpler approach, empirical approach to the starlight heating.
Within a single ``pixel'' (which may include $\sim$$10^4\pc^2$ of
disk area) the dust may be exposed to a wide range
of starlight intensities, ranging from the general starlight background
in the diffuse ISM, to high intensities found in star-forming regions.
The DL07 model adopts
a parameterized distribution function for
the starlight heating rate: the dust within a given pixel
is assumed to be subject to starlight heating rates
ranging from $U=\Umin$ to a peak value $U_{\rm max}=10^7$,
with an intensity distribution given by Equation (\ref{eq:dM/dU}).
When fitting the dust model to the data, we in effect use 
the dust grains as photometers to determine
the intensity of starlight in the regions where dust is present.
The parameter $\Umin$ is interpreted as being the 
starlight heating rate in the diffuse ISM.
The mean (weighted by dust mass) starlight heating rate within
a pixel is $\langle U\rangle$.

Figure \ref{fig:Umin}(a) is a map of $\Umin$ for M31.
The $\Umin$ parameter is strongly peaked at the
center.  Figure \ref{fig:Umin}(b)
shows a generally smooth decline of $\Umin$
with increasing galactocentric
radius $R$, declining from $\Umin\approx 25$ in the central 200\,pc
(at S350 resolution)
to $\Umin\approx 0.2$ at $R=16\kpc$.
Beyond $16\kpc$, $\Umin$ obtained from single-pixel modeling
(triangles)
begins to differ from $\Umin$
obtained from annular photometry (circles).

\begin{figure}[t]
\begin{center}
\includegraphics[width=\figwidth,angle=0,
                 clip=true,trim=0.6cm 0.6cm 1.6cm 8.1cm]
                {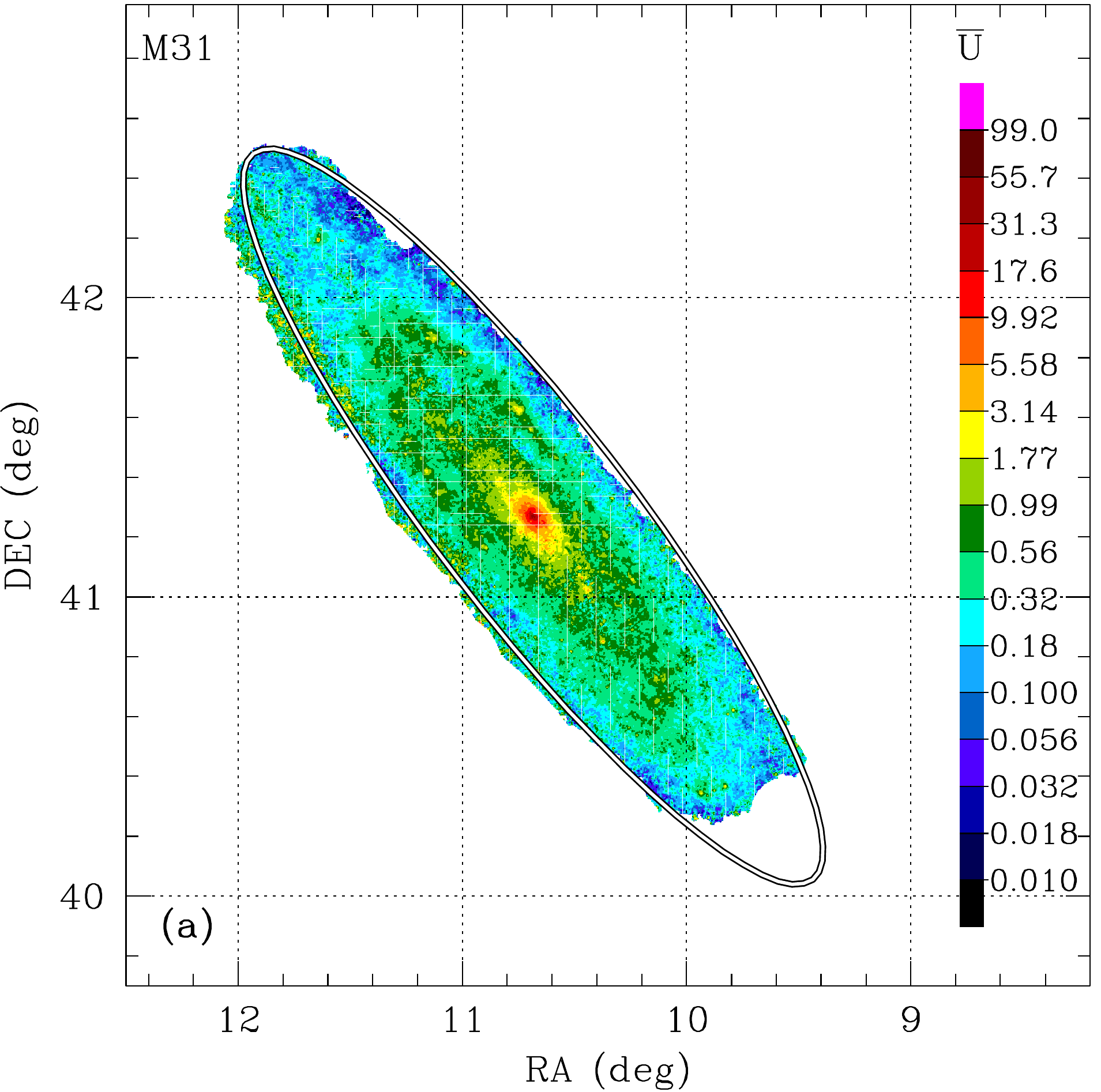}
\includegraphics[width=\figwidth,angle=0,
                 clip=true,trim=0.5cm 4.5cm 0.5cm 2.8cm]
                {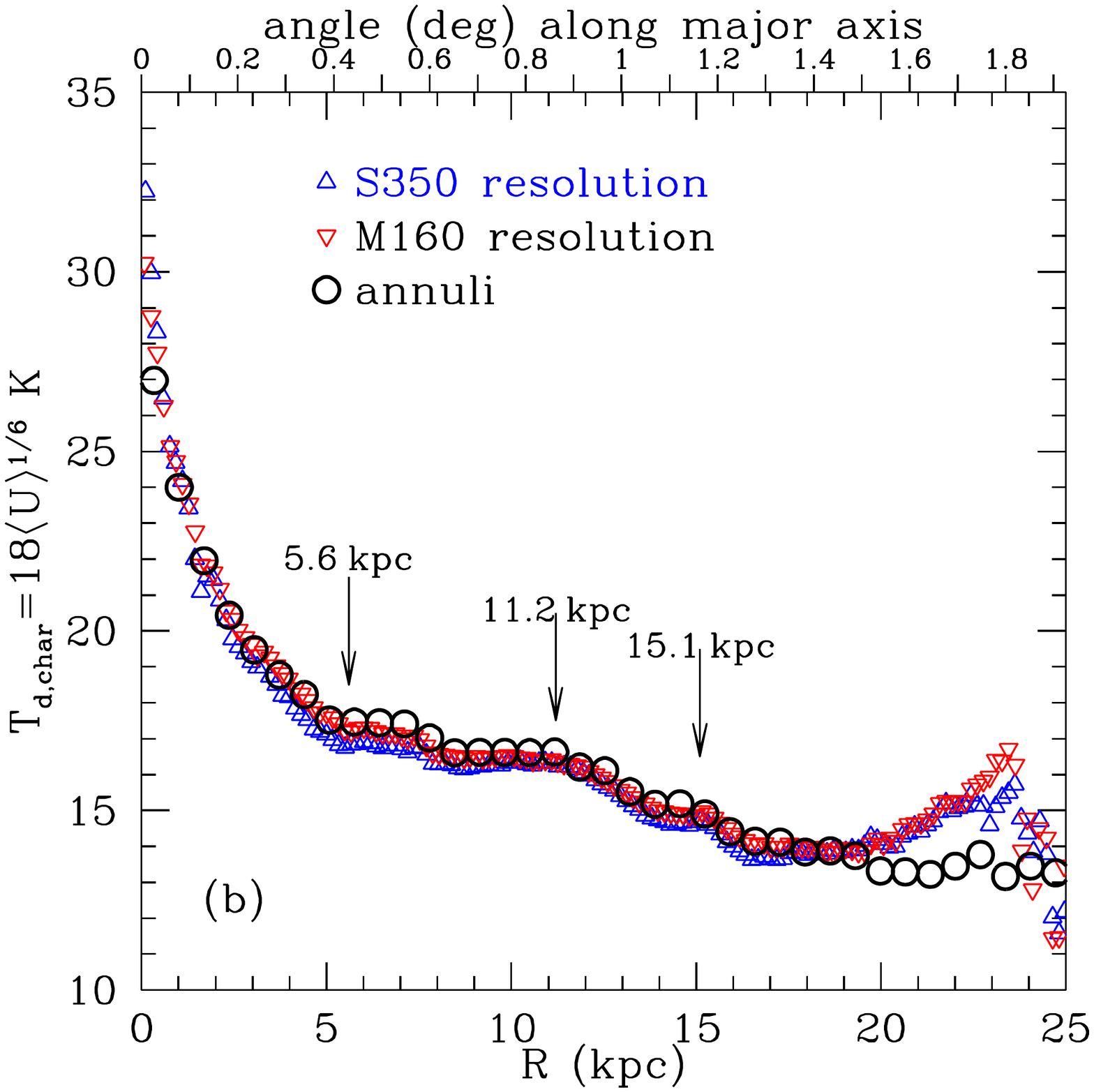}
\caption{\label{fig:Ubar_map}
         \footnotesize
         Left: map of mean starlight heating rate parameter $\langle U\rangle$ 
         \newtext{$=\bar{U}$} in M31 at S350 resolution.
         Right: radial profile of characteristic dust temperature.  
         Locations of dust
         surface density maxima are indicated.
        }
\end{center}
\end{figure}
For starlight with the 
\oldtext{MMP83 spectrum,}
\newtext{interstellar radiation field spectrum of
\citet[][hereafter MMP83]{Mathis+Mezger+Panagia_1983}} 
the dust heating rate parameter
$U$ is
\beq
U = \frac{u_\star}{8.6\times10^{-13}\erg\cm^{-3}}
~~~,
\eeq
where $u_\star$ is the starlight energy density.
For the DL07 model,
the characteristic grain temperature (of the grains dominating the
emission at $\lambda > 100\micron$) is related to the heating rate
parameter $U$ as
\beq \label{eq:Td}
\Tdchar\approx 18\, U^{1/6} \K
~~~.
\eeq
This is only a representative temperature -- dust grains of different
sizes and compositions illuminated by a single radiation field 
have different steady-state temperatures, and very small grains undergo
temperature fluctuations due to the quantized heating by stellar photons.
Figure \ref{fig:Ubar_map}(b)
shows $\Tdchar$ as a function of radius.

\begin{figure}[t]
\begin{center}
\includegraphics[width=\figwidth,angle=0,
                 clip=true,trim=0.5cm 4.5cm 0.5cm 2.8cm]
                {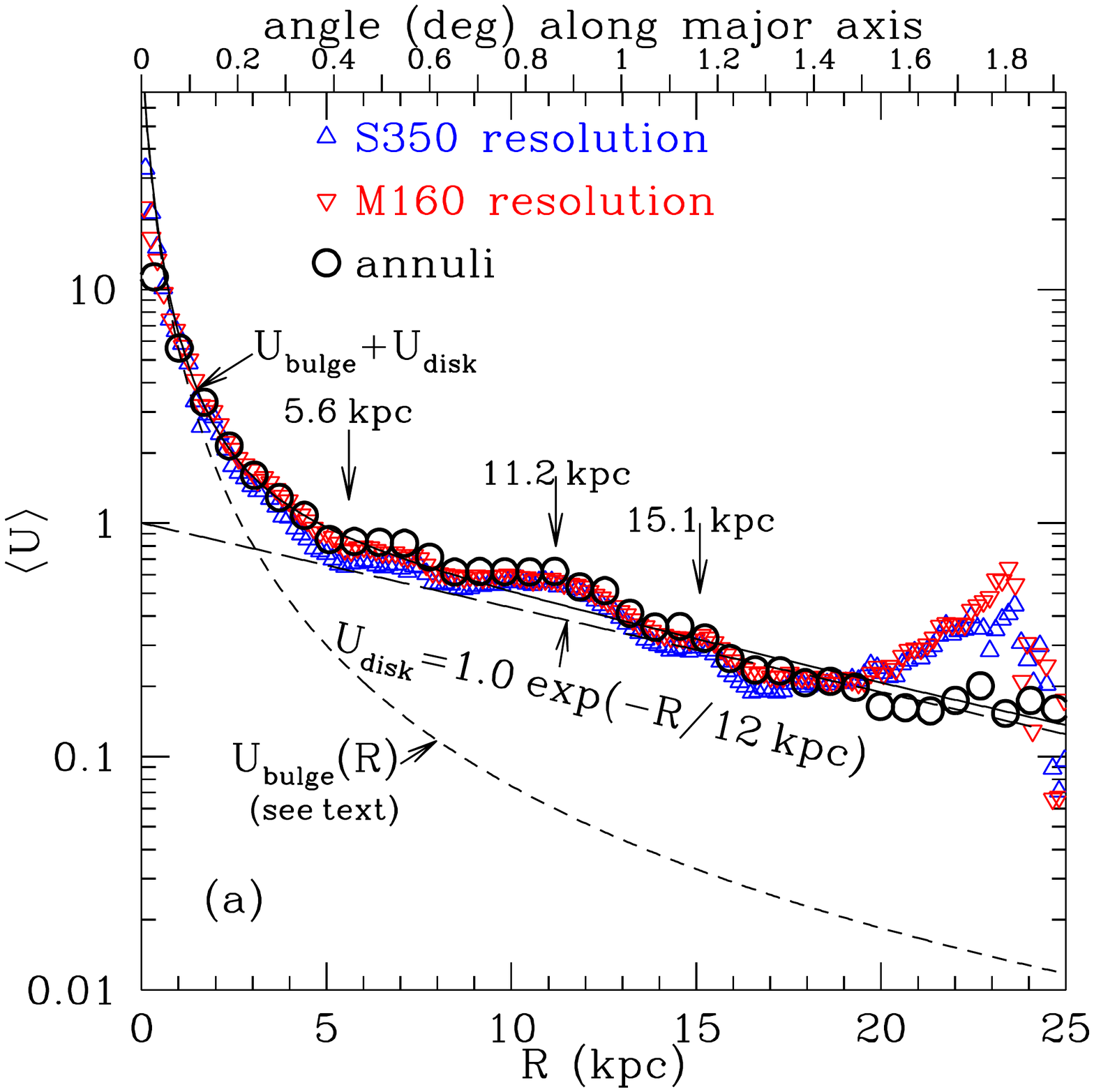}
\includegraphics[width=\figwidth,angle=0,
                 clip=true,trim=0.5cm 4.5cm 0.5cm 2.8cm]
                {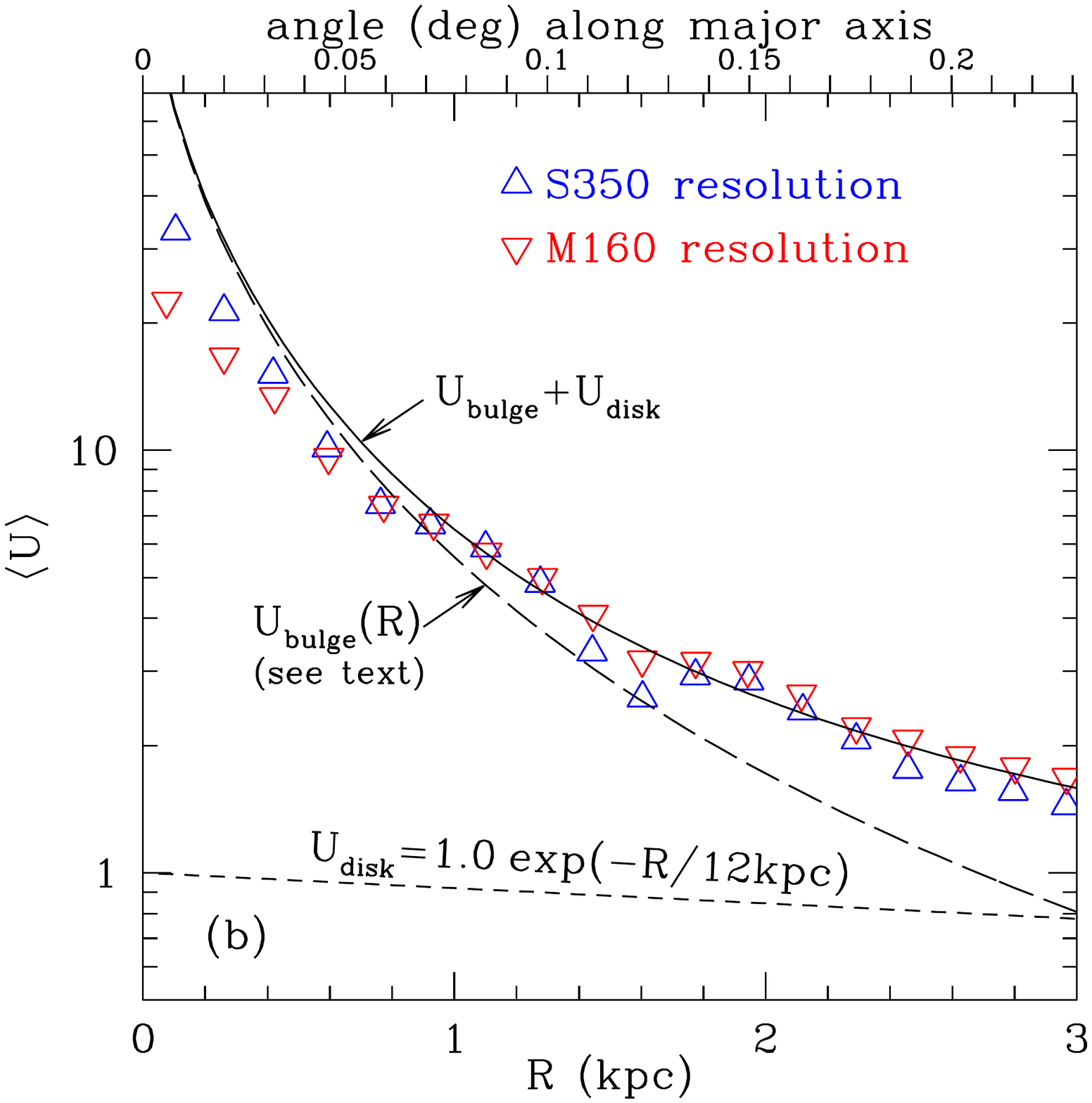}
\caption{\label{fig:Ubar_ctr}
         \footnotesize
         (a) Radial profile of $\langle U\rangle$ in M31
         from modeling the IR SED
         at S350 resolution and at M160 resolution (triangles),
         and using annular photometry (circles).
         The dashed curves show the
         estimated contributions to the heating
         from disk and bulge stars (see text).
         (b) Radial profile of $\langle U\rangle$ in the central 3 kpc.
        }
\end{center}
\end{figure}

The mean starlight heating rate $\langle U\rangle$ is shown in Figure
\ref{fig:Ubar_map}a.
At S350 resolution, the center has $\langle U\rangle\approx 32$
(see Figure \ref{fig:Ubar_map}), corresponding to
$\Tdchar\approx 33\K$.
At $R=15\kpc$ we find $U\approx 0.3$, or $\Tdchar\approx 15\K$,
and at $R=20\kpc$ we find $U\approx 0.25$, or $\Tdchar\approx14\K$.

Near the center, our $\Tdchar\approx33\K$ is
in close agreement with $\Td\approx35\K$ estimated
by \citet{Groves+Krause+Sandstrom+etal_2012},
who modeled the $100-500\micron$ SED using a modified blackbody,
$F_\nu\propto\nu^2 B_\nu(\Td)$ in $\Delta R=230\pc$ annuli for
$0<R<15\kpc$.
At $R=15\kpc$, our $\Tdchar=15\K$ is somewhat higher than the value
$12.5\K$ found by \citet{Groves+Krause+Sandstrom+etal_2012}.

How does the starlight heating rate inferred from the dust IR emission
compare with what is expected?
At $R>5\kpc$, the observed mean heating rate $\langle U\rangle$
in Figure \ref{fig:Ubar_map}(a) is presumably primarily due to starlight
from disk stars.
In the star-forming parts of the disk, the radiation field heating
the dust is from a mixture of young and old stars,
modified by dust attenuation; 
$\langle U\rangle_{\rm disk}$ includes the contribution
of both young and old stars.  Estimating the distribution of
$U$ values seen by the dust would be a challenging radiative transfer
problem even if we knew the three-dimensional
distributions of stars and dust.
The observed $\langle U\rangle$ in Figure \ref{fig:Ubar_ctr}(a)
shows an approximately exponential decline with increasing $R$ out to 20$\kpc$,
which can be approximated by 
\beq \label{eq:Udisk0}
\langle U\rangle_{\rm disk} \approx 1.0 \exp(-R/12\kpc)
~~~.
\eeq
This is plotted in Figures \ref{fig:Ubar_ctr}; we will use it to estimate
the contribution of disk starlight to $\langle U\rangle$
in the central regions.

Near the center of M31, the radiation from the stellar bulge
population becomes dominant.  The distribution of bulge luminosity,
and the resulting energy density of starlight from the bulge, is
discussed by \citet{Groves+Krause+Sandstrom+etal_2012}, and elaborated
further in Appendix \ref{app:bulge}.  From a three-dimensional model for
the stellar bulge, the energy density of bulge
starlight is estimated to be 
\beq \label{eq:ustar_bulge}
u_{\star,{\rm bulge}}^{(0)}(R) = 5.1\times10^{-11} I(R/r_b) \erg\cm^{-3}
~~~,
\eeq
where $I(x)$ is given by Equation (\ref{eq:I(x)_eval}),
$r_b=0.58\kpc$ is the core radius of the bulge
(see Appendix \ref{app:bulge})\footnote{%
   Our estimate for $u_{\star,{\rm bulge}}$ (Equation (\ref{eq:ustar_bulge}))
   is slightly below the value
   $u_{\star,{\rm bulge}}(R)=5.7\times10^{-11} I(R/r_b)\erg\cm^{-3}$
   obtained by
   \citet{Groves+Krause+Sandstrom+etal_2012}.}
and the superscript $(0)$ indicates that it is a theoretical estimate,
with dust extinction neglected.
For the starlight spectrum of the bulge, we estimate
\beq \label{eq:U_bulge}
U_{\rm bulge}^{(0)} = 
\frac{u_{\star,{\rm bulge}}^{(0)}}{1.7\times10^{-12}\erg\cm^{-3}}
= 30\,I(R/r_b)
~~~.
\eeq
Equation (\ref{eq:U_bulge}) gives 
$U_{\rm bulge}^{(0)}(R=1\kpc)=7.9$, whereas our estimated heating rate from
the IR SED at
$R=1\kpc$ is $\langle U\rangle\approx 7.5$ (see Figure \ref{fig:Ubar_ctr}),
which presumably includes a contribution from
disk stars,
which is estimated from Equation (\ref{eq:Udisk0}) to be $\sim0.92$.  Thus
at $R=1\kpc$ we might estimate the heating contribution of the bulge stars to
be $\sim6.5$, 80\% of our theoretical estimate $U_{\rm bulge}^{(0)}$ at
this radius.
At smaller radii we find that $U_{\rm bulge}^{(0)}$ exceeds the
inferred $\langle U\rangle$ by $\sim$$30\%$.

There are a number of possible explanations for the discrepancy between
the theoretical estimate $U_{\rm bulge}^{(0)}$ and $\langle U\rangle$
estimated from modeling the IR SED:
(1) Perhaps the
bulge starlight has simply been overestimated by $\sim$$30\%$.
(2) The dust grains that we are using as ``photometers'' are likely
to be mainly located in clouds distributed in a thin disk; 
internal extinction could lower the starlight heating rate 
below the optically-thin estimate (Equation \ref{eq:U_bulge}).
(3) Some of the dust is presumably above and below the disk plane; thus
the observed emission from the center will include emission from
dust that is actually at larger radii, projected onto the center.
(4) To estimate $\langle U\rangle$ from the infrared observations, 
we have assumed the dust grains to have the
properties of dust in the DL07 model.  The dust near the center of M31
may well differ from the solar-neighborhood dust on which the DL07
is based.
If the dust grains 
at $R\ltsim 2\kpc$ have a {\it lower} ratio of
(optical absorption cross section)/(FIR absorption cross section)
than the DL07 model grains, a given radiation field will
heat them to a lower temperature than the DL07 dust.
A 30\% reduction in the ratio
(optical absorption cross section)/(FIR absorption cross section)
would be sufficient to remove the discrepancy.
Such a reduction in the ratio of optical absorption to FIR emission
cross section would occur if the grain radii in the center were larger
by $\sim30\%$.

To approximate our observed heating rate $\langle U\rangle$, we will
adopt the spatial profile expected for the bulge starlight, but will
scale down the heating rate by a factor $0.7$.  
To this we add our empirical estimate (Equation (\ref{eq:Udisk0}))
for the heating due to the
disk stars.
Thus:
\beqa \label{eq:Umodel}
\langle U\rangle &=& U_{\rm bulge} + \langle U\rangle_{\rm disk}
\\ \label{eq:Ubulge}
U_{\rm bulge} &=& 0.7 U_{\rm bulge}^{(0)} = 21\, I(R/r_b)  ~~~,~~~r_b=0.58\kpc
\\ \label{eq:Udisk}
\langle U\rangle_{\rm disk} &\approx& 1.0\exp(-R/12\kpc)
~~~.
\eeqa
This estimate of $U$ is plotted in Figures \ref{fig:Ubar_map} and
\ref{fig:Ubar_ctr}.
The agreement is good, except at $R<0.4\kpc$ (see Figure \ref{fig:Ubar_ctr})
where the observed dust heating rates fall somewhat
below Equation (\ref{eq:Umodel}).
The deviation at $R< 0.4\kpc$ is likely due to the limited spatial resolution
of the observations.
Even at S350 resolution, the FWHM of the PSF
corresponds to $\Delta R=0.4\kpc$ along the minor axis.
In addition, as already noted above,
out-of-plane dust, projected along the line-of-sight, can also
act to lower the ``observed'' dust temperatures near the center.


Using modified blackbody fits,
\citet{Groves+Krause+Sandstrom+etal_2012} found dust
temperatures in the center that are $\sim$15\% lower than the $T_{\rm d,char}$
values found here.  Their lower dust temperature estimates
may be due to use of an earlier
version of the data reduction pipeline and calibration factors, and
possibly also to use of a modified blackbody rather than the
multicomponent physical grain model used here.

To summarize, we find relatively 
good agreement between the observed dust
emission spectrum and that which would be expected for heating by
a combination of the bulge starlight and a disk heating component.
The observed dust temperatures are
only slightly below what would be expected.
The inferred heating rate from the bulge stars is only $\sim$30\% lower than
predicted for a simple model of the bulge starlight.
The 30\% discrepancy in heating rate corresponds to only a 5\%
discrepancy in grain temperature.
Given that there are a number of effects that could account for such
a discrepancy, this agreement is gratifying.

The dust temperature $T$ depends on the starlight heating rate parameter $U$
and on 
the ratio $\langle C_{\rm abs}\rangle_\star(a)/\langle C_{\rm abs}(a)\rangle_T$,
where $\langle C_{\rm abs}(a)\rangle_\star$ is the
dust absorption cross section averaged over the spectrum of the
illuminating starlight for a grain of radius $a$, 
and $\langle C_{\rm abs}(a)\rangle_T$
is the Planck-averaged absorption cross section for grain temperature $T$.
The fact that the observed dust temperature is within 5\% of the
predicted dust temperature indicates that the actual values of
$\langle C_{\rm abs}\rangle_\star/\langle C_{\rm abs}\rangle_T$
are close to the values in the DL07 grain model.
This builds confidence in our grain model, and hence in the
dust mass estimates, which
are proportional to $\rho a/\langle C_{\rm abs}(a)\rangle_T$.
Given the extreme environmental differences,
it is remarkable that the dust properties near the center of
M31 appear to be so similar to 
values inferred for dust in the solar neighborhood.

\section{PAH Abundance
         \label{sec:pah}}

The intensity in the IRAC bands includes both direct starlight
and emission from dust population.
The observed intensities at $\lambda \geq 3.6\micron$
are modeled as the sum of a stellar
component (modeled as a 5000K blackbody)
plus a nonstellar component $(F_\nu)_{\rm ns}$ contributed primarily
by PAHs.
In practice, the stellar component is determined by the IRAC3.6 and
IRAC4.5 photometry; subtraction of the stellar component typically
leaves a positive ``nonstellar'' residual in IRAC5.8 and IRAC8.0 that can
be reproduced by varying the PAH abundance in the dust model.

The parameter $\qpah$ in the DL07 model is defined to be
the fraction of the
total dust mass contributed by PAHs containing fewer than $10^3$ C atoms.
In the model fitting, $\qpah$ is essentially proportional to the ratio of
the power in the 6.2 and 7.7$\micron$ PAH emission features 
divided by the total power radiated by the dust.  
When only IRAC photometry is available for $\lambda < 20\micron$,
the $\qpah$ parameter is, in practice, 
proportional to the nonstellar contribution to
IRAC8.0 divided by the aggregate 70--350$\micron$ luminosity.
For a fixed starlight spectrum, $F(8\micron)$ is 
approximately proportional to $\qpah$ and to the total infrared power,
$F_{\rm TIR}$:
\beq \label{eq:qpah_estimator}
(\nu F_\nu)_{\rm IRAC8,ns}\approx A_\star \qpah F_{\rm TIR}
~~~,
\eeq
where $(\nu F_\nu)_{\rm IRAC8,ns}$ is the non-stellar contribution to
the IRAC 8.0$\micron$ band,
and the dimensionless coefficient 
$A_\star$ depends on the spectrum of the illuminating starlight.
\citet{Draine+Li_2007} show that
$A_{\rm MMP83} \approx 4.72$ for the MMP83 spectrum of the starlight in the
solar neighborhood.

Varying the spectrum of the starlight can
cause $A_\star$ to change, for two reasons.
(1) The PAH absorption
cross section depends on wavelength differently from the
absorption 
of overall grain mixture ($\propto F_{\rm TIR}$), hence
the fraction of the starlight power that is absorbed by PAHs 
will depend on the
spectrum.
(2) The PAH emission spectrum is the result of single-photon heating,
hence the fraction of energy absorbed by PAHs that is reradiated in 
the IRAC8.0 band depends on the illuminating spectrum.
As an example of the dependence of $A_\star$ on the illuminating spectrum,
\citet{Draine_2011b} 
showed that $A_{20{\rm kK}}\approx 7.1 = 1.5A_{\rm MMP83}$
for a $20$kK blackbody cut off at $13.6\eV$.

Draine \& Li (2014, in preparation) 
calculated the emission from the DL07 dust model
for illumination by starlight with the spectrum of the M31 bulge
population, finding
$A_{\rm bulge}\approx 1.95$.
For a mixed spectrum, where the overall dust heating rate is
$U=U_{\rm bulge}+U_{\rm MMP83}$, the effective value is 
the dust luminosity-weighted mean
\beq \label{eq:Astar}
A_\star = 
\frac{A_{\rm bulge}U_{\rm bulge}+A_{\rm MMP83}U_{\rm MMP83}}
     {U_{\rm bulge} + U_{\rm MMP83}}
=
\frac{1.95 U_{\rm bulge}+4.72 U_{\rm MMP83}}
               {U_{\rm bulge}+U_{\rm MMP83}}
~~~.
\eeq
To correct the estimate of $\qpah$ made assuming the MMP83 radiation field,
we will take
\beq \label{eq:qpahcorr}
\left(\qpah\right)_{\rm corr}
= 
\frac{4.72}{A_\star} \times \left(\qpah\right)_{\rm MMP83}
\eeq
where $(\qpah)_{\rm MMP83}$ is the value of $\qpah$ estimated assuming
the dust to be heated by starlight with the MMP83 spectrum, with
$U_{\rm bulge}$ and $U_{\rm MMP83}$ given by Equations (\ref{eq:Ubulge})
and (\ref{eq:Udisk}).

\begin{figure}[t]
\begin{center}
\includegraphics[width=\figwidth,angle=0,
                 clip=true,trim=0.6cm 0.6cm 1.6cm 8.1cm]
                {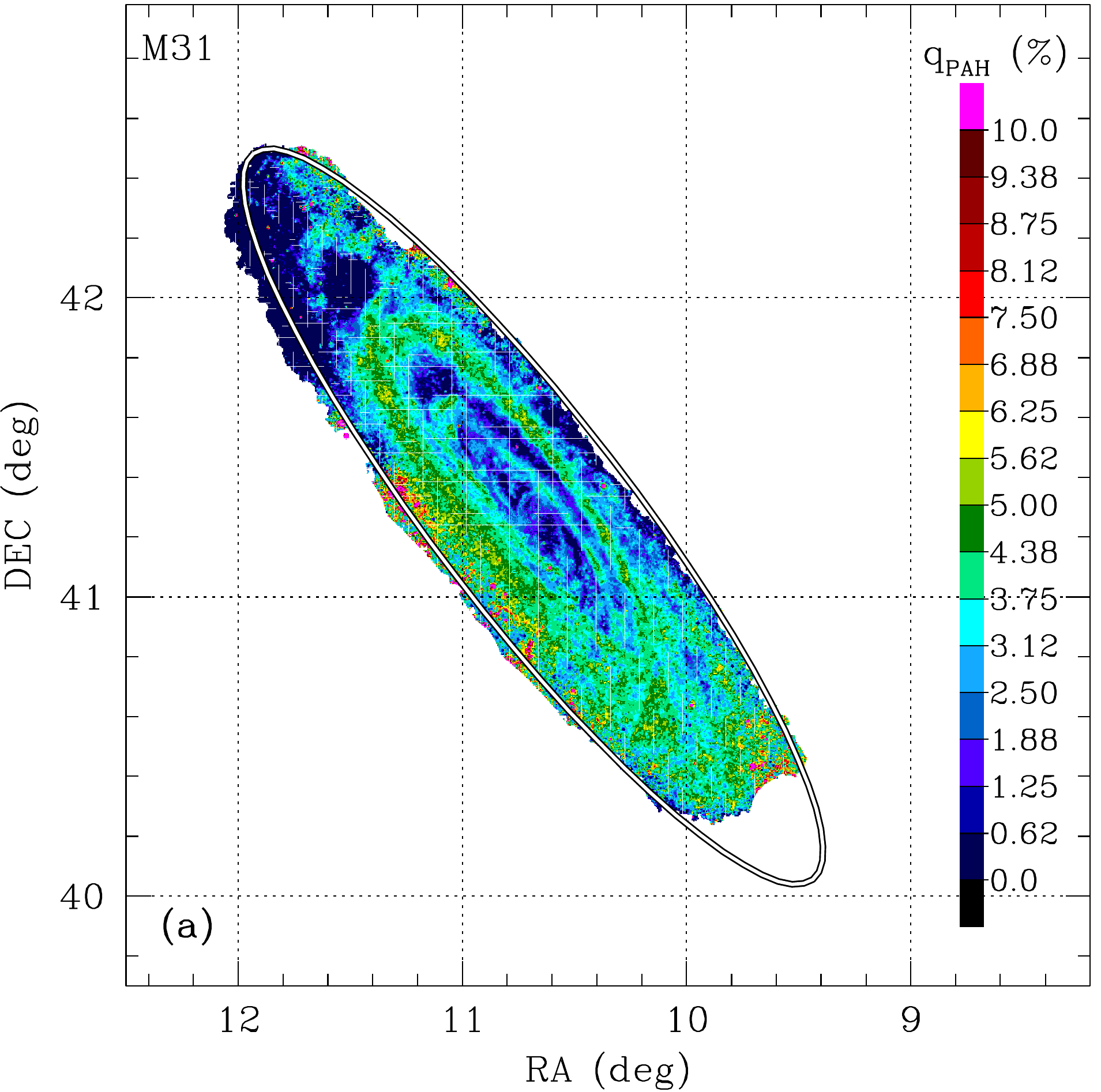}
\includegraphics[width=\figwidth,angle=0,
                 clip=true,trim=0.5cm 4.5cm 0.5cm 2.8cm]
                {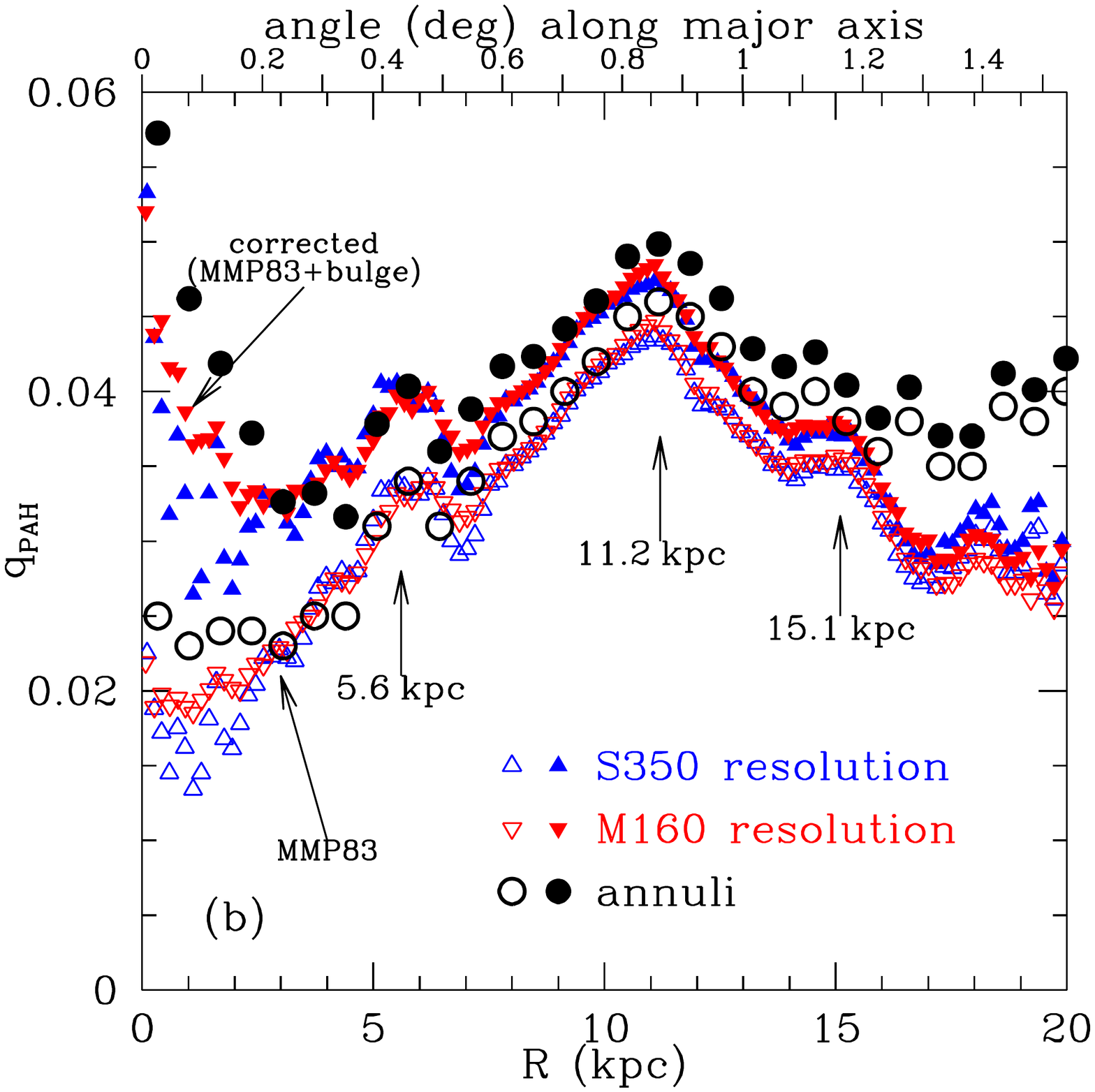}
\caption{\label{fig:qpah_map}
         \footnotesize
         (a) Map of PAH abundance parameter $(q_{\rm PAH})_{\rm MMP83}$ 
         in M31 at S350 resolution.
         The low values of $\qpah$ in the NE may be due to
         problems with background subtraction in IRAC5.8 and IRAC8.0
         (see text).
         (b) Radial profiles: $(\qpah)_{\rm MMP83}$ (open symbols)
         and $(\qpah)_{\rm corr}$ from
         Equation (\ref{eq:qpahcorr}) (filled symbols).
        }
\end{center}
\end{figure}

A map of $(\qpah)_{\rm MMP83}$ at 
S350 resolution is shown in Figure \ref{fig:qpah_map}(a).
Figure \ref{fig:qpah_map}(b) shows the radial profile of $(\qpah)_{\rm MMP83}$
and $(\qpah)_{\rm corr}$.
Interestingly, $\qpah$ appears to peak in the 11\,kpc ring, attaining a value
$(\qpah)_{\rm corr}\approx 0.049$ that is close to the value
$\qpah\approx 0.045$ estimated
for the diffuse ISM in the solar neighborhood.

In Figure \ref{fig:qpah_map},
$(\qpah)_{\rm MMP83}$ declines as one approaches the center,
reaching a value $(\qpah)_{\rm MMP83}\approx 0.02$ in the central regions.
However, this decline is largely an artifact of assuming that the
spectrum of the illuminating starlight is independent of radius, which
is incorrect -- in the central regions the starlight is dominated by
light from the bulge stars, which is much redder than the MMP83 spectrum.

Figure \ref{fig:qpah_map}(b) shows that when we allow for the starlight
being increasingly dominated by an old stellar population as we move
to the center, $(\qpah)_{\rm corr}$ shows only limited variation as
we move from $R=11\kpc$ to the central kpc.
Evidently the balance between PAH formation and destruction in the ISM
remains relatively constant from the central kpc out to $R=20\kpc$.
There does appear to be a systematic radial decline in $\qpah$ for
$R>11\kpc$, but this again might be an artifact of a radial gradient in the
spectrum of the starlight as one moves from the 11.2\,kpc ring -- where
star formation is active -- to outer regions where there appears to be
little contemporary star formation.

At $R\gtsim20\kpc$ $\qpah$ estimated from the $\Delta R=677\pc$ annuli
appears to rise.
However, we suspect this to be an artifact of imperfect background subtraction
in the $5.8$ and $8.0\micron$ images.
Although background subtraction works well in other bands out to 
$\sim$$25\kpc$, it appears less successful for the IRAC 5.8$\micron$
and $8.0\micron$ bands.
The difficulty with background subtraction may be due to systematic
effects on the IRAC detectors, including an effect referred to as
``banding'', multiplexor ``bleeding'', and scattered light
\citep{Hora+Fazio+Allen+etal_2004}.
The SAGE-SMC survey \citep{Gordon+Meixner+Meade+etal_2011}
was able to minimize these problems by
combining images taken with very different roll angles together
with custom processing techniques, but we are simply using the
M31 images from \citet{Barmby+Ashby+Bianchi+etal_2006}.
It is also possible that the Galactic foreground has structure arising from 
variations in PAH abundance or ionization state
on $\sim$$0.5^\circ$ scales, which would not be identified by the
background subtraction procedures used here.

\section{Dust Properties
         \label{sec:dust_properties}}

In the present work, we attempt to reproduce the observed SED in
each pixel using the DL07 dust model and a parameterized distribution
of starlight intensities.  Above we have examined the values of the
dust modeling parameters, such as the dust mass, $\qpah$, and properties
of the starlight intensity distribution.  Here we compare the models
with observations to see how well the model actually reproduces the
data, and whether any systematic deviations are present that
indicate systematic problems with the modeling.

\begin{figure}[t]
\begin{center}
\includegraphics[width=\figwidth,angle=0,
                 clip=true,trim=0.6cm 0.6cm 1.6cm 8.1cm]
                {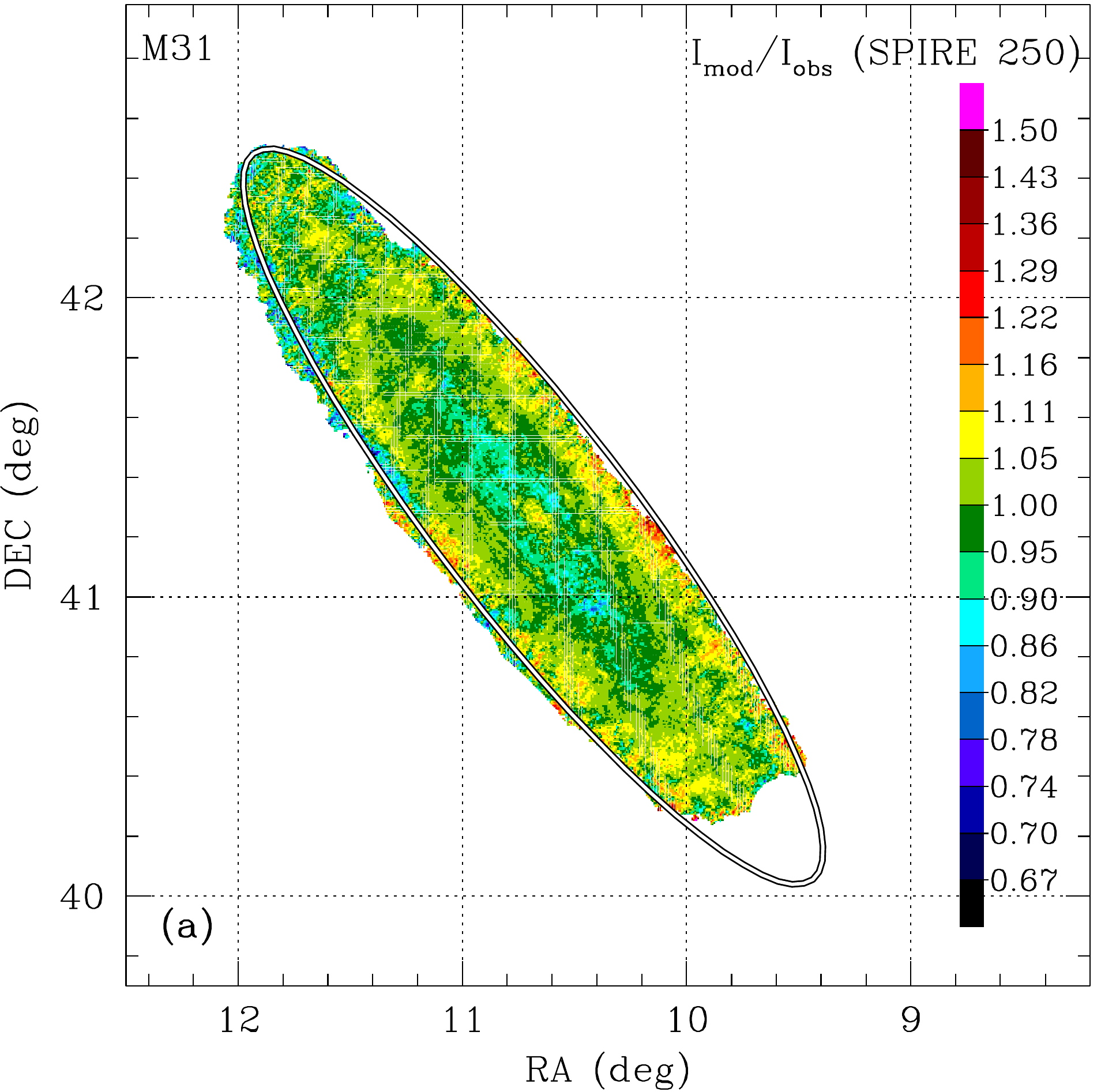}
\includegraphics[width=\figwidth,angle=0,
                 clip=true,trim=0.6cm 0.6cm 1.6cm 8.1cm]
                {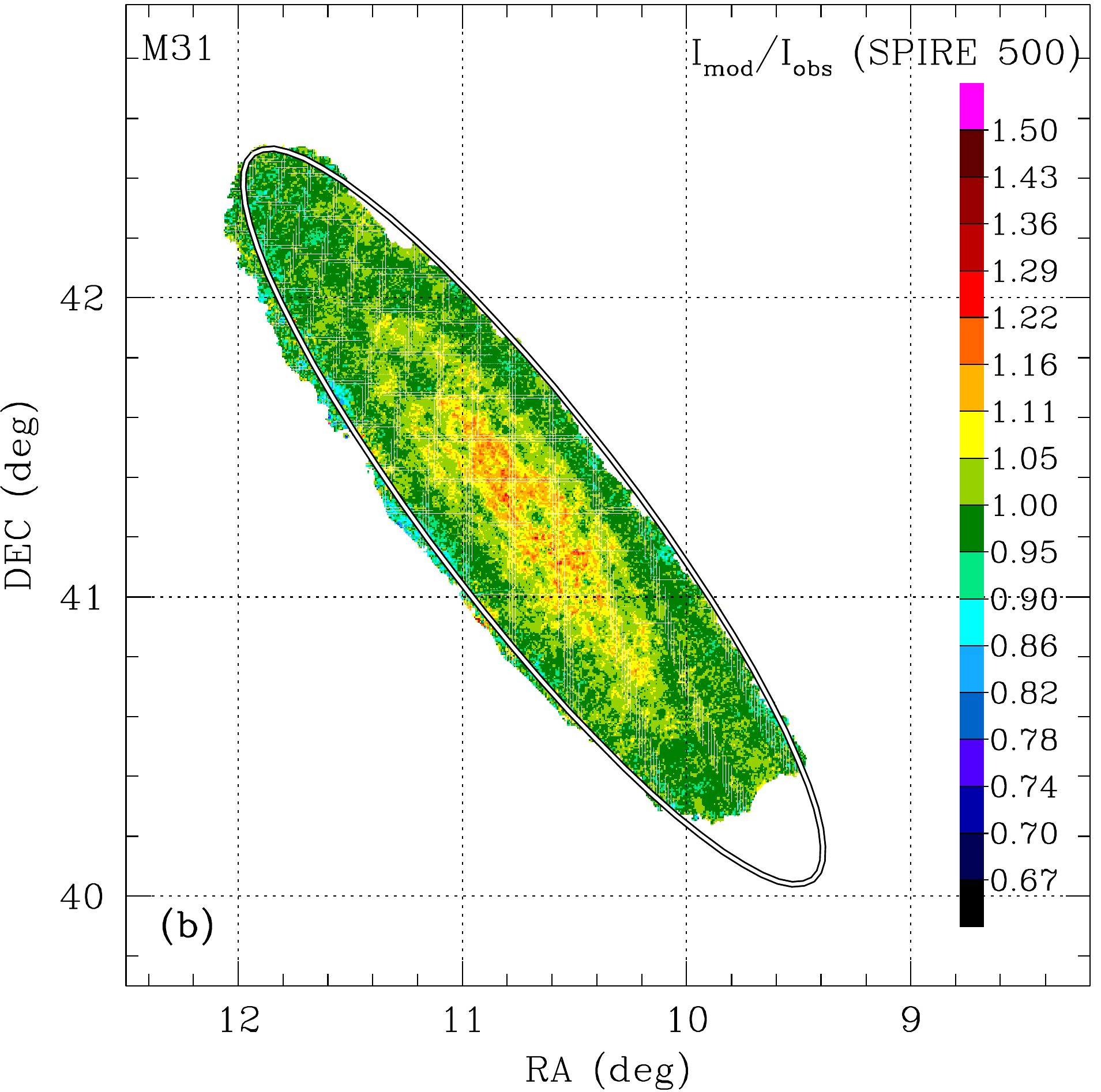}
\caption{\label{fig:S500_ratio}
         \footnotesize
         For M160 resolution modeling:
         ratio of the model intensity divided by the observed
         intensity for the SPIRE $250$ and $500\micron$ bands.
         The DL07 model successfully fits the $500\micron$ emission out
         to the edge of the ``galaxy mask'', at $R\approx20\kpc$.
         In the inner regions, the DL07 model tends to 
         underpredict SPIRE $250\micron$ and overpredict
         SPIRE $500\micron$ (see also Figure \ref{fig:S500/S250}),
         but the deviations are only at the $\sim$10\% level.
                  }
\end{center}
\end{figure}

\begin{figure}[t]
\begin{center}
\includegraphics[width=\figwidth,angle=0,
                 clip=true,trim=0.5cm 4.5cm 0.5cm 2.8cm]
                {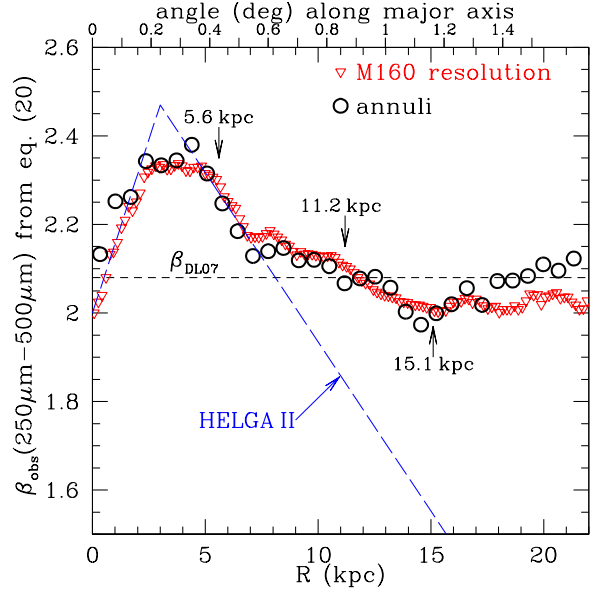}
\caption{\label{fig:S500/S250}
         \footnotesize
         Opacity spectral index
         $\beta_{\rm obs}$ from Equation (\ref{eq:Delta_beta})
         obtained from modeling at M160 resolution (triangles) and
         using annular photometry (circles).
         The DL07 model 
         overpredicts the SPIRE500/SPIRE250 band
         ratio by a factor $> 1.2$ between $R\approx2\kpc$ and $R\approx6\kpc$,
         corresponding to $\beta_{\rm obs}> \beta_{\rm DL07} + 0.26$.
         For $R\gtsim10\kpc$ the DL07 model agrees with the observations to
         within 10\%, and the model appears to have nearly the correct
         value of $\beta$.
         Also shown is the
         HELGA\,II \citep{Smith+Eales+Gomez+etal_2012} result for
         $\beta$ (see text).
         }
\end{center}
\end{figure}

Aside from minor effects associated with variation in the PAH abundance
when $\qpah$ is allowed to vary,
the composition (amorphous silicate and graphitic carbon)
is assumed to be the same everywhere
in the DL07 models used to fit the infrared emission.
Further, the dust opacity is assumed to be
independent of temperature.
The DL07 dust model used here has an opacity with a fixed dependence
on frequency at long wavelengths ($\lambda \gtsim 50\micron$).
Here we compare the model to the observed emission from M31
to look for residuals that might be indicative of differences between
the model and the actual dust in M31.

We will focus on the behavior of the dust opacity in the 250--500$\micron$
region.
Figure \ref{fig:S500_ratio} shows the ratio of model to observation
at SPIRE250 and SPIRE500.
The first impression is that the model is good: the ratio of model to
observation is generally between 0.86 and 1.16, which seems good in view
of noise in the observations, uncertainties in calibration, and
the general uncertainties in the adopted dust opacities.
However, systematic
trends are evident: in the central regions of M31 (excluding the
center itself) the model tends to
be low at SPIRE250, and high at SPIRE500.

Let $\beta\equiv \ln\left[\kappa(250\micron)/\kappa(500\micron)\right]/\ln 2$ be
the effective power-law index of the dust opacity between 250 and 500$\micron$.
For the DL07 model (see Table \ref{tab:renormalized_DL07})
this ratio is $\beta_{\rm DL07}=2.08$.
If the fitted dust temperatures were left unchanged, the
500/250 flux ratio could be brought into agreement with observations 
if $\beta$ were changed to
\beq
\label{eq:Delta_beta}
\beta_{\rm obs} = \beta_{\rm DL07} + 
\frac{\ln\left([I_\nu(250\micron)/I_\nu(500\micron)]_{\rm obs}/
         [I_\nu(250\micron)/I_\nu(500\micron)]_{\rm model}\right)}
     {\ln(500/250)}
~~~.
\eeq
Figure \ref{fig:S500/S250} shows $\beta_{\rm obs}$ as a function of
$R$ for $R<22\kpc$, using only M160 resolution pixels with
$\SigLd>\SigLdmin$.
For $1 < R < 10\kpc$, $\beta_{\rm obs}$ is larger than
$\beta_{\rm DL07}$, indicating that the opacity between 250 and 500$\micron$
should fall more rapidly than $\nu^{2.08}$.

The dust opacity ratio $\kappa(250\micron)/\kappa(500\micron)$
could vary with location because the dust composition is varying,
or it could conceivably result from variations of the grain temperature,
if the dust opacities are temperature-dependent.
The DL07 model assumes the dust opacities to be independent of
temperature.  However, some materials do exhibit temperature-dependent
opacities in the laboratory
\citep[e.g.,][]{Mennella+Brucato+Colangeli+etal_1998,
                Boudet+Mutschke+Nayral+etal_2005,
                Coupeaud+Demyk+Meny+etal_2011},
and such behavior is expected in some models of amorphous
solids \citep{Meny+Gromov+Boudet+etal_2007}.
A number of studies have claimed that interstellar dust opacities
are temperature-dependent
\citep{Dupac+Bernard+Boudet+etal_2003,
       Paradis+Veneziani+Noriega-Crespo+etal_2010,
       Paradis+Bernard+Meny+Gromov_2011,
       Liang+Fixsen+Gold_2012}
although apparent $\beta - T$
correlations 
can arise from both observational noise
and line-of-sight
temperature variations
\citep{Shetty+Kauffmann+Schnee+Goodman_2009,
       Shetty+Kauffmann+Schnee+etal_2009,
       Kelly+Shetty+Stutz+etal_2012}.
The ``two level system'' (TLS) model 
\citep{Meny+Gromov+Boudet+etal_2007,Paradis+Bernard+Meny+Gromov_2011} 
predicts that increasing dust temperature $T$ should lead to a
{\it lower} value of $\beta$.  However, in M31 it appears
that the 1--10\,kpc regions where the dust is warmer than the outer
disk have a {\it larger} value of $\beta$
than the value in the outer disk, and the center -- where the dust is
hottest -- has essentially the same value of $\beta$ as the relatively
cold dust at $R\approx20\kpc$.  The observed variations in $\beta$
do not seem to be consistent with what would be expected for the
TLS model, unless one allows for substantial radial variations in the
TLS model parameters themselves.
Rather than attributing the apparent changes in $\beta$ to temperature,
it would appear instead that the dust composition must be varying
with radius in M31.

The HELGA\,II collaboration \citep{Smith+Eales+Gomez+etal_2012} reported
a sharp change in the dust properties at $R\approx3\kpc$, based on the
value of the dust emissivity index $\beta$ obtained from their modified
blackbody fits.
Interior to $3\kpc$, the inferred dust temperature decreased from
$T=27.5\K$ at the center to $T=16.8\K$ at $R=3.0\kpc$, 
with $\beta$ simultaneously rising from $2$ to 2.5; beyond $R=3\kpc$,
they found the temperature to be rising and $\beta$ falling with 
increasing $R$, reaching $T=18\K$ and $\beta=1.5$ at $R=15\kpc$.
The decrease in $\beta$ at large $R$ was invoked to account for an
apparent ``500$\micron$ excess''.

The HELGA\,II result for $\beta$, shown in Figure \ref{fig:S500/S250},
is similar to the present study for $R \ltsim 7\kpc$, but we find
very different behavior for $R\gtsim 7\kpc$.
We see no evidence of a 500$\micron$ excess at large $R$ -- quite the
contrary, the DL07 model tends to slightly {\it over}predict SPIRE500
in the outer regions of M31.

The HELGA\,II analysis was based on observations of M31 that were shallower
than the data used here, and also
used an earlier calibration of the three SPIRE bands.  It is possible that
the differences between the HELGA\,II results and those of the present
study may be due in part to differences in signal/noise,
calibration or data reduction
pipeline.  
In the
outer regions results are also sensitive to background
subtraction.

If we were over- or under-subtracting the IR backgrounds,
we would obtain very low or very high dust/gas ratios.
The fact that we obtain sensibly-behaved dust/H ratios out to $R=25\kpc$
(see Figures \ref{fig:dust_to_gas_pixels} and
\ref{fig:dust_to_gas_vs_R})
is evidence that the
automatic background subtraction algorithm used here 
\citep{Aniano+Draine+Calzetti+etal_2012} -- which is based only on the
IR imaging, and makes no use of \ion{H}{1}~21\,cm or CO emission --
is working well, at least out to $R=25\kpc$.

Our overall conclusion is that the DL07 dust properties appear to
provide a generally good match to the observations of M31, except for
the need for a modest increase in the opacity index $\beta$
in the 2--6$\kpc$ region.  It would be of interest to apply this same
approach to other galaxies (e.g., M33) to see if similar
variations in $\beta_{\rm obs}$ are found.

\section{Summary
         \label{sec:summary}}

The principal conclusions of this work are as follows:
\begin{enumerate}
\item Consistent with previous observations, 
we find that the dust mass surface density in M31 peaks in two rings, at
$R=5.6\kpc$ and $R=11.2\kpc$, with a third ring seen at $R\approx15.1\kpc$.

\item We find a total dust mass $\Md=(5.4\pm1.1)\times10^7\Msol$
within $R=25\kpc$.
95\% of this dust mass lies within $R=21\kpc$, with the dust
surface density peaking at $R=11.2\kpc$.

\item The dust/H mass ratio exhibits a smooth radial decline with
increasing $R$, from $\sim$$0.027$ at the center to $\sim$$0.0027$ at
$R=25\kpc$.

\item The dust/H mass ratio 
      parallels measurements of O/H in \ion{H}{2} regions,
      consistent with a constant fraction of the refractory elements
      Mg, Si, and Fe being in dust.
      Based on our estimated dust/H mass ratio, we infer that the metallicity
      $Z/Z_\odot$ varies from $\sim$$3$ at $R=0$ to $\sim$$0.3$ at $R=25\kpc$ --
      see Figure \ref{fig:dust_to_gas_vs_R}(b) and 
      Equation (\ref{eq:Z_vs_R}).

\item The starlight heating rate parameter $\langle U\rangle$ shows a
nearly monotonic decline with galactocentric radius, from
$\langle U\rangle\approx 50$ at the center (at S350 resolution) to
$\langle U\rangle\approx 0.2$ at $R\approx 20\kpc$.

\item We confirm the finding of \citet{Groves+Krause+Sandstrom+etal_2012}
      that the dust heating in the central 2~kpc is dominated by
      light from the stellar bulge.
      We find that the starlight heating rates 
      inferred from the observed IR emission 
      are consistent with the heating rates
      expected from the bulge starlight.
      It is remarkable that the dust properties near the center of
      M31 appear to be similar to the properties of dust in the
      solar neighborhood.

\item After taking into account variation in the spectrum of the
starlight, we find the PAH abundance $\qpah$ to be
approximately constant from the center of M31 out to $\sim$20\,kpc.
There is some indication of decline with $R$ for $R>11\kpc$.
The global value of $\qpah=0.039$ for $R<17\kpc$.

\item While the DL07 dust model generally provides a good fit to the
observed SED, there are some systematic deviations.
The
dust at $R\approx 1-6\kpc$ appears to have an opacity
spectral index $2.2 < \beta <  2.33$ in the 250--500$\micron$ wavelength
range, whereas the dust at $R>7\kpc$ has $\beta\approx2.08$, consistent
with the dust in the DL07 model.
We are in approximate agreement with the
radial variation of $\beta$ found by 
HELGA\,II \citep{Smith+Eales+Gomez+etal_2012} in the central $\sim$$7\kpc$,
but do not confirm their finding of low $\beta$ values for
$R > 8\kpc$.

\item At large radii $R\gtsim10\kpc$ the DL07 model, with
fixed opacity, is consistent with the observed photometry, and
returns dust masses that are consistent with the observations of
the gas and metallicity.

\end{enumerate}

\acknowledgements
We thank
Edvige Corbelli,
\newtext{Stephen Eales,}
\newnewtext{Jeremy Goodman,}
\newtext{Matthew Smith,
and
the anonymous referee}
for helpful comments.
This work was supported in part by NSF grant AST 1008570.
G.A.\ acknowledges support from European Research Council grant
ERC-267934.

\begin{appendix}

\section{\label{app:PACS_vs_MIPS}
         PACS versus MIPS}
\begin{figure}[t]
\begin{center}
\includegraphics[width=\figwidth,angle=0,
                 clip=true,trim=0.6cm 0.6cm 1.6cm 8.1cm]
                {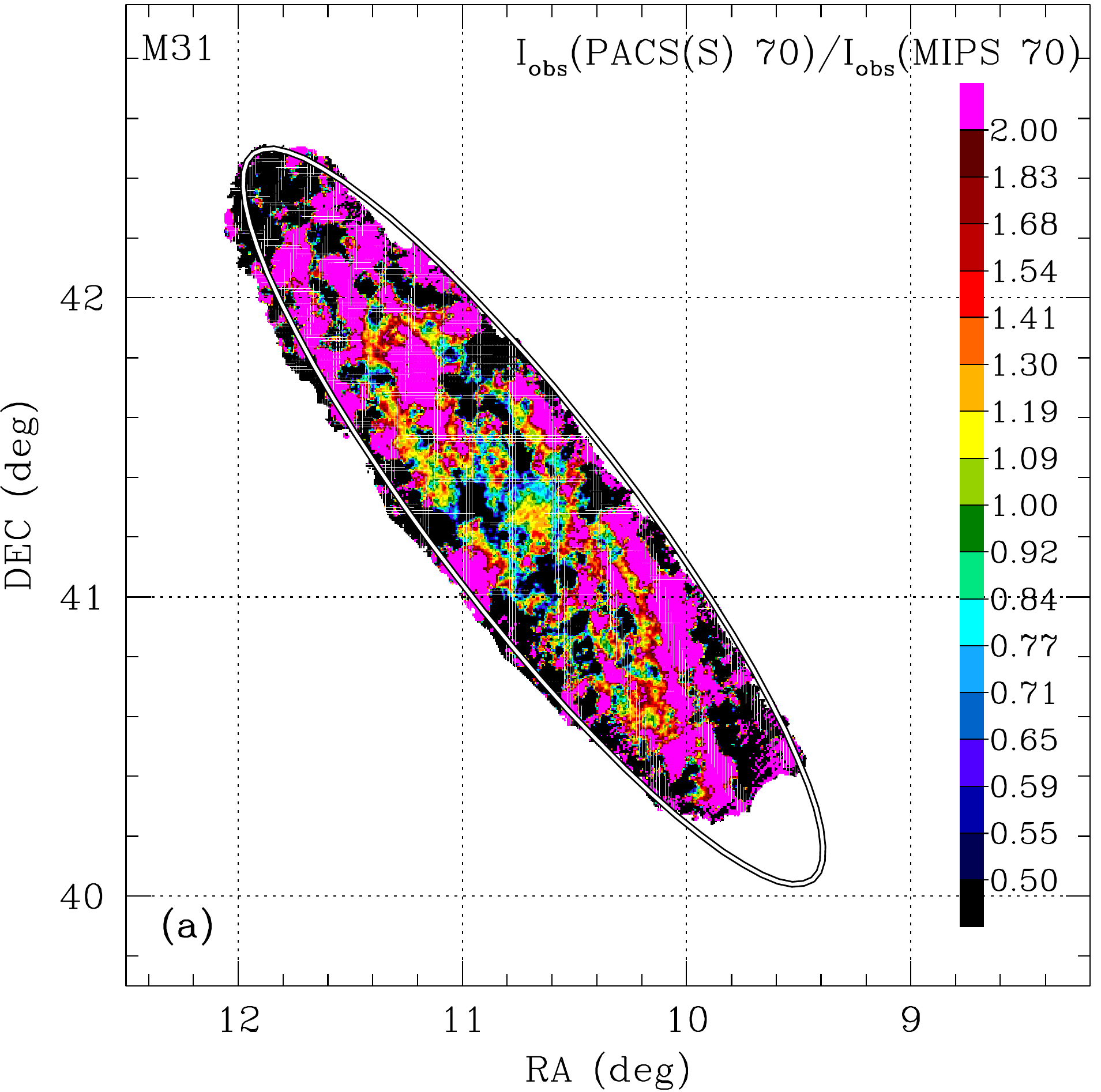}
\includegraphics[width=\figwidth,angle=0,
                 clip=true,trim=0.6cm 0.6cm 1.6cm 8.1cm]
                {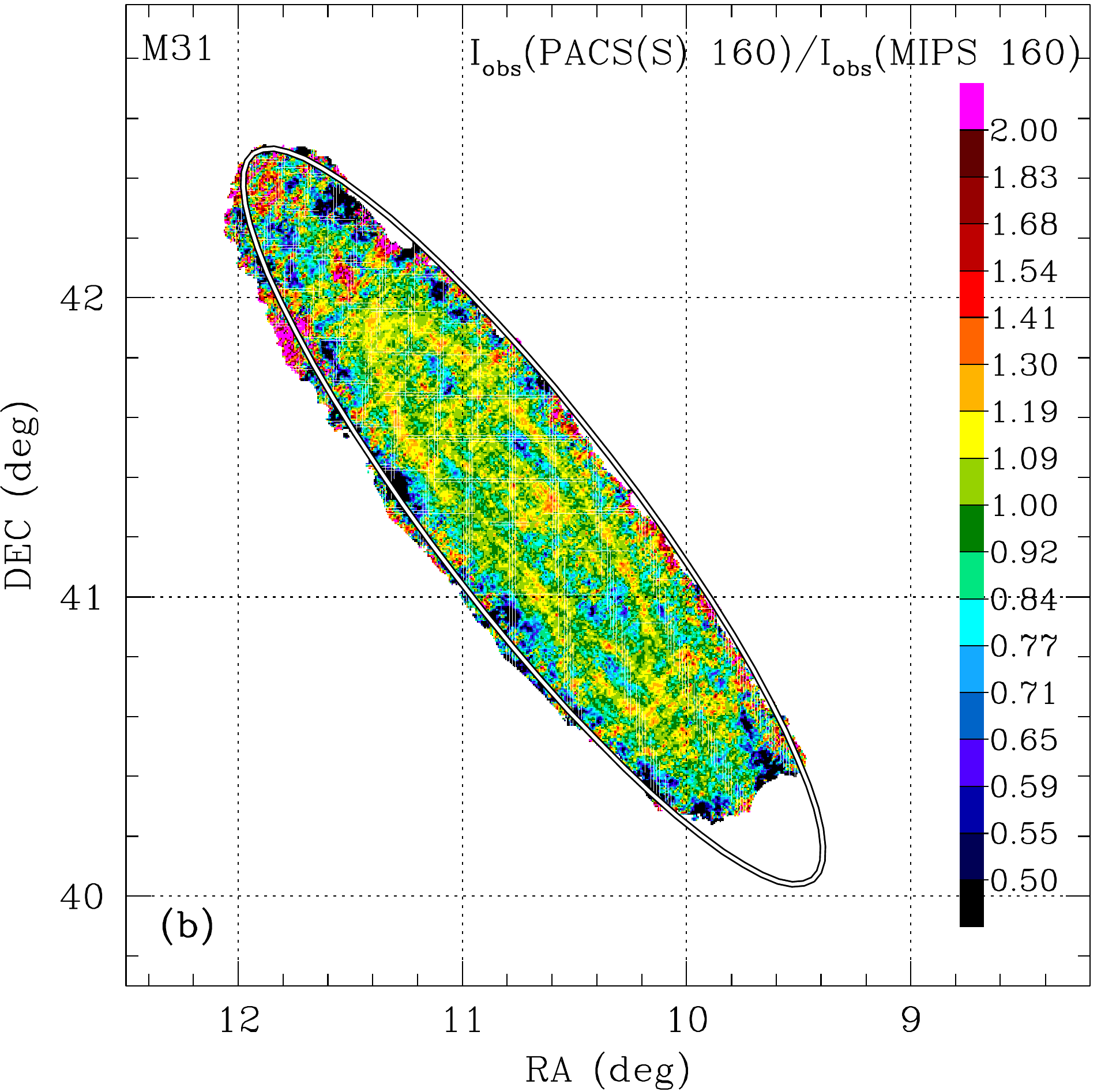}
\caption{\label{fig:PACS_vs_MIPS}
         \footnotesize
         Ratios of images convolved to the MIPS160 PSF, displayed
         within the ``galaxy mask'' where the signal/noise is high
         enough that dust modeling is possible.
         Left: ratio of PACS70/MIPS70.
         Right: ratio of PACS160/MIPS160.
         }
\end{center}
\end{figure}
MIPS and PACS have two wavelengths in common: $70\micron$ and $160\micron$.
While the filter response functions are not identical, they are similar
enough that we would expect MIPS and PACS to measure very similar
flux densities for smooth SEDs.
However, comparisons of MIPS and PACS imaging often shows discrepancies
that are much larger than expected --
see, e.g., the cases of NGC~628 and NGC~6946
\citep{Aniano+Draine+Calzetti+etal_2012}.

Figure \ref{fig:PACS_vs_MIPS} shows the ratio of PACS/MIPS photometry
of M31 in the 70 and 160$\micron$ bands, 
after first convolving each image to the common
resolution of the MIPS160 PSF.

The PACS70/MIPS70 comparison is particularly striking: while
there are a few regions near the center with PACS70/MIPS70 $\sim$0.7,
many regions have PACS70/MIPS70$>2$.  The fact that this occurs
over extended regions makes it clear that this is not simply a consequence
of random noise.
The fact that it is non-uniform over the image makes it clear that
it is not simply a result of calibration.
Background subtraction is of critical importance, but we have employed the
same background estimation procedures 
\citep{Aniano+Draine+Calzetti+etal_2012}
for both MIPS and PACS.
It is now thought that MIPS70 suffers from sub-linear behavior in high-surface
brightness regions, but such high surface brightnesses are found only
at the center of M31, whereas we see high values of PACS70/MIPS70
occurring across the disk.
The reason for the photometric discrepancy
is unclear, and we therefore opt to use both MIPS70 and PACS70 data in
our model-fitting.

The PACS160/MIPS160 comparison is much more satisfactory than the
PACS70/MIPS70 comparison, but there are still many regions -- particularly
along the $11\kpc$ ring -- where PACS160/MIPS160 $>1.4$.
This difference is much larger than the claimed uncertainties in
the observed intensities at these locations.
MIPS160 is now thought to have sublinear response for
$I_\nu\gtsim50\MJy\sr^{-1}$
\citep{Paladini+Linz+Altieri+Ali_2012}, 
but most of M31 is below this value.
This again emphasizes the value of being able to include the MIPS160 data
in the analysis, possible only if one degrades all the other
imaging to the MIPS160 PSF (FWHM $39\arcsec$).

\section{Heating by Bulge Starlight
         \label{app:bulge}}
\citet{Groves+Krause+Sandstrom+etal_2012} have discussed the contribution
of M31's bulge stars to the heating of dust.
For the adopted bulge stellar luminosity density profile
\citep{Geehan+Fardal+Babul+Guhathakurta_2006}
\beq
\rho_\star(R) = \frac{L_{\rm bulge}}{2\pi r_b^3}
\frac{1}{(R/r_b)(1+R/r_b)^3}
~~~,
\eeq
the starlight energy density due to bulge stars 
(neglecting possible absorption or scattering by dust) is
\beqa
u_{\star,\rm bulge}(R) &=& \frac{L_{\rm bulge}}{4\pi r_b^2 c} ~ I(R/r_b)
\\ \label{eq:I(x)_def}
I(x) &\equiv& \frac{1}{x}\int_0^\infty 
\frac{\ln\left(\frac{x+y}{|x-y|}\right)}{(1+y)^3}dy
\\ \label{eq:I(x)_eval}
&=& \frac{1}{x^2-1} + \frac{\ln(1+x)}{2x(1+x)^2} - \frac{2\ln x}{(x^2-1)^2}
~~~.
\eeqa
$I(x)$ is logarithmically divergent for $x\rightarrow 0$.
Near $x=1$,
\beq
I(x=1+\epsilon) = \frac{1}{2} + \frac{\ln2}{8} -0.778(x-1) + O((x-1)^2)
~~~.
\eeq
The projected luminosity interior to radius $R$ is
\beqa
L(<R) &=& \int_0^R \rho_\star(r) 4\pi r^2 dr
+ \int_R^\infty \rho_\star(r) \left[1-\sqrt{1-(R/r)^2}\right]
4\pi r^2 dr
\\
&=&
L_{\rm bulge}
\left[1 - 
2\alpha\int_0^{\pi/2}\frac{\cos^2\theta d\theta}{(1+\alpha\sin\theta)^3}
\right] 
~~~~~~\alpha\equiv \frac{r_b}{R}
\\
&=& 0.4711 L_{\rm bulge} ~~~~~{\rm for}~\alpha=0.61
\eeqa
We obtain $L_{\rm bulge}$ by integrating 
the average of the intrinsic and reddened
spectrum obtained by \citet{Groves+Krause+Sandstrom+etal_2012}
for $L(< 1\kpc)$ over the wavelength interval $[0.0912\micron,4\micron]$.
\citet{Groves+Krause+Sandstrom+etal_2012} assumed $D=780\kpc$ and
$r_b=0.61\kpc$.
Corrected to $D=744\kpc$, we obtain
$L_{\rm bulge}(0.0912-4\micron)
=1.60\times10^{10}\Lsol$,
with $r_b=0.58\kpc$.  Thus, the 
$0.0912\micron < \lambda < 4\micron$ energy density due to the
bulge stars is
\beq
u_{\star,{\rm bulge}}(R) = 5.07\times10^{-11} I(R/r_b) \erg\cm^{-3}
~~~.
\eeq
This energy density of bulge starlight gives a theoretical estimate
for the dust heating rate,
normalized to the heating rate for the MMP83 radiation field,
\beq
U_{\rm bulge}^{(0)}(R) = 30 I(R/r_b)
~~~.
\eeq
For $R=1\kpc$ this gives $U_{\rm bulge}^{(0)}(1\kpc) = 7.9$.
This is our zero-th order estimate for the dust heating rate
in the central few kpc of M31.
In Section \ref{sec:starlight} we show that this theoretical estimate
for the dust heating rate agrees well with the dust heating rate
inferred from the IR SED.

\end{appendix}

\bibliography{/u/draine/work/bib/btdrefs}

\begin{thebibliography}{}

\bibitem[\protect\citeauthoryear{{Aniano} et~al.}{{Aniano}
  et~al.}{2012}]{Aniano+Draine+Calzetti+etal_2012}
{Aniano}, G., {Draine}, B.~T., {Calzetti}, D., et~al. 2012, \apj, 756, 46

\bibitem[\protect\citeauthoryear{{Aniano} et~al.}{{Aniano}
  et~al.}{2011}]{Aniano+Draine+Gordon+Sandstrom_2011}
{Aniano}, G., {Draine}, B.~T., {Gordon}, K.~D.,  \& {Sandstrom}, K.~M. 2011,
  \pasp, 123, 1218

\bibitem[\protect\citeauthoryear{{Barmby} et~al.}{{Barmby}
  et~al.}{2006}]{Barmby+Ashby+Bianchi+etal_2006}
{Barmby}, P., {Ashby}, M.~L.~N., {Bianchi}, L., et~al. 2006, \apjl, 650, L45

\bibitem[\protect\citeauthoryear{{Block} et~al.}{{Block}
  et~al.}{2006}]{Block+Bournaud+Combes+etal_2006}
{Block}, D.~L., {Bournaud}, F., {Combes}, F., et~al. 2006, \nature, 443, 832

\bibitem[\protect\citeauthoryear{{Bogd{\'a}n} \& {Gilfanov}}{{Bogd{\'a}n} \&
  {Gilfanov}}{2008}]{Bogdan+Gilfanov_2008}
{Bogd{\'a}n}, {\'A}.,  \& {Gilfanov}, M. 2008, \mnras, 388, 56

\bibitem[\protect\citeauthoryear{{Bolatto}, {Wolfire}, \& {Leroy}}{{Bolatto}
  et~al.}{2013}]{Bolatto+Wolfire+Leroy_2013}
{Bolatto}, A.~D., {Wolfire}, M.,  \& {Leroy}, A.~K. 2013, \araa, 51, 207

\bibitem[\protect\citeauthoryear{{Boudet} et~al.}{{Boudet}
  et~al.}{2005}]{Boudet+Mutschke+Nayral+etal_2005}
{Boudet}, N., {Mutschke}, H., {Nayral}, C., et~al. 2005, \apj, 633, 272

\bibitem[\protect\citeauthoryear{{Braun} et~al.}{{Braun}
  et~al.}{2009}]{Braun+Thilker+Walterbos+Corbelli_2009}
{Braun}, R., {Thilker}, D.~A., {Walterbos}, R.~A.~M.,  \& {Corbelli}, E. 2009,
  \apj, 695, 937

\bibitem[\protect\citeauthoryear{{Chemin}, {Carignan}, \& {Foster}}{{Chemin}
  et~al.}{2009}]{Chemin+Carignan+Foster_2009}
{Chemin}, L., {Carignan}, C.,  \& {Foster}, T. 2009, \apj, 705, 1395

\bibitem[\protect\citeauthoryear{{Ciardullo} et~al.}{{Ciardullo}
  et~al.}{1988}]{Ciardullo+Rubin+Jacoby+etal_1988}
{Ciardullo}, R., {Rubin}, V.~C., {Jacoby}, G.~H., {Ford}, H.~C.,  \& {Ford},
  W.~K., Jr. 1988, \aj, 95, 438

\bibitem[\protect\citeauthoryear{{Corbelli} et~al.}{{Corbelli}
  et~al.}{2010}]{Corbelli+Lorenzoni+Walterbos+etal_2010}
{Corbelli}, E., {Lorenzoni}, S., {Walterbos}, R., {Braun}, R.,  \& {Thilker},
  D. 2010, \aap, 511, A89

\bibitem[\protect\citeauthoryear{{Coupeaud} et~al.}{{Coupeaud}
  et~al.}{2011}]{Coupeaud+Demyk+Meny+etal_2011}
{Coupeaud}, A., {Demyk}, K., {Meny}, C., et~al. 2011, \aap, 535, A124

\bibitem[\protect\citeauthoryear{{Crane}, {Dickel}, \& {Cowan}}{{Crane}
  et~al.}{1992}]{Crane+Dickel+Cowan_1992}
{Crane}, P.~C., {Dickel}, J.~R.,  \& {Cowan}, J.~J. 1992, \apjl, 390, L9

\bibitem[\protect\citeauthoryear{{Dalcanton} et~al.}{{Dalcanton}
  et~al.}{2012}]{Dalcanton+Williams+Lang+etal_2012}
{Dalcanton}, J.~J., {Williams}, B.~F., {Lang}, D., et~al. 2012, \apjs, 200, 18

\bibitem[\protect\citeauthoryear{{de Vaucouleurs} et~al.}{{de Vaucouleurs}
  et~al.}{1991}]{deVaucoleurs+deVaucoleurs+Corwin+etal_1991}
{de Vaucouleurs}, G., {de Vaucouleurs}, A., {Corwin}, H.~G., Jr., et~al. 1991,
  {Third Reference Catalogue of Bright Galaxies} (New York: Springer)

\bibitem[\protect\citeauthoryear{{Devereux} et~al.}{{Devereux}
  et~al.}{1994}]{Devereux+Price+Wells+Duric_1994}
{Devereux}, N.~A., {Price}, R., {Wells}, L.~A.,  \& {Duric}, N. 1994, \aj, 108,
  1667

\bibitem[\protect\citeauthoryear{{Draine}}{{Draine}}{2003}]{Draine_2003a}
{Draine}, B.~T. 2003, \araa, 41, 241

\bibitem[\protect\citeauthoryear{{Draine}}{{Draine}}{2011a}]{Draine_2011b}
{Draine}, B.~T. 2011a, in EAS Publications Series, Vol.~46, PAHs and the
  Universe, ed. C.~{Joblin} \& A.~G.~G.~M. {Tielens}, 29

\bibitem[\protect\citeauthoryear{{Draine}}{{Draine}}{2011b}]{Draine_2011a}
{Draine}, B.~T. 2011b, {Physics of the Interstellar and Intergalactic Medium}
  (Princeton, NJ: Princeton Univ.\ Press)

\bibitem[\protect\citeauthoryear{{Draine} et~al.}{{Draine}
  et~al.}{2007}]{Draine+Dale+Bendo+etal_2007}
{Draine}, B.~T., {Dale}, D.~A., {Bendo}, G., et~al. 2007, \apj, 663, 866

\bibitem[\protect\citeauthoryear{{Draine} \& {Li}}{{Draine} \&
  {Li}}{2001}]{Draine+Li_2001}
{Draine}, B.~T.,  \& {Li}, A. 2001, \apj, 551, 807

\bibitem[\protect\citeauthoryear{{Draine} \& {Li}}{{Draine} \&
  {Li}}{2007}]{Draine+Li_2007}
{Draine}, B.~T.,  \& {Li}, A. 2007, \apj, 657, 810

\bibitem[\protect\citeauthoryear{{Dupac} et~al.}{{Dupac}
  et~al.}{2003}]{Dupac+Bernard+Boudet+etal_2003}
{Dupac}, X., {Bernard}, J.-P., {Boudet}, N., et~al. 2003, \aap, 404, L11

\bibitem[\protect\citeauthoryear{{Fazio} et~al.}{{Fazio}
  et~al.}{2004}]{Fazio+Hora+Allen+etal_2004}
{Fazio}, G.~G., {Hora}, J.~L., {Allen}, L.~E., et~al. 2004, \apjs, 154, 10

\bibitem[\protect\citeauthoryear{{Fritz} et~al.}{{Fritz}
  et~al.}{2012}]{Fritz+Gentile+Smith+etal_2012}
{Fritz}, J., {Gentile}, G., {Smith}, M.~W.~L., et~al. 2012, \aap, 546, A34

\bibitem[\protect\citeauthoryear{{Geehan} et~al.}{{Geehan}
  et~al.}{2006}]{Geehan+Fardal+Babul+Guhathakurta_2006}
{Geehan}, J.~J., {Fardal}, M.~A., {Babul}, A.,  \& {Guhathakurta}, P. 2006,
  \mnras, 366, 996

\bibitem[\protect\citeauthoryear{{Gordon} et~al.}{{Gordon}
  et~al.}{2006}]{Gordon+Bailin+Engelbracht+etal_2006}
{Gordon}, K.~D., {Bailin}, J., {Engelbracht}, C.~W., et~al. 2006, \apjl, 638,
  L87

\bibitem[\protect\citeauthoryear{{Gordon} et~al.}{{Gordon}
  et~al.}{2011}]{Gordon+Meixner+Meade+etal_2011}
{Gordon}, K.~D., {Meixner}, M., {Meade}, M.~R., et~al. 2011, \aj, 142, 102

\bibitem[\protect\citeauthoryear{{Griffin} et~al.}{{Griffin}
  et~al.}{2010}]{Griffin+Abergel+Abreu+etal_2010}
{Griffin}, M.~J., {Abergel}, A., {Abreu}, A., et~al. 2010, \aap, 518, L3

\bibitem[\protect\citeauthoryear{{Griffin} et~al.}{{Griffin}
  et~al.}{2013}]{Griffin+North+Schulz+etal_2013}
{Griffin}, M.~J., {North}, C.~E., {Schulz}, B., et~al. 2013, \mnras, 1306.1778

\bibitem[\protect\citeauthoryear{{Groves} et~al.}{{Groves}
  et~al.}{2012}]{Groves+Krause+Sandstrom+etal_2012}
{Groves}, B., {Krause}, O., {Sandstrom}, K., et~al. 2012, \mnras, 426, 892

\bibitem[\protect\citeauthoryear{{Haas} et~al.}{{Haas}
  et~al.}{1998}]{Haas+Lemke+Stickel+etal_1998}
{Haas}, M., {Lemke}, D., {Stickel}, M., et~al. 1998, \aap, 338, L33

\bibitem[\protect\citeauthoryear{{Habing} et~al.}{{Habing}
  et~al.}{1984}]{Habing+Miley+Young+etal_1984}
{Habing}, H.~J., {Miley}, G., {Young}, E., et~al. 1984, \apjl, 278, L59

\bibitem[\protect\citeauthoryear{{Hora} et~al.}{{Hora}
  et~al.}{2004}]{Hora+Fazio+Allen+etal_2004}
{Hora}, J.~L., {Fazio}, G.~G., {Allen}, L.~E., et~al. 2004, in Society of
  Photo-Optical Instrumentation Engineers (SPIE) Conference Series, Vol. 5487,
  77

\bibitem[\protect\citeauthoryear{{Kelly} et~al.}{{Kelly}
  et~al.}{2012}]{Kelly+Shetty+Stutz+etal_2012}
{Kelly}, B.~C., {Shetty}, R., {Stutz}, A.~M., et~al. 2012, \apj, 752, 55

\bibitem[\protect\citeauthoryear{{Kennicutt} et~al.}{{Kennicutt}
  et~al.}{2011}]{Kennicutt+Calzetti+Aniano+etal_2011}
{Kennicutt}, R.~C., {Calzetti}, D., {Aniano}, G., et~al. 2011, \pasp, 123, 1347

\bibitem[\protect\citeauthoryear{{Leroy} et~al.}{{Leroy}
  et~al.}{2011}]{Leroy+Bolatto+Gordon+etal_2011}
{Leroy}, A.~K., {Bolatto}, A., {Gordon}, K., et~al. 2011, \apj, 737, 12

\bibitem[\protect\citeauthoryear{{Lewis} et~al.}{{Lewis}
  et~al.}{2013}]{Lewis+Braun+McConnachie+etal_2013}
{Lewis}, G.~F., {Braun}, R., {McConnachie}, A.~W., et~al. 2013, \apj, 763, 4

\bibitem[\protect\citeauthoryear{{Li} \& {Draine}}{{Li} \&
  {Draine}}{2001}]{Li+Draine_2001b}
{Li}, A.,  \& {Draine}, B.~T. 2001, \apj, 554, 778

\bibitem[\protect\citeauthoryear{{Liang}, {Fixsen}, \& {Gold}}{{Liang}
  et~al.}{2012}]{Liang+Fixsen+Gold_2012}
{Liang}, Z., {Fixsen}, D.~J.,  \& {Gold}, B. 2012, submitted to \mnras
  (arXiv:1201.0060)

\bibitem[\protect\citeauthoryear{{Mathis}, {Mezger}, \& {Panagia}}{{Mathis}
  et~al.}{1983}]{Mathis+Mezger+Panagia_1983}
{Mathis}, J.~S., {Mezger}, P.~G.,  \& {Panagia}, N. 1983, \aap, 128, 212

\bibitem[\protect\citeauthoryear{{Mennella} et~al.}{{Mennella}
  et~al.}{1998}]{Mennella+Brucato+Colangeli+etal_1998}
{Mennella}, V., {Brucato}, J.~R., {Colangeli}, L., et~al. 1998, \apj, 496, 1058

\bibitem[\protect\citeauthoryear{{Meny} et~al.}{{Meny}
  et~al.}{2007}]{Meny+Gromov+Boudet+etal_2007}
{Meny}, C., {Gromov}, V., {Boudet}, N., et~al. 2007, \aap, 468, 171

\bibitem[\protect\citeauthoryear{{Neugent}, {Massey}, \& {Georgy}}{{Neugent}
  et~al.}{2012}]{Neugent+Massey+Georgy_2012}
{Neugent}, K.~F., {Massey}, P.,  \& {Georgy}, C. 2012, \apj, 759, 11

\bibitem[\protect\citeauthoryear{{Nieten} et~al.}{{Nieten}
  et~al.}{2006}]{Nieten+Neininger+Guelin+etal_2006}
{Nieten}, C., {Neininger}, N., {Gu{\'e}lin}, M., et~al. 2006, \aap, 453, 459

\bibitem[\protect\citeauthoryear{{Paladini} et~al.}{{Paladini}
  et~al.}{2012}]{Paladini+Linz+Altieri+Ali_2012}
{Paladini}, R., {Linz}, H., {Altieri}, B.,  \& {Ali}, B. 2012, PACS ICC
  Document, PICC-NHSC-TR-034

\bibitem[\protect\citeauthoryear{{Paradis} et~al.}{{Paradis}
  et~al.}{2011}]{Paradis+Bernard+Meny+Gromov_2011}
{Paradis}, D., {Bernard}, J.~P., {M{\'e}ny}, C.,  \& {Gromov}, V. 2011, \aap,
  534, A118

\bibitem[\protect\citeauthoryear{{Paradis} et~al.}{{Paradis}
  et~al.}{2010}]{Paradis+Veneziani+Noriega-Crespo+etal_2010}
{Paradis}, D., {Veneziani}, M., {Noriega-Crespo}, A., et~al. 2010, \aap, 520,
  L8

\bibitem[\protect\citeauthoryear{{Pilbratt} et~al.}{{Pilbratt}
  et~al.}{2010}]{Pilbratt+Riedinger+Passvogel+etal_2010}
{Pilbratt}, G.~L., {Riedinger}, J.~R., {Passvogel}, T., et~al. 2010, \aap, 518,
  L1

\bibitem[\protect\citeauthoryear{{Planck Collaboration} et~al.}{{Planck
  Collaboration} et~al.}{2013a}]{Planck_dust_gal_plane_2013}
{Planck Collaboration}, {Ade}, P.~A.~R., {Aghanim}, N., et~al. 2013a, ArXiv
  e-prints, 1307.6815

\bibitem[\protect\citeauthoryear{{Planck Collaboration} et~al.}{{Planck
  Collaboration} et~al.}{2013b}]{Planck_overview_2013}
{Planck Collaboration}, {Ade}, P.~A.~R., {Aghanim}, N., et~al. 2013b,
  arXiv:1303.5062, 1303.5062

\bibitem[\protect\citeauthoryear{{Planck Collaboration} et~al.}{{Planck
  Collaboration} et~al.}{2013c}]{Planck_component_sep_2013}
{Planck Collaboration}, {Ade}, P.~A.~R., {Aghanim}, N., et~al. 2013c, ArXiv
  e-prints, 1303.5072

\bibitem[\protect\citeauthoryear{{Planck Collaboration} et~al.}{{Planck
  Collaboration} et~al.}{2011}]{Planck_dust_2011}
{Planck Collaboration}, {Ade}, P.~A.~R., {Aghanim}, N., et~al. 2011, \aap, 536,
  A19

\bibitem[\protect\citeauthoryear{{Poglitsch} et~al.}{{Poglitsch}
  et~al.}{2010}]{Poglitsch+Waelkens+Geis+etal_2010}
{Poglitsch}, A., {Waelkens}, C., {Geis}, N., et~al. 2010, \aap, 518, L2

\bibitem[\protect\citeauthoryear{{Popescu} \& {Tuffs}}{{Popescu} \&
  {Tuffs}}{2013}]{Popescu+Tuffs_2013}
{Popescu}, C.~C.,  \& {Tuffs}, R.~J. 2013, ArXiv e-prints, 1305.0232

\bibitem[\protect\citeauthoryear{{Reach} et~al.}{{Reach}
  et~al.}{2005}]{Reach+Megeath+Cohen+etal_2005}
{Reach}, W.~T., {Megeath}, S.~T., {Cohen}, M., et~al. 2005, \pasp, 117, 978

\bibitem[\protect\citeauthoryear{{Rieke} et~al.}{{Rieke}
  et~al.}{2004}]{Rieke+Young+Engelbracht+etal_2004}
{Rieke}, G.~H., {Young}, E.~T., {Engelbracht}, C.~W., et~al. 2004, \apjs, 154,
  25

\bibitem[\protect\citeauthoryear{{Roussel}}{{Roussel}}{2013}]{Roussel_2013}
{Roussel}, H. 2013, \pasp, 125, 1126

\bibitem[\protect\citeauthoryear{{Shetty} et~al.}{{Shetty}
  et~al.}{2009a}]{Shetty+Kauffmann+Schnee+Goodman_2009}
{Shetty}, R., {Kauffmann}, J., {Schnee}, S.,  \& {Goodman}, A.~A. 2009a, \apj,
  696, 676

\bibitem[\protect\citeauthoryear{{Shetty} et~al.}{{Shetty}
  et~al.}{2009b}]{Shetty+Kauffmann+Schnee+etal_2009}
{Shetty}, R., {Kauffmann}, J., {Schnee}, S., {Goodman}, A.~A.,  \& {Ercolano},
  B. 2009b, \apj, 696, 2234

\bibitem[\protect\citeauthoryear{{Smith} et~al.}{{Smith}
  et~al.}{2012}]{Smith+Eales+Gomez+etal_2012}
{Smith}, M.~W.~L., {Eales}, S.~A., {Gomez}, H.~L., et~al. 2012, \apj, 756, 40

\bibitem[\protect\citeauthoryear{{Tabatabaei} \& {Berkhuijsen}}{{Tabatabaei} \&
  {Berkhuijsen}}{2010}]{Tabatabaei+Berkhuijsen_2010}
{Tabatabaei}, F.~S.,  \& {Berkhuijsen}, E.~M. 2010, \aap, 517, A77

\bibitem[\protect\citeauthoryear{{Vilardell} et~al.}{{Vilardell}
  et~al.}{2010}]{Vilardell+Ribas+Jerdi+etal_2010}
{Vilardell}, F., {Ribas}, I., {Jordi}, C., {Fitzpatrick}, E.~L.,  \& {Guinan},
  E.~F. 2010, \aap, 509, A70

\bibitem[\protect\citeauthoryear{{Weingartner} \& {Draine}}{{Weingartner} \&
  {Draine}}{2001}]{Weingartner+Draine_2001a}
{Weingartner}, J.~C.,  \& {Draine}, B.~T. 2001, \apj, 548, 296

\bibitem[\protect\citeauthoryear{{Werner} et~al.}{{Werner}
  et~al.}{2004}]{Werner+Roellig+Low+etal_2004}
{Werner}, M.~W., {Roellig}, T.~L., {Low}, F.~J., et~al. 2004, \apjs, 154, 1

\bibitem[\protect\citeauthoryear{{Xu} \& {Helou}}{{Xu} \&
  {Helou}}{1996}]{Xu+Helou_1996}
{Xu}, C.,  \& {Helou}, G. 1996, \apj, 456, 163

\bibitem[\protect\citeauthoryear{{Zurita} \& {Bresolin}}{{Zurita} \&
  {Bresolin}}{2012}]{Zurita+Bresolin_2012}
{Zurita}, A.,  \& {Bresolin}, F. 2012, \mnras, 427, 1463

\end{thebibliography}

\end{document}